\documentclass[12pt]{article}
\pdfoutput=1
\usepackage{jheppub}
\usepackage{amsmath}
\usepackage{array}

\usepackage{multirow}
\usepackage{subfigure}
\usepackage{longtable}
\usepackage{arydshln}
\usepackage{afterpage}

\allowdisplaybreaks 

\newcommand{\smallminus}{{\rm\rule[2.4pt]{6pt}{0.65pt}}}

\newcommand{\mi}{\, \smallminus\,}

\newcommand{\la}{\langle}
\newcommand{\ra}{\rangle}

\def\NeqFour{{\cal N} = 4}

\def\fig#1{Fig.~\ref{fig:#1}}

\def\sect#1{Sect.~\ref{sec:#1}}
\def\subsect#1{Sect.~\ref{subsec:#1}}
\def\eqn#1{Eq.~(\ref{eqn:#1})}
\def\Eqn#1{Eq.~(\ref{eqn:#1})}

\def\tab#1{Table~\ref{tab:#1}}
\def\Tab#1{Table~\ref{tab:#1}}

\def\dlog{d {\rm log}}

\def\la{\langle}
\def\ra{\rangle}
\def\PT{{\rm PT}}

\def\lra{\leftrightarrow}
\def\f{{\tilde f}}

\newcommand{\ab}[1]{\langle #1 \rangle}     
\newcommand{\sqb}[1]{[ #1 ]}
\newcommand{\aMs}[3]{\langle #1|#2|#3]}  		

\newcommand{\sab}[1]{s_{#1}}

\newcommand{\lam}[1]{\lambda_{#1} }
\newcommand{\lamt}[1]{\widetilde \lambda_{#1} }

\newcommand{\twhite}[1]{\textcolor{white}{#1}}

\title{Evidence for a Nonplanar Amplituhedron}

\author{Zvi Bern$^1$,}
\author{Enrico Herrmann$^2$,}
\author{Sean Litsey$^1$,}
\author{James Stankowicz$^1$,}
\author{Jaroslav Trnka$^{2,3}$}

\affiliation{$^1$ Department of Physics and Astronomy, UCLA, Los
Angeles, CA 90095}
\affiliation{$^2$ Walter Burke Institute for Theoretical Physics, \\
California Institute of Technology, Pasadena, CA 91125}
\affiliation{$^3$ Center for Quantum Mathematics and Physics (QMAP),\\ 
Department of Physics, University of California, Davis, CA 95616}

\emailAdd{bern@physics.ucla.edu, eherrmann@caltech.edu, slitsey@ucla.edu,
jjstankowicz@ucla.edu, trnka@ucdavis.edu}

\abstract{The scattering amplitudes of planar $\NeqFour$
  super-Yang-Mills exhibit a number of remarkable analytic structures,
  including dual conformal symmetry and logarithmic singularities of
  integrands.  The amplituhedron is a geometric construction of the
  integrand that incorporates these structures.  This geometric
  construction further implies the amplitude is fully specified by
  constraining it to vanish on spurious residues.  By writing the
  amplitude in a $\dlog$ basis, we provide nontrivial evidence that
  these analytic properties and ``zero conditions'' carry over into
  the nonplanar sector.  This suggests that the concept of the
  amplituhedron can be extended to the nonplanar sector of
  $\NeqFour$ super-Yang-Mills theory. 
	}

\preprint{
\begin{flushright}UCLA/15/TEP/103 \hfill  CALT-TH-2015-064 \end{flushright}
}

\begin{document}
\setcounter{tocdepth}{2}
\maketitle

\section{Introduction}

Recent years have seen an enormous advance in our understanding of
scattering amplitudes in $\mathcal{N}=4$ super-Yang-Mills (SYM)
theory.  Most progress has been made in the planar sector, with
many calculations of new amplitudes with large numbers of loops and
legs both at the integrand and integrated levels now available.  Discoveries of
new structures and symmetries have led to the
development of deep theoretical frameworks which greatly aid new
computations while also connecting to new areas of mathematics.

In the planar theory there are a number of recently discovered structures,
including dual conformal symmetry~\cite{DualConformalMagic,
  Alday:2007hr, Drummond:2008vq}, Yangian
symmetry~\cite{Drummond:2009fd}, integrability~\cite{Beisert:2003yb,
  Beisert:2006ez}, a dual interpretation in terms of Wilson
loops~\cite{Drummond:2007aua, Brandhuber:2007yx, DualConfWI,
  Mason:2010yk, CaronHuot:2010ek, Alday:2010zy}, amplitudes at finite
coupling using OPE~\cite{Basso:2013vsa, Basso:2014hfa, Basso:2015uxa},
hexagon bootstrap~\cite{Dixon:2013eka, Dixon:2014iba, Dixon:2015iva},
and symbols and cluster polylogarithmics~\cite{Goncharov:2010jf,
  Golden:2013xva, Drummond:2014ffa, Parker:2015cia}, as well as a
variety of other structures.  More recently, scattering amplitudes
were reformulated using on-shell diagrams and the positive
Grassmannian~\cite{OnshellDiagrams, ArkaniHamed:2009dn,
  ArkaniHamed:2009vw, Mason:2009qx, ArkaniHamed:2009dg,
  ArkaniHamed:2009sx,ArkaniHamed:2010kv} (see related work in
Ref.~\cite{Huang:2013owa, Huang:2014xza, Kim:2014hva,
  ElvangGrassmannian}). This reformulation fits
nicely into the geometric concept of the
amplituhedron~\cite{Arkani-Hamed:2013jha} (see also
Refs.~\cite{Arkani-Hamed:2013kca, Franco:2014csa,Bai:2014cna,
  Arkani-Hamed:2014dca, Lam:2014jda, Bai:2015qoa,Ferro:2015grk}), and
makes connections to active areas of research in algebraic geometry
and combinatorics (see e.g. Refs.~\cite{Lusztig, postnikov,
  postnikov2, lauren, goncharov, knutson}).

In this paper, we investigate how some of these properties carry over
to the nonplanar sector.  A basic difficulty in the nonplanar sector
is that it is currently unclear how to define a unique integrand,
largely due to the lack of global variables with which to describe a
nonplanar integrand.  Such ambiguities greatly obscure the desired
structures that might be hiding in the amplitude.  In addition, we
lose Yangian symmetry and presumably any associated integrability
constraints, as well as the connection between amplitudes and Wilson
loops. Naively we also lose the ability to construct amplitudes using
on-shell diagrams, the positive Grassmannian and the amplituhedron.

Nevertheless, one might suspect that many features of the planar
theory can be extended to the full theory including nonplanar
contributions.  In particular, the conjectured duality between color
and kinematics~\cite{BCJ,BCJLoop} suggests that nonplanar integrands
are obtainable directly from planar ones, and hence properties of the
nonplanar theory should be related to properties of the planar sector.  However,
it is not a priori obvious which features can be carried over.

The dual formulation of planar $\mathcal{N} = 4$ super-Yang-Mills
scattering amplitudes using on-shell diagrams and the positive
Grassmannian makes manifest that the integrand has only logarithmic
singularities, and can be written in a {\it $\dlog$ form}.
Furthermore, the integrand has no poles at infinity as a consequence
of dual conformal symmetry.  Recently, Arkani-Hamed, Bourjaily,
Cachazo and one of the authors conjectured the same singularity
properties hold to all loop orders for all maximally helicity violating (MHV) amplitudes
in the nonplanar sector as well~\cite{Log}.  In a previous
paper~\cite{ThreeLoopPaper}, we confirmed this explicitly for the full
three-loop four-point integrand of $\mathcal{N} = 4$ SYM by finding a
basis of diagram integrands where each term manifests these properties.
We also conjectured that to all loop orders the constraints give us
the key analytic information contained in dual conformal symmetry.
Additional evidence for this was provided from studies of the four-
and five-loop amplitudes.  These results then offer concrete evidence
that analytic structures present in the planar amplitudes do
indeed carry over to the nonplanar sector of the theory.

Now we take this further and show that in the planar case dual
conformal invariance is equivalent to integrands with (i) no poles at
infinity, and (ii) special values of leading singularities (maximal
codimension residues).  In the MHV sector, property (ii) and
superconformal invariance imply that leading singularities are
necessarily $\pm 1$ times the usual Parke-Taylor
factor~\cite{Parke:1986gb,Mangano:1987xk}.  Moreover, the existence of
a dual formulation using on-shell diagrams and the positive
Grassmannian implies that (iii) integrands have only logarithmic
singularities. While (i) and (iii) can be directly conjectured also
for nonplanar amplitudes, property (ii) must be modified. As proven in
Ref.~\cite{NonPlanarOnshellToAppear} for both planar and nonplanar
cases, the leading singularities are linear combinations of
Parke-Taylor factors with different orderings and with coefficients
$\pm 1$.  This set of conditions was first conjectured in \cite{Log},
and here we give a more detailed argument as to why the content of dual
conformal symmetry is exhausted by this set of conditions.  We also
provide direct nontrivial evidence showing they hold for the two-loop
five-point amplitude and the three-loop four-point amplitude.

The main purpose of this paper is to present evidence for the
amplituhedron concept~\cite{Arkani-Hamed:2013jha} beyond the planar
limit. The amplituhedron is defined in
momentum twistor variables which intrinsically require cyclic ordering of
amplitudes, making direct nonplanar tests in these variables impossible. However, we can test specific
implications even for nonplanar amplitudes.  In
Ref.~\cite{Arkani-Hamed:2014dca}, Arkani-Hamed, Hodges and one of the
authors argued that the existence of the ``dual'' amplituhedron
implies certain positivity conditions of amplitude integrands. Indeed,
these conditions were proven analytically for some simple cases and
numerically in a large number of examples. (Interestingly, these
conditions appear to hold even post-integration
\cite{Arkani-Hamed:2014dca,WithLance}). The dual amplituhedron 
can be interpreted as a geometric region of which the amplitude is literally a volume,
in contrast to the original definition where the amplitude is a form with logarithmic singularities
on the boundaries of the amplituhedron space.
This implies a very interesting
property when the integrand is combined into a single rational function: its
numerator represents a codimension one surface which lies outside the
dual amplituhedron space. The surface is simply described as a
polynomial in momentum twistor variables and therefore can be fully
determined by the zeros of the polynomial, which correspond to points violating
positivity conditions defining the amplituhedron. A nontrivial statement 
implied by the amplituhedron geometry is that \textit{all} these zeros can be 
interpreted as cuts where the amplitude vanishes. 

This leads to a concrete feature that can be tested even in a diagrammatic
representation of a nonplanar amplitude:
\begin{center}
\framebox[1.03\width]{
$\begin{array}{c} 
		\mbox{The integrand should be determined entirely from homogeneous conditions,} \\ 
		\mbox{up to an overall normalization.}
\end{array}
$}
\end{center}
Concretely, by ``homogeneous conditions'' we mean the conditions of no poles at
infinity, only logarithmic singularities, and also unitarity cuts that
vanish. That is, in the unitarity method, the only required cut equations
are the ones where one side of the equation is 
zero, as opposed to a nontrivial kinematical function.  These zeros
occur either because the amplitude vanishes on a particular branch of
the cut solutions or because the cut is spurious%
\footnote{A spurious cut is one that exposes a non-physical singularity,
i.e.~a singularity that is not present in the full amplitude.}. This conjecture has
exciting implications because this feature is
closely related to the underlying geometry in the planar sector,
suggesting that the nonplanar contributions to
amplitudes admit a similar structure.

To test this conjecture we use the three-loop four-point and two-loop five-point
nonplanar amplitudes as nontrivial examples.  A key assumption is that the
desired properties can all be made manifest diagram-by-diagram~\cite{ThreeLoopPaper}.  
While it is unknown if this
assumption holds for all amplitudes at all loop orders, at the relatively low
loop orders that we work our results confirm that this is a good
hypothesis.  The three-loop four-point integrand was first obtained
in Ref.~\cite{GravityThreeLoop}, while the two-loop five-point integrand
was first calculated in Ref.~\cite{HenrikJJ} in a format that makes
the duality between color and kinematics manifest.
Here we construct different representations that make manifest that the
amplitudes have only logarithmic singularities and no poles at
infinity. These representations are then compatible with the notion that there exists a nonplanar
analogue of dual conformal symmetry and a geometric formulation of nonplanar
amplitudes.  We organize the amplitudes in terms of basis integrands that
have only $\pm 1$ leading singularities.  The coefficient of these
integrals in the amplitudes are then simply sums of Parke-Taylor
factors, as proved in Ref.~\cite{NonPlanarOnshellToAppear}.  We also
show that homogeneous conditions are sufficient to determine
both amplitudes up to an overall factor, as expected if a nonplanar 
analog of the amplituhedron were to exist. 

This paper is organized as follows.  In \sect{DualPicture} we
summarize properties connected to the amplituhedron picture of
amplitudes in planar $\NeqFour$ SYM.  
Then in \sect{NonPlanar} we turn to a discussion of
properties of nonplanar amplitudes, showing in various examples that
the consequences of dual conformal invariance 
and the logarithmic singularity condition do carry over to the
nonplanar sector. Finally, in \sect{Zeros} we give evidence for a geometric interpretation
of the amplitude by showing that the coefficients in the diagrammatic
expansion are determined by zero conditions. In
\sect{Conclusion} we give our conclusions.

\section{Dual Picture for Planar Integrands}
\label{sec:DualPicture}

In this section we summarize known properties of planar amplitudes in
${\cal N}=4$ SYM theory that we wish to carry beyond the planar
limit to amplitudes of the full theory.  We emphasize those features associated
with the amplituhedron construction.  In the planar case, we strip the
amplitude of color factors.  Later when we deal with the nonplanar
case, we restore them.

The classic representation of scattering amplitudes uses Feynman
diagrams.  At loop level the diagrams can be expressed in terms of
scalar and tensor integrals.  We can then write the amplitude as\footnote{In 
general we drop overall factors of $1/(2\pi)^D$ and couplings from the
amplitude, since these play no role in our discussion.}
\begin{equation}
{\cal M} =  \sum_j d_j \int d{\cal I}^j \,,
\label{eqn:gen1}
\end{equation}
where the sum is over a set of basis integrands $d{\cal I}^j$ and
$d_j$ are functions of the momenta of external particles, hereafter
called kinematical functions. In general the integrations should be
performed in $D=4-2\epsilon$ dimensions as a means for regulating both
infrared and ultraviolet divergences.  While the integrand can contain
pieces that differ between four dimensions and $D$ dimensions, in the
present paper we ignore any potential contributions proportional to
$(-2\epsilon)$ components of loop momenta.  At four-points we do not
expect any such contribution through at least six
loops~\cite{DDimCheck}, but they can enter at lower loop orders
as the number of legs increases~\cite{Bern:2008ap}.  
We will not deal with such contributions in this paper,
but we expect that they can be treated systematically as corrections
to any uncovered four-dimensional structure.

In ${\cal N}=4$ SYM we can split off an MHV prefactor, including the 
supermomentum conserving delta function $\delta^8(Q)$, from all $d_j$,
\begin{align}
\label{eqn:PT}
  \PT(1234 \cdots n) = \frac{\delta^8(Q)}
            {\ab{12}\ab{23}\ab{34}\cdots \ab{n1}}\,,
\end{align}
which defines a Parke-Taylor factor~\cite{Parke:1986gb,Mangano:1987xk}.
Usually, we describe the $d{\cal I}^j$ in terms of local integrals that
share the same Feynman propagators as corresponding Feynman diagrams.
However, in the planar sector of the amplitude we do not need to rely
on those diagrams.  Instead we can choose {\it dual coordinates} $k_i = x_i
-x_{i\mi1}$ to encode external kinematics, as well as analogously defined $y_j$ for different
loop momenta. The variables are
associated with the faces of each diagram, are globally defined for
all diagrams, and allow us to define a unique integrand by
appropriately symmetrizing over the faces~\cite{ArkaniHamed:2010kv}.
With these variables, we can sum all diagrams under one integration symbol and
write an $L$-loop amplitude as
\begin{equation}
{\cal M}\sim \int d{\cal I}(x_i,y_j) 
	 = \int d^4y_1\,d^4y_2\dots d^4y_L\,\,{\cal I}(x_i,y_j)\,,
\end{equation}
where $d{\cal I}$ is the integrand form and ${\cal I}$ is the unique
{\it integrand} of the scattering amplitude. 
The integrand form $d{\cal I}$ for the $n$-point amplitude is a
unique rational function with many extraordinary properties that we
will review in this section. Particularly effective ways of constructing 
the integrand are unitarity cut methods~\cite{Bern:1994zx,Bern:1994cg,MaximalCuts} 
or BCFW recursion relations~\cite{BCFW,ArkaniHamed:2010kv}.

\subsection{Dual conformal symmetry}
\label{subsec:DCIplanar}

A key property of ${\cal N}=4$ SYM planar amplitudes is that they
possess {\it dual conformal symmetry}~\cite{DualConformalMagic,
  Alday:2007hr, Drummond:2008vq}. This symmetry acts like ordinary
conformal symmetry on the dual variables $x_i$ and $y_j$ mentioned
above. This can be supersymmetrically extended to a dual \textit{super}conformal
symmetry, and in combination with the ordinary superconformal symmetry it
closes into the infinite dimensional Yangian
symmetry~\cite{Drummond:2009fd}. This is a symmetry of tree-level
amplitudes, and at loop level is a symmetry of quantities such as
the integrand $d{\cal I}$, and IR safe quantities like ratio
functions~\cite{Drummond:2008vq}.

We are interested in understanding the implications of dual conformal
symmetry on the analytic structure of the amplitude.  Good variables for doing so
are the momentum
twistor variables $Z_i$, introduced in Ref.~\cite{Hodges}. These are points in
complex projective space $\mathbb{CP}^3$ and are related to the spinor
helicity variables $\lambda_i \equiv |i\rangle$, $\widetilde{\lambda}_i\equiv |i]$ via
\begin{equation}
Z_i = \left(\begin{array}{c} \lambda_i\\ \mu_i\end{array}\right)\qquad
  \mbox{where}\quad \mu_i^{\dot{a}} = x_i^{a\dot{a}} \lambda_{i,a}\,,
\end{equation}
where $x_i^{a\dot a}$ are the dual variables defined above in spinor
indices. The set of $n$ on-shell
external momenta are then described by $n$ momentum twistors $Z_i$,
$i=1,2,\dots,n$. Momentum twistors are unconstrained variables and
trivialize momentum conservation, which is a quadratic condition on the
$\lambda_i$, $\widetilde{\lambda}_i$ spinors. Each off-shell loop
momentum $\ell_i$ is equivalent to a point $y_i$ in dual momentum space, 
which in turn is represented by a line $Z_{A_i}Z_{B_i}$ in momentum
twistor space.

Dual conformal symmetry acts as $SL(4)$ on $Z_i$, and we can
construct invariants from a contraction of four different $Z$'s,
\begin{equation}
\la ijkl\ra \equiv \la Z_iZ_jZ_kZ_l\ra =
 \epsilon_{\alpha\beta\rho\sigma}Z_i^\alpha Z_j^\beta Z_k^\rho Z_l^\sigma \,.
\end{equation}
Any dual conformal invariant can be written using these four-brackets. The contractions of spinor helicity variables $\lambda$ can
be written as
\begin{equation}
\la ij\ra \equiv \epsilon_{ab}\lambda_i^a\lambda_j^b =
\epsilon_{\alpha\beta\rho\sigma}Z_i^\alpha Z_j^\beta I^{\rho\sigma}\,,
\end{equation}
where $I^{\rho\sigma}$ is the {\it infinity twistor} defined in
Ref.~\cite{Hodges}.  An expression containing $I^{\rho \sigma}$ breaks dual conformal
symmetry because $I^{\rho \sigma}$ does not transform as a tensor. There is a simple
dictionary between momentum space and momentum twistor invariants; we
refer the reader to Ref.~\cite{Hodges} for details.

A simple example of a dual conformal invariant integrand is the zero-mass box,
\begin{equation}
d{\cal I} = \frac{d^4\ell\,(k_1+k_2)^2(k_2+k_3)^2}{\ell^2(\ell - k_1)^2(\ell - k_1 - k_2)^2(\ell+k_4)^2}
 = \frac{\la AB\,d^2A\ra\la AB\,d^2B\ra\la 1234\ra^2}
{\la AB12\ra\la AB23\ra\la AB34\ra\la AB41\ra} \,.
\label{eqn:box}
\end{equation}
This represents the full
one-loop four-point integrand form in $\NeqFour$ SYM. Note that the integrand in 
\eqn{box} is completely projective in all variables $Z$, and the infinity twistor is
absent in this expression. This is true for any dual conformal
invariant integrand. 

This brings us to a key question we would like to answer here: 
\begin{center}
{\it
  What is the content of dual conformal symmetry for momentum-space
  integrands?}  
\end{center}
In momentum twistor space the answer is obvious: the infinity twistor
$I^{\rho \sigma}$ is absent.  Suppose instead the infinity twistor is present.  What
is the implication in momentum space?  The first trivial case is when
the prefactor of the integrand is not chosen properly. For example, if
the factor $(k_2+k_3)^2$ in the numerator of the zero-mass box in \eqn{box} is
replaced with say $(k_1 + k_2)^2$, this will introduce a dependence on
$I$, signaling broken dual conformal invariance. In this case,
the only dependence on the infinity twistor is through four-brackets
$\ab{ijI}$ involving only external variables. The presence of these is
easily avoided by correctly normalizing $d\mathcal{I}$.

The nontrivial interesting cases occur when the infinity twistor
appears in combination with the line $Z_AZ_B$ that represents a loop
momentum, e.g.~$\la ABI\ra$.  In this case no prefactor depending
only on external kinematics can fix it, and the integrand form
necessarily violates dual conformal symmetry.  The factor $\ab{ABI}$
(or its powers) can appear either in the numerator or the
denominator. If it is in the denominator, the integrand has a spurious
singularity at $\la ABI\ra=0$.  In momentum space this corresponds to
sending $\ell\rightarrow\infty$. To see this, consider a simple
example: the one-loop triangle given by
\begin{equation}
d{\cal I} = \frac{d^4\ell\,(k_1+k_2)^2}{\ell^2(\ell - k_1)^2(\ell - k_1 - k_2)^2} 
= \frac{\la AB\,d^2A\ra\la AB\,d^2B
\ra\la 1234\ra\la23I\ra}{\la AB12\ra\la AB23\ra\la AB34\ra\la ABI\ra} \,.
\end{equation}
If we parametrize the loop momentum as $\ell = \alpha
\lambda_1\widetilde{\lambda}_1 + \beta \lambda_2\widetilde{\lambda}_2
+ \gamma \lambda_1\widetilde{\lambda}_2 + \delta
\lambda_2\widetilde{\lambda}_1$ and send $\gamma\rightarrow\infty$
while keeping $\gamma\delta={\rm finite}$, there is a pole which
corresponds to $\ell\rightarrow\infty$.  Bubble integrals even have a
double pole at infinity, which corresponds to a double pole $\la
ABI\ra^2$ when written in momentum twistor space.

If the $\la ABI\ra$ factor is in the numerator there is a problem with the
values of leading singularities. For an $L$-loop integrand these are $4L$-dimensional 
residues that are just rational functions of external kinematics \cite{Cachazo:2008vp}. 
If the integrand form is dual conformal invariant, all its leading singularities are dual conformal cross
ratios (defined in Ref.~\cite{Drummond:2007bm}). A special case is when they are all $\pm1$, as for the box
integrand in \eqn{box}. 

If the integrand has $\la ABI\ra$ in the numerator,
the values of leading singularities, denoted $LS(\cdot)$, depend on $\la (AB)^*I\ra$,
\begin{equation}
LS(d{\cal I}) = \la (AB)^\ast I\ra \cdot {\cal F}(Z_i,\la ab\ra)\,,
\end{equation}
where $(AB)^*$ is the position of the line $AB$ with the leading
singularity solution substituted in. The function ${\cal F}$ is dual
conformal invariant up to some two-brackets of external twistors $\la ab\ra$ from
normalization. For one particular leading singularity we can choose the normalization
of $d{\cal I}$ and therefore force ${\cal F}$ to cancel $\la (AB)^\ast I\ra $,
restoring dual conformal symmetry. However, different leading
singularities -- of which each integrand has at least two by the residue theorem -- are
located at different $(AB)^\ast$ so that it is not possible to simultaneously normalize 
all leading singularities correctly using only external data. 
As a result, some of the leading singularities necessarily are not dual conformal invariant.
A simple example is the scalar one-loop pentagon,
\begin{align}
d{\cal I} &=\frac{d^4\ell\,(k_1+k_2)^2(k_2+k_3)^2(k_3+k_4)^2}
{\ell^2(\ell-k_1)^2(\ell-k_1-k_2)^2(\ell-k_1-k_2-k_3)^2(\ell+k_5)^2}\nonumber\\ 
&\hspace{4cm}= \frac{\la AB\,d^2A\ra\la AB\,d^2B\ra\la ABI\ra
\la 1234\ra\la 2345\ra\la 5123\ra}
{\la AB12\ra\la AB23\ra\la AB34\ra\la AB45\ra\la AB51\ra\la 23I\ra}\,,
\label{eqn:penta}
\end{align}
which is not dual conformal invariant, as implied by the appearance 
of the infinity twistor. The numerator of this pentagon can be modified 
to a chiral version studied in Ref.~\cite{ArkaniHamed:2010gh}, which restores dual
conformal symmetry.

Based on these considerations, we can summarize the content of
dual conformal symmetry of individual integrands in momentum space
in two conditions:
\begin{enumerate}
\item There are no poles as $\ell\rightarrow\infty$.
\item All leading singularities are dual conformal cross ratios.
\end{enumerate}
Any integrand that satisfies these properties necessarily
is dual conformal invariant.

In the context of integrands for MHV amplitudes in planar ${\cal N}=4$ SYM, if we 
strip off the MHV tree-level amplitude, i.e. the Parke-Taylor factor 
${\rm PT}(123\dots n)$ \eqn{PT},
\begin{equation}
{\cal M} = {\rm PT}(123\dots n)\,\int d{\cal I}\,,
\end{equation}
then the integrand $d{\cal I}$ is dual conformal invariant satisfying both 
properties above. There are even stronger constraints: superconformal 
symmetry requires that all leading singularities
are holomorphic functions~\cite{Witten:2003nn} of $\lam{i}$'s alone. 
The only functions that are holomorphic, satisfy property 2 above, and
have the correct mass dimension and little-group weight are pure numbers. In the 
normalization conventions adopted here, they are $\pm 1$ or 0. While we do not
have a direct formulation of dual conformal symmetry in the nonplanar sector, 
we shall find analogous analytic structures in the amplitudes for all the examples 
we study. The role of the Parke-Taylor factor will have to be modified slightly however.

\subsection{On-shell diagrams}
                                                                       
\begin{figure}[tb]
\centering
\begin{tabular}
{>{\centering\arraybackslash}m{0.30\textwidth}
>{\centering\arraybackslash}m{0.30\textwidth}
}
\includegraphics[scale=1.3]{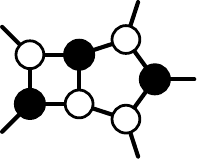} &
\includegraphics[scale=1.3]{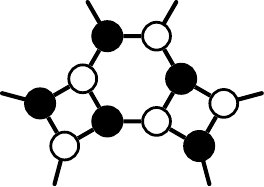}
\end{tabular}
\caption[]{
Sample on-shell diagrams.  The black and white dots respectively represent MHV
and $\overline{\text{MHV}}$ three-point amplitudes. Black lines are on-shell
particles.
     }
\label{fig:OnShellDiagrams}
\end{figure}

{\it On-shell diagrams} provide another novel representation of the
integrand~\cite{OnshellDiagrams}. These are diagrams with black and white vertices connected 
by lines, as illustrated in \fig{OnShellDiagrams}. Black vertices represent MHV three-point
amplitudes, white vertices $\overline{\text{MHV}}$ three-point
amplitudes, and all lines, both internal and external, represent
on-shell particles. There are two indices associated with any on-shell diagram:
the number of external legs $n$ and the helicity index $k$.
The $k$-index is defined as
\begin{equation}
k = \sum_V k_V - P\,,
\end{equation} 
where the sum is over all vertices $V$, $k_V$ is the $k$-count of the
tree-level amplitude in a given vertex, and
$P$ is the number of on-shell internal propagators.
Black and white vertices have $k_B = 2$ and $k_W = 1$, respectively.
As an example, the first diagram in \fig{OnShellDiagrams}
has $k= ( 2 + 2 + 2 ) + (1 + 1 + 1 + 1) - 8 = 2 $.
This $k$ corresponds to the total number of external negative helicities.

The values of the diagrams are computed by integrating over the phase space $d\Omega_i$
of on-shell internal particles the product of tree-level amplitudes
$A_j$ for each vertex
\begin{equation}
d{\Omega} = \prod_{i} \int d\Omega_i \prod_{j} A_{j}\,.
\label{eqn:GenOn}
\end{equation}
An on-shell diagram may be interpreted as a specific generalized
unitarity cut of an amplitude.  In this interpretation, the internal
lines of an on-shell diagram represent cut propagators. The on-shell
diagram represents a nonvanishing valid cut of the amplitude only if
the labels $n,k$ of the on-shell diagram coincide with the same labels
of the amplitude.

A very different way to describe and calculate planar on-shell
diagrams is as cells of a positive Grassmannian
$G_+(k,n)$~\cite{OnshellDiagrams}. For each diagram we define
variables $\alpha_j$ associated with edges or faces of the diagram. Using certain
rules~\cite{OnshellDiagrams}, we build a $(k\times n)$ matrix $C$ with positive main minors --
a cell in the positive Grassmannian. Then the value of the diagram is
given by a logarithmic form in the variables of the diagram, multiplied
by a delta function which connects the $C$ matrix with external
variables (ordinary momenta or momentum twistors),%
\footnote{We suppress wedge notation for forms throughout: $ dx dy \equiv dx \wedge dy$.}
\begin{equation}
d\Omega = \frac{d\alpha_1}{\alpha_1}\frac{d\alpha_2}{\alpha_2}
\frac{d\alpha_3}{\alpha_3}\dots \frac{d\alpha_m}{\alpha_m}\,\,\,\delta(C\cdot Z) \,.
\label{eqn:onshell}
\end{equation}
This is known as a ``$\dlog$ form'' since all singularities have the
structure $\dlog\,\alpha_i\equiv d\alpha_i/\alpha_i$.  For further
details we refer the reader to Ref.~\cite{OnshellDiagrams}.
 
Since the planar integrand can be expressed as a sum of these on-shell diagrams via recursion relations 
\cite{OnshellDiagrams}, all its singularities are also logarithmic. That is, if we approach a
singularity of the amplitude for $\alpha_j\rightarrow 0$ the integrand
develops a pole
\begin{equation}
d{\cal I} \xrightarrow{\alpha_j=0} \frac{d\alpha_j}{\alpha_j}\,
d\widetilde{\cal I}\quad\mbox{where $d\widetilde{\cal I}$ does not 
depend on $\alpha_j$.}
\label{eqn:dlog}
\end{equation}
This property is not at all obvious in more traditional
diagrammatic representations of scattering amplitudes.

The on-shell diagrams are individually both dual conformal
and Yangian invariant and therefore are good building blocks
that make both symmetries manifest. On the other hand, rewriting the
variables $\alpha_j$ in terms of momenta results in 
spurious poles which only cancel in the sum over all contributions.

While \eqn{dlog} holds for all planar $\NeqFour$ integrands for all helicities, in general
the variables $\alpha_j$ are variables of on-shell diagrams
that are nontrivially related to the loop and external variables
through the delta function $\delta(C\cdot Z)$.  For MHV, NMHV (next-to-MHV) and
N$^2$MHV (next-to-next-to-MHV), this change of variables implies that the integrand
also has logarithmic singularities directly in momentum space. For
higher N$^m$MHV amplitudes with $m>2$,
the fermionic Grassmann variables enter in the change of variables so that the integrand 
is not a $\dlog$ form in momentum variables directly.
In this paper, we only deal with
the case of MHV amplitudes, so that the $\dlog$ structure is straightforwardly
visible in momentum space.  As conjectured in Ref.~\cite{Log}, the same 
properties hold at the nonplanar level.

\subsection*{Pure integrand diagrams}

In the MHV sector, we can check the $\dlog$ property for
individual momentum-space planar diagrams with only Feynman propagators.  
In this check, we consider different cuts%
\footnote{We use the words ``cuts'' and ``residues'' interchangeably throughout this paper.}
 of a diagram and probe whether \eqn{dlog} is always
valid in momentum space.  If so, its integrand form indeed has logarithmic
singularities and can in principle be written as a sum of $\dlog$ forms
\begin{equation}
d{\cal I}^j = \sum_k b_k\,\,\dlog\,f_1^{(k)}\,\,
\dlog\,f_2^{(k)}\dots \dlog\,f_{4L}^{(k)}\,,
\label{eqn:dlog2}
\end{equation}
where $f_m^{(k)}$ are some functions of external and loop
momenta. Constraining these integrands to be dual conformal invariant
further enforces that the functions $\dlog\,f_m^{(k)}$ never generate a
pole if any of the loop momenta approach infinity,
$\ell_i\rightarrow\infty$. In addition, for appropriately normalized
diagrams the coefficients $b_k$ are all equal to $\pm1$.  A form
$d{\cal I}^j$ with all these properties is called a {\it pure
  integrand form}.  A simple example of such a form is the box integrand
in \eqn{box} which can be expressed explicitly as a single $\dlog$
form~\cite{OnshellDiagrams}. More complicated $\dlog$ integrands have been used
to write explicit expressions for one-loop and two-loop planar
integrands for all multiplicities \cite{Bourjaily:2013mma,Bourjaily:2015jna}.
Whenever the amplitude is built from  
$d{\cal I}^j$'s that are individually pure integrands, we will refer to such an expansion as a 
{\it pure integrand representation} of the amplitude,
and to the set of $d{\cal I}^j$'s as a {\it pure integrand basis}.

We can now expand the $n$-point planar MHV integrand with Parke-Taylor tree amplitudes factored out
as a sum of pure integrands,
\begin{equation}
d{\cal I} = \sum_j a_j\, d{\cal I}^{j}\,.
\label{eqn:exp1}
\end{equation}
The existence of a diagram basis of pure integrands $d{\cal I}^j$ with only local poles 
is a conjecture. There is no guarantee that we can fix the $a_j$ coefficients of this ansatz to
match the integrand of the amplitude; it might have been necessary to
use non-pure integrands where unwanted singularities cancel between diagrams. 
Presently, it seems that pure integrands are sufficient
up to relatively high loop order. The
coefficients must all be $a_j=\pm1,0$ based on the requirements of
superconformal and dual conformal symmetry. Their precise values
are determined by calculating leading singularities or other 
unitarity cuts.  

We note that the representation in \eqn{exp1} does not make the full Yangian
symmetry manifest, as there is a tension between this symmetry and locality. 
However, the representation does make manifest both dual conformal symmetry and logarithmic singularities.

\subsection{Zero conditions from the amplituhedron}

With on-shell diagrams, scattering amplitudes are built from
abstract mathematical objects with no reference to spacetime
dynamics. This is an important step towards finding a new
description of physics where locality and unitarity are not
fundamental, but rather are derived from geometric properties of
amplitudes. The on-shell diagrams individually have this flavor, but
the particular sum that gives the amplitude is dictated by recursion
relations that are based on unitarity properties. A procedure that
dictates which particular sum of on-shell
diagrams gives the amplitude without reference to unitarity would
therefore be an improvement on recursion relations.
The amplituhedron exactly has this property~\cite{Arkani-Hamed:2013jha} 
as it is a self-consistent geometric
definition of the planar integrand. Here we will not need 
the details of this object, just some of its basic properties.

We focus mainly on the fact that the integrand
of scattering amplitudes is defined as a differential form $d\Omega$
with logarithmic singularities on the boundaries of the amplituhedron
space. This space is defined as a certain map of the positive
Grassmannian through the matrix of positive (bosonized) external data
$Z$ for the tree-level case, and its generalization to loops. 
A particular representation of the amplitude in terms of
on-shell diagrams provides a triangulation of this space, but the
definition of the amplituhedron is independent of any particular triangulation.

The underlying assumptions in this construction are \textit{logarithmic singularities},
in terms of which the form $d\Omega$ is defined, and \textit{dual conformal
symmetry}, which is manifest in momentum twistor space and generalizations thereof. 
All other properties of the integrand,
including locality and unitarity, are derived from the amplituhedron 
geometry. This gives a complete definition of the integrand in a
geometric language; yet, as mentioned in 
Ref.~\cite{Arkani-Hamed:2014dca}, it is desirable
to find another formulation which calculates the integrand as a volume
of an object rather than as a differential form with special properties.
In search of this {\it dual amplituhedron} it was conjectured in
Ref.~\cite{Arkani-Hamed:2014dca} and checked in a variety of cases that the
integrand ${\cal I}$ (without the measure) is positive when evaluated
inside the amplituhedron. This is exactly the property we expect
to be true for a volume function. If we write ${\cal I}$ as a numerator divided by
all local poles,
\begin{equation}
{\cal I} = \frac{N}
{\prod {\rm\, (local\,\,poles)}}\,,
\label{eqn:numden}
\end{equation}
then, since $N$ is a polynomial in the loop variables $(AB)_j$ 
(and for non-MHV cases also in other objects), it must be completely fixed by its zeros (roots). 
An interesting conjecture is that the zeros of $N$ have two simple interpretations:
\begin{itemize}
\item The zeros correspond to forbidden cuts generated by the denominator;
  geometrically these are points outside the
  amplituhedron. 
\item The zeros cancel higher poles in the denominator
  to ensure that all singularities are logarithmic.
\end{itemize}

This should be true for all singularities of the integrand, both in external
and loop variables. In the context of MHV amplitudes however, only the loop
part is nontrivial. As an example, we can write the MHV one-loop
integrand in the following way,
\begin{equation}
{\cal I} = \frac{N(AB,Z_i)}
{\la AB12\ra\la AB23\ra\la AB34\ra\dots \la ABn1\ra}\,,
\label{eqn:oneloop}
\end{equation}
where $N(AB,Z_i)$ is a degree $n\mi4$ polynomial in $AB$ with
proper little group weights in $Z_i$. In this case the denominator
generates only logarithmic poles on the cuts, and the numerator $N$ is 
completely fixed (up to an overall constant) only by
requiring that it vanishes on all forbidden cuts. There are two types
of forbidden cut solutions for MHV amplitudes:
\begin{itemize}
\item {\it Unphysical cut solutions}: all helicity amplitudes vanish.
In the on-shell diagram representation: no on-shell diagram exists.
\item {\it Non-MHV cut solutions:} only MHV
  amplitudes vanish while other helicity amplitudes can be non-zero.
  In the on-shell diagram representation: the corresponding on-shell diagram has $k\neq2$.
\end{itemize}
A simple example of the first case is the collinear cut $\ell=\alpha
k_1$ followed by cutting another propagator $(\ell - k_1 - k_2 - k_3)^2$
of the pentagon integral in \eqn{penta}. 
In momentum twistor geometry this corresponds to
$\la AB12\ra=\la AB23\ra = \la AB45\ra=0$ (as well as setting a Jacobian to zero) which localizes $Z_A=Z_2$,
$Z_B = (123)\cap(45)$. This is an example of an unphysical cut which vanishes 
for all amplitudes including MHV, and the numerator $N$ in \eqn{oneloop} 
vanishes for this choice of $Z_A$, $Z_B$.
                                                                       
\begin{figure}[tb]
\begin{center}
\includegraphics[scale=0.3]{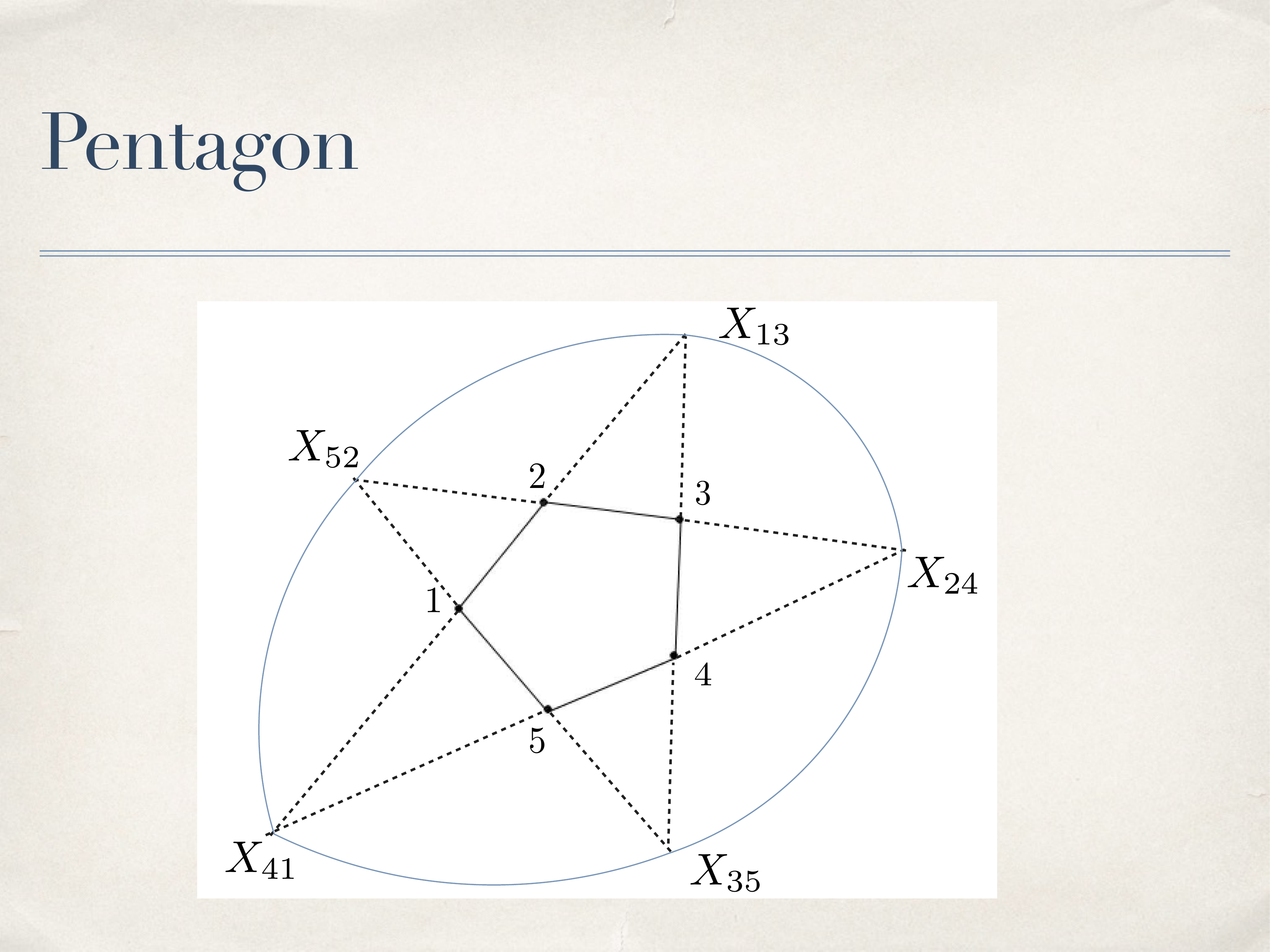}
\caption[]{A simple amplituhedron example~\cite{Arkani-Hamed:2014dca}.
The area of the pentagon formed by the black solid line is the amplituhedron.
The points $X_{i,i+2}$ define zeros of a numerator, so the conic given by the outer (blue) solid 
lines connecting the points represents the numerator.}
\label{fig:Amplituhedron}
\end{center}
\end{figure}

All forbidden cuts correspond to points outside the amplituhedron and
therefore we can think about $N$ as a codimension one surface
outside the amplituhedron. The amplituhedron and the surface $N$ can only touch on lower dimensional
boundaries. This is completely consistent with the picture of the amplitude
being the actual volume of the dual amplituhedron, making a clear distinction between
inside and outside of the space.

Consider the simple example discussed in Ref.~\cite{Arkani-Hamed:2014dca} and shown in
\fig{Amplituhedron}.  In this case the amplituhedron is the area of
the pentagon.
The numerator $N$ is given by the conic that
passes through five given points cyclically labeled by the
$X_{i,i+2}$.
These points correspond to ``unphysical" singularities of
the form $d{\Omega}$. Knowing the positions of the $X_{i,i+2}$ fully fixes the numerator $N$ 
as there is a unique conic passing through five points. Knowing $N$
fixes the integrand ${\cal I}$, per \eqn{numden}. Note that all five $X_{i,i+2}$ are outside the amplituhedron
(in this case the pentagon). The existence of a zero
surface outside the amplituhedron in this example directly leads to a geometric
construction of the integrand. The same happens
for more complicated amplituhedra, which
may lack such an intuitive visualization.

Now let us go several steps back and consider the standard expansion for 
the integrand, \eqn{exp1}, in momentum space as the starting point, and 
think about the zero conditions as coming from physics (unphysical cuts) 
rather than geometry (forbidden boundaries). We can reformulate the 
conjecture about fixing $N$ in \eqn{numden} in terms of unknown 
coefficients $a_j$ in the expansion in \eqn{exp1}:
\begin{center}
\vskip -.3cm
\framebox[1.1\width]{\twhite{$\Big|$}All coefficients $a_j$ are fixed by {\it zero conditions}, up to an overall normalization.}
\end{center}
By zero conditions we mean both unphysical and non-MHV cuts (as defined above)
for which the integrand vanishes, $0 = d{\cal I}|_{\rm cuts}$.
The overall normalization just means the overall scale of the amplitude is one undetermined 
coefficient of the $a_j$, which may be fixed by one non-zero condition.

Assuming the 
integrand may be expanded as in \eqn{exp1}
automatically assumes the presence of
only logarithmic singularities as well as the cancellation of some unphysical cuts,
viz.~those which do not correspond to planar diagrams. On one hand, we can think about this
conjecture as a reduced version of the one stated in
Ref.~\cite{Arkani-Hamed:2014dca} where both logarithmic singularities and 
diagram-like cuts were nontrivial conditions on the numerator $N$ of the 
planar integrand \eqn{numden}. 
On the other hand, a (dual) amplituhedron exactly implies our conjecture about zero conditions
given the diagrammatic expansion of the integrand in \eqn{exp1}.
And most importantly, our new conjecture is now formulated in a language which 
allows us to carry it over to the nonplanar sector later in the paper.

\begin{figure}[tb]
\begin{center}
\includegraphics[scale=1]{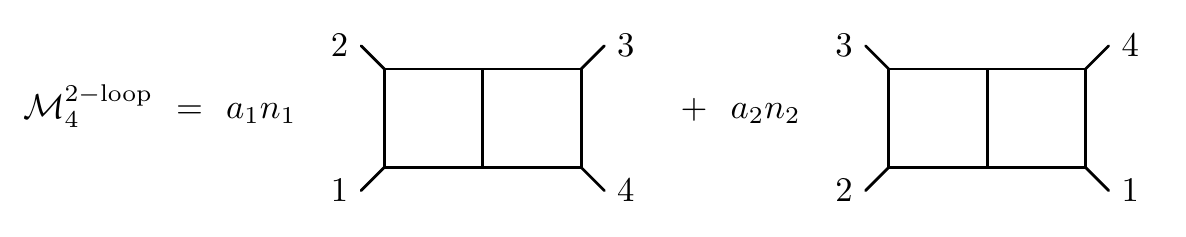}
\caption[]{The planar two-loop four-point amplitude can be represented in terms of
double-box diagrams.}
\vskip -.2 cm 
\label{fig:TwoLoopPlanar}
\end{center}
\end{figure}

A first simple example that illustrates our zero conditions conjecture
is the planar two-loop four-point amplitude~\cite{BRY}, which can be
represented using the diagrammatic expansion in \fig{TwoLoopPlanar}.
The diagrams represent the denominators of individual integrands and
their unit leading singularity normalizations are $n_1=s^2t$,
$n_2=st^2$. The overall planar Parke-Taylor factor $\PT(1234)$ is
suppressed.  We can consider a simple non-MHV cut on which the
amplitude should vanish and relate the coefficients as $a_1=a_2$, which is
indeed correct. We will elaborate on this example in section
\ref{sec:Zeros} in the context of nonplanar amplitudes where more
diagrams contribute.

\section{Nonplanar Amplitudes}
\label{sec:NonPlanar}

As already noted, there is an essential difference between the planar
and nonplanar sectors. In the nonplanar case, it is not known how to
construct a unique integrand prior to integration.  This is a direct
consequence of the lack of global variables.  Without those, the
choice of variables in one nonplanar diagram relative to the choice in
another diagram is arbitrary.  This is a nontrivial obstruction to
carrying over the planar amplituhedron construction directly to the
full amplitude.

Here we circumvent this problem and follow the same strategy as in
Refs.~\cite{Log,ThreeLoopPaper}, which is to consider diagrams as
individual objects and to impose all desired properties
diagram-by-diagram. These elements then form a basis for the complete
amplitude and give us a representation in terms of a linear
combination of said objects. Each integral is furthermore dressed by
color factors $c_j$ and with some kinematical coefficients $d_{j}$ that
need to be determined,
\begin{equation}
{\cal M}= \sum_j d_{j} c_j \int d{\cal I}^j\,.
\label{eqn:gen2}
\end{equation}

The individual pieces $d{\cal I}^j$ interpreted as integrand
forms are not really well defined because of the arbitrariness in
their choice of variables, and they become well-defined only when
integrated over loop momenta. However, we can still impose nontrivial
requirements on the singularity structure of individual diagrams as
was done in Refs.~\cite{Log,ThreeLoopPaper}.  This is because unitarity
cuts of the amplitude impose constraints in terms of a well defined 
set of cut momenta, just as they do
in the planar sector. This implies that the integrand forms
$d{\cal I}^j$ are interesting in their own right and that we can
systematically study their properties with the tools at hand. In
particular, we will see concrete examples where MHV integrands may be expanded
in a pure integrand basis.

\subsection{Nonplanar conjectures}

In the context of ${\cal N}=4$ SYM it is natural to propose
the following properties of the ``integrand'' even in nonplanar cases:
\begin{itemize}
\item[(i)] The integrand has only logarithmic singularities.
\item[(ii)] The integrand has no poles at infinity.
\item[(iii)] The leading singularities of the integrand all take on special values.
\end{itemize}

The presence of only logarithmic
singularities (i) would be an indication of the ``volume'' interpretation of
nonplanar amplitudes.  We will give more detailed evidence for such an
interpretation in the next section.
Demonstrating properties (ii) and (iii) would provide nontrivial evidence
for the existence of an analog of dual conformal symmetry for full
${\cal N}=4$ SYM amplitudes, including the nonplanar sector.
Since we lack nonplanar momentum twistor variables
we cannot formulate an analogous symmetry directly,
yet the basic constraints of properties (ii) and (iii) on nonplanar amplitudes would be
identical to the constraints of dual conformal symmetry on planar amplitudes.

\begin{figure}[tb]
\begin{center}
\includegraphics[scale=1.0]{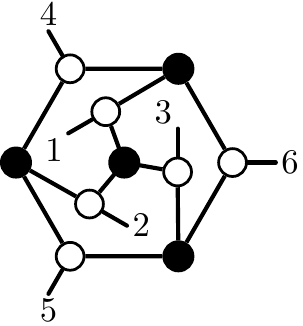}
\caption[]{Example of a nonplanar on-shell diagram. 
}
\label{fig:Nonplanar}
\end{center}
\end{figure}

The first property (i) can be directly linked to the properties of
on-shell diagrams, which are well-defined beyond the planar
sector~\cite{NonPlanarOnshellToAppear, Franco:2014csa}. 
Nonplanar on-shell diagrams, one of which is illustrated in \fig{Nonplanar}, 
are calculated following the same rules as in the planar case~\cite{OnshellDiagrams}.  
In particular, they are given by the same logarithmic form \eqn{onshell}, where the
$C$-matrix is now some cell in the (not necessarily positive) Grassmannian
$G(k,n)$. However, the singularities are again logarithmic and for MHV,
NMHV, and N$^2$MHV amplitudes; this property holds directly in momentum
space like in the planar case.  At present it is not known whether this
is a property of the full amplitude, including nonplanar
contributions. Unlike the planar case, we do not currently have an on-shell
diagram representation of the amplitude since it is
not known how to unambiguously implement recursion relations.
If such a construction exists then the amplitude would share the properties 
of the on-shell diagrams, including their singularity structure. 
Therefore it is very natural to conjecture that the full amplitude 
indeed has only logarithmic singularities~\cite{Log}.

Because there is no global definition for the integrand, it is reasonable
to assume that there exist $d{\cal I}^j$ as in \eqn{gen2}
such that each $d{\cal I}^j$  has only logarithmic singularities~\cite{ThreeLoopPaper}.
That is, we assume that there exists
a $\dlog$ representation \eqn{dlog2} for each diagram,
\begin{align}
\label{eqn:nondlog}
d{\cal I}^j = \sum_{k \vspace{-.2cm}} b_k\ \dlog f_1^{(k)}\ \dlog f_2^{(k)} \dots \ \dlog f_{4L}^{(k)}\,,
\end{align}
\vskip -.2cm \noindent
where $f_i^{(k)}$ are some functions of external and loop momenta and
the coefficients $b_k$ are numerical coefficients independent of external kinematics.

In the planar sector, the other two properties (ii) and (iii) are
closely related to dual conformal symmetry. As discussed
in \subsect{DCIplanar}, the exact constraints of dual conformal
symmetry on MHV amplitudes are that the amplitudes have
unit leading singularities (when combined with ordinary
superconformal symmetry and stripped off Parke-Taylor factor)
and no poles at infinity.
Property (ii) can be directly carried over to any nonplanar
integrand, in particular it would imply that the $\dlog$ forms in
\eqn{nondlog} never generate a pole as $\ell\rightarrow\infty$. As for
property (iii), the value of leading singularities cannot be directly translated to the
nonplanar case, since there is no single overall Parke-Taylor factor to
strip off.  Superconformal invariance only allows us to write leading singularities as any
holomorphic function ${\cal F}_n(\lambda)$, but as proven in
Ref.~\cite{NonPlanarOnshellToAppear}, the only allowed functions are
\begin{equation}
{\cal F}_n = \sum_\sigma a_\sigma \PT_\sigma
\label{eqn:MHV2}\,,
\end{equation}
where $a_\sigma=(\pm1,0)$ and PT stands for a Parke-Taylor 
factor with a given ordering,
\begin{equation}
\PT_\sigma \equiv \PT(\sigma_1 \sigma_2 \sigma_3 \dots \sigma_n) = 
\frac{\delta^8(Q)} {\la \sigma_1\sigma_2\ra\la \sigma_2\sigma_3
\ra\dots\la \sigma_n\sigma_1\ra}\,.
\label{eqn:PTDefinition}
\end{equation}
The sum over $\sigma$ runs over the Parke-Taylor amplitudes
independent under the Kleiss-Kuijf relations~\cite{KleissKuijf}.
There are additional relations between the amplitudes, but those
introduce ratios of kinematic invariants~\cite{BCJ} --- which
introduce spurious poles in external kinematics since they involve
$\tilde \lambda$ --- and so we will not make use of them here.

As an example, consider the on-shell diagram from \fig{Nonplanar} above, which
is equal to the sum of seven Parke-Taylor factors (see Eq.~(3.11) of
Ref.~\cite{NonPlanarOnshellToAppear}),
\begin{align}
{\cal F}_6 &= \PT(123456)+\PT(124563)+\PT(142563)+\PT(145623)\nonumber\\
				   &\hspace{3.5cm}+\PT(146235)+\PT(146253)+\PT(162345) \,.
\end{align}
This is a nontrivial property since there exist many holomorphic functions 
${\cal F}_n(\lambda)$ for $n\geq 6$ which are not of the form of \eqn{MHV2}. 

Analogously to how it works in the planar sector, we can define a
pure integrand to take the form \eqn{dlog2},
so that the integrand has unit logarithmic
singularities with no poles at infinity. Putting together the results
from Refs.~\cite{Log,ThreeLoopPaper,NonPlanarOnshellToAppear}, our
conjecture is that all MHV amplitudes in ${\cal N}=4$ SYM theory can
be written as
\begin{equation}
{\cal M}=   \sum_{k,\sigma,j} a_{\sigma,k,j}\,c_k\, \PT_\sigma \int d{\cal I}^j \,,
\label{eqn:gen3}
\end{equation}
where $a_{\sigma,k,j}$ are numerical rational coefficients and $d{\cal
  I}^j$ are pure integrands with leading singularities $(\pm1,0)$.
The $\PT_{\sigma}$ are as in \eqn{PTDefinition}, 
and $c_k$ are color factors. For contributions with the maximum
number of propagators, the unique color factors can be read off
directly from the corresponding diagrams, but contact term
contributions may have multiple contributing color factors.
The $a_{\sigma,k,j}$ coefficients
are such that, up to sums of Parke-Taylor factors, the leading
singularities of the amplitude are normalized to be $(\pm 1,0)$, reflecting a known
property of the amplitude.

\subsubsection{\it Uniqueness and total derivatives}

There is an important question about the uniqueness of our result. 
The standard wisdom is that the final amplitude ${\cal M}$ is a unique 
object while the planar integrand $d{\cal I}$ is ambiguous, as we can add 
any total derivative $d{\cal I}^{\text{tot}}$,
\begin{equation}
\int d{\cal I}^{\text{tot}} = 0\,,
\end{equation}
that leaves ${\cal M}$ invariant. Note that this is not true in our way 
of constructing the integrand, which relies on matching the cuts of the 
amplitude. This was sharply stated in Ref.~\cite{ArkaniHamed:2010kv}: 
there is only one function which satisfies all constraints (logarithmic singularities, 
dual conformal symmetry) and cut conditions. Any total derivative $d{\cal I}^{\text{tot}}$ 
would violate one or the other. In other words, if we demand dual conformal invariance and 
logarithmic singularities then any integrand would necessarily contribute to some of the 
cuts; the integrand therefore cannot be left undetected by all cuts. It does not matter if it integrates 
to zero or not, its coefficient is completely fixed by cut conditions.

The same is true in the case of nonplanar amplitudes in general. In practice, 
our bases of pure integrands for all examples in the following subsections are complete. 
The pure integrand representation does not distinguish between forms that do integrate to zero and
those that do not. 
Therefore, once the cuts are matched, the pure integrand basis does not miss any total derivatives 
that satisfy our constraints, and thus we cannot add any terms like 
$\int d{\cal I}^{\text{tot}}$ to our amplitude. In fact, some linear combination 
of the basis elements $d{\cal I}^{j}$ in \eqn{gen3} might be total derivatives, 
but the linear combination must contribute to the amplitude prior to integration with fixed coefficients
to match all cuts. There is no freedom to change this coefficient to some other 
value. As a result, like in the planar sector, the nonplanar result is unique 
once we impose all constraints. 

\medskip

In the remainder of this section, we explicitly demonstrate that
the two-loop four-point, three-loop four-point, and two-loop five-point amplitudes
may be written in this pure integrand expansion.  In \sect{Zeros}, we
furthermore demonstrate that the coefficients $a_{\sigma,k,j}$ are
all determined from homogeneous information.

\subsection{Two-loop four-point amplitude}
\label{subsec:TwoLoop4ptAmpl}

\begin{figure}[tb]
\begin{center}
\begin{tabular}
{
>{\centering\arraybackslash}m{0.30\textwidth}
>{\centering\arraybackslash}m{0.30\textwidth}
}
\includegraphics[scale=.15]{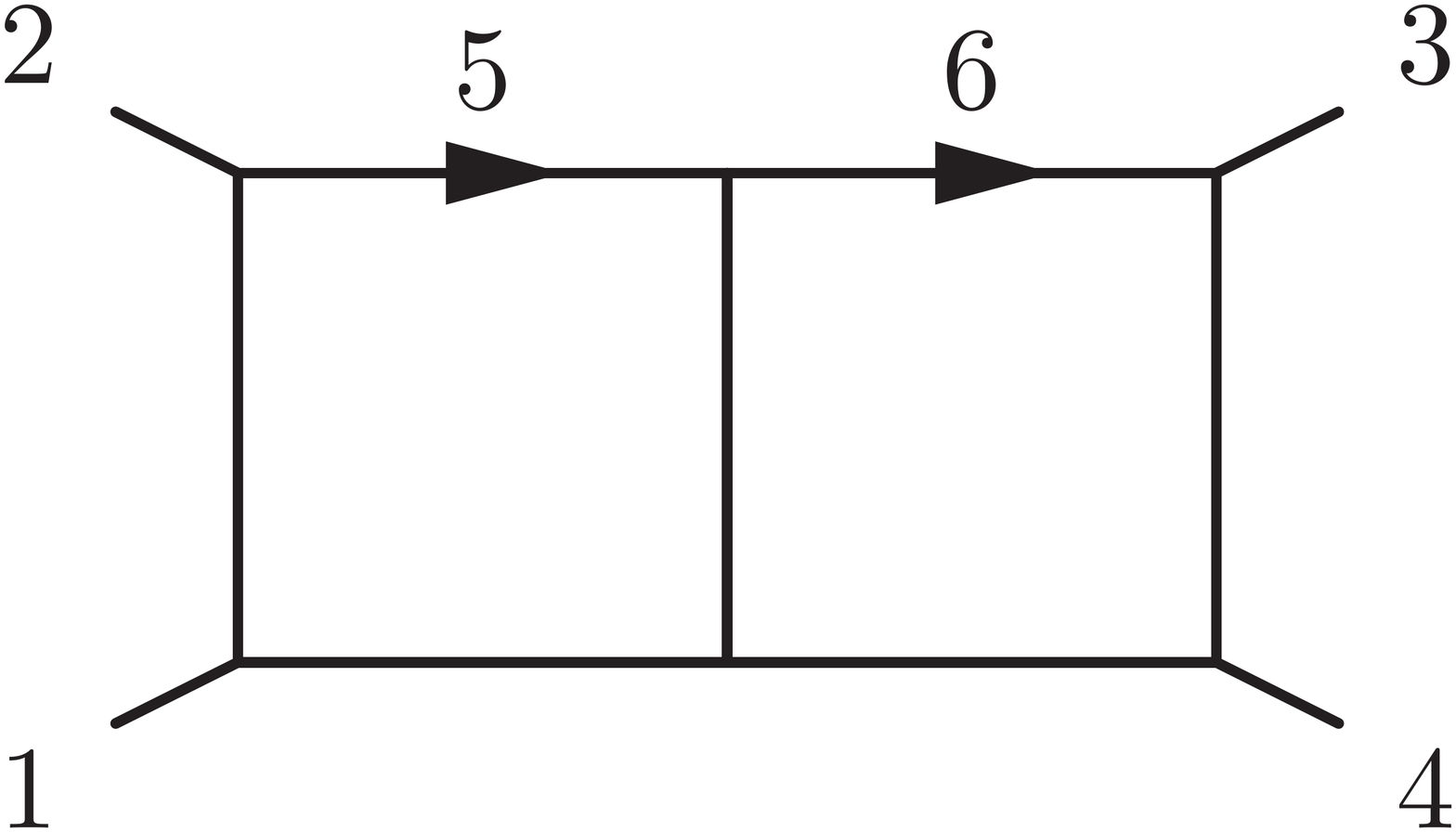} &
\includegraphics[trim={0cm 3cm 0cm 3cm},clip,scale=.15]{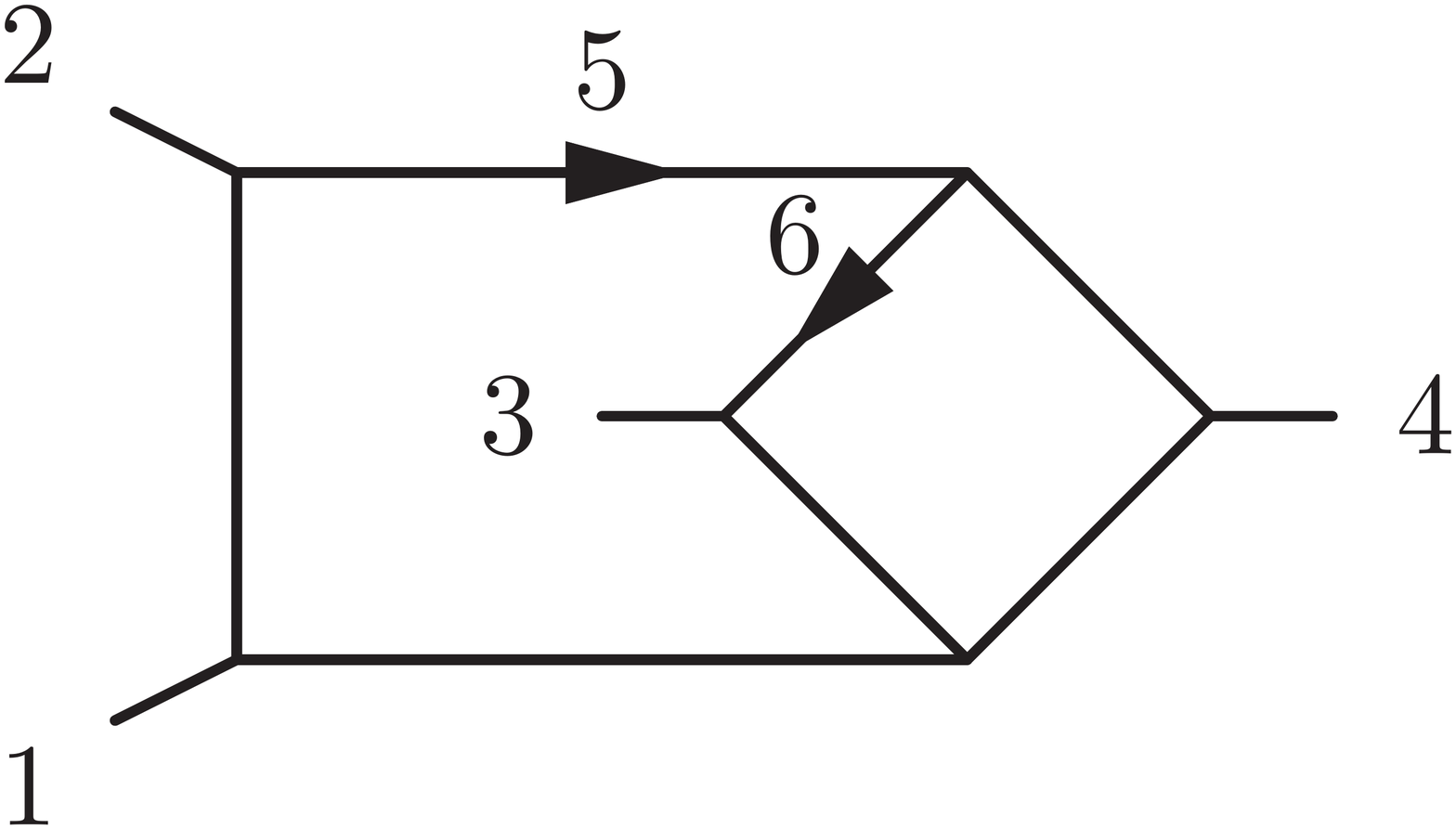} 	\\
(p) & (np) 		\\
\end{tabular}
\end{center}
\vspace{-.3cm}
\caption{The integrals appearing in the two-loop four-point amplitude of $\NeqFour$ SYM theory.  }
\label{fig:2loop4ptParents}
\end{figure}

The simplest multi-loop example is the two-loop four-point amplitude, 
which was first obtained in Refs.~\cite{BRY,BDDPR}. In
Ref.~\cite{Log} these results were reorganized in terms of individual
integrals with only logarithmic singularities and no poles at
infinity. There are two topologies: planar and nonplanar
double boxes, as illustrated in \fig{2loop4ptParents}.  The numerators
for the planar and nonplanar double box integrals with these
properties are
\begin{equation}
\widetilde N^{\rm (p)} = s\,, \hskip 1 cm 
\widetilde N^{\rm (np)} = (\ell_5-k_3)^2 + (\ell_5-k_4)^2\,,
\label{eqn:OlddLog}
\end{equation}
up to overall factors independent of loop momentum.
The integrand $d{\cal I}^{\rm (np)}$ with this numerator has logarithmic
singularities and no poles at infinity, but it is not a pure
integrand. That is, the leading singularities are not all $\pm
1$ but also contain ratios of the form, $\pm u/t$.  The kinematic
invariants $s= (k_1 + k_2)^2$, $t = (k_2 + k_3)^2$ and $u = (k_1 + k_3)^2$
are the usual Mandelstam invariants.

Here we want to decompose the $\tilde N$ numerators so that the resulting integrands $d{\cal I}^j$
are pure, and express the amplitude in terms of the resulting pure integrand basis.  In practice,
we do this by retaining (with respect to Ref.~\cite{Log}) the permutation invariant
function $\mathcal{K} = st\PT(1234)=su\PT(1243)$, and by requiring each
basis integrand to have correct mass dimension --- six in this case --- and unit leading singularities $\pm1$. 
This gives us three basis elements:
\begin{equation}
N^{\rm (p)}=s^2t\,, \qquad 
N^{\rm (np)}_{1} = su(\ell_5-k_3)^2\,,\qquad 
N^{\rm (np)}_{2} = st(\ell_5-k_4)^2\,.
\label{eqn:num1}
\end{equation}
The two nonplanar basis integrals are related by the symmetry of the
diagram, but to maintain unit leading singularities
we keep the terms distinct. The corresponding pure integrand forms
$d{\cal I}^{\rm (p)}$, $d{\cal I}^{\rm (np)}_{1}$, $d{\cal I}^{\rm (np)}_{2}$ 
are obtained by including the integration measure and the
appropriate propagators that can be read off from
\fig{2loop4ptParents}

We note that for the planar double box, an explicit $\dlog$ form is known~\cite{Log},
\begin{align}
d{\cal I}^{\rm (p)}
            =\ &\dlog \frac{\ell_5^2}{(\ell_5-\ell_5^\ast)^2}\,
	       \dlog \frac{(\ell_5+k_2)^2}{(\ell_5-\ell_5^\ast)^2}\,
	       \dlog \frac{(\ell_5+k_1+k_2)^2}{(\ell_5-\ell_5^\ast)^2}\,
	       \dlog \frac{(\ell_5-k_3)^2}{(\ell_5-\ell_5^\ast)^2}\nonumber\\
	      &\times \dlog\frac{(\ell_5-\ell_6)^2}{(\ell_6-\ell^\ast_6)^2}\,
	       \dlog \frac{\ell_6^2}{(\ell_6-\ell^\ast_6)^2} \,
	       \dlog \frac{(\ell_6-k_3)^2}{(\ell_6-\ell^\ast_6)^2}\,
	       \dlog \frac{(\ell_6-k_3-k_4)^2}{(\ell_6-\ell^\ast_6)^2} \,,
\end{align}
where
\begin{equation}
\ell_5^\ast = -\frac{\ab{12}}{\ab{13}}\lam{3}\lamt{2}\,, \qquad 
\ell_6^\ast = k_3 + \frac{(\ell_5-k_3)^2}{\la 4|\ell_5|3]}\lam{4} \lamt{3} \,,
\end{equation}
denote one of the solutions to the on-shell conditions. 
Ref.~\cite{Log} gave the $\dlog$ form for the nonplanar double box
with numerators as in  \eqn{num1} as a sum of four $\dlog$ forms with
prefactors (leading to different Parke-Taylor factors). This
representation has the advantage that it naturally separates parity
even and odd pieces. In Ref.~\cite{ThreeLoopPaper} this was rewritten
in a way that manifestly splits into unit leading
singularity pieces, so that there are single $\dlog$ forms
corresponding to each of the nonplanar numerators $N^{\rm (np)}_{1}$
and $N^{\rm (np)}_{2}$. As usual, we suppress the wedge notation and write,
\begin{equation}
d{\cal I}^{\rm (np)}_{1} = d\Omega_1 \ d\Omega_{2,(1)},\qquad 
d{\cal I}^{\rm (np)}_{2} = d\Omega_1 \ d\Omega_{2,(2)}\,.
\label{eqn:dlogNPBox}
\end{equation}
More explicitly, these forms are
\begin{align}
d\Omega_1^{\twhite{(1)}} &= \dlog \frac{\ell_6^2}{(\ell_6-\ell_6^\ast)^2} \,
			     \dlog \frac{(\ell_6-k_3)^2}{(\ell_6-\ell_6^\ast)^2}\,
			     \dlog \frac{(\ell_6-\ell_5)^2}{(\ell_6-\ell_6^\ast)^2}\,
			     \dlog \frac{(\ell_6-\ell_5+k_4)^2}{(\ell_6-\ell_6^\ast)^2}\,,\nonumber\\
d\Omega_{2,(1)} &= \dlog \frac{\ell^2_5}{\aMs{4}{\ell_5}{3}}\,
		    \dlog \frac{(\ell_5+k_2)^2}{\aMs{4}{\ell_5}{3}} \,
		    \dlog \frac{(\ell_5+k_1+k_2)^2}{\aMs{3}{\ell_5}{4}}\,
		    \dlog \frac{(\ell_5-\ell_{5,1}^{\ast})^2}{\aMs{3}{\ell_5}{4}}\,,
		    \nonumber\\
d\Omega_{2,(2)} &= \dlog \frac{\ell^2_5}{\aMs{3}{\ell_5}{4}}\,
		    \dlog \frac{(\ell_5+k_2)^2}{\aMs{3}{\ell_5}{4}}\,
		    \dlog \frac{(\ell_5+k_1+k_2)^2}{\aMs{4}{\ell_5}{3}}\,
		    \dlog \frac{(\ell_5-\ell_{5,2}^{\ast})^2}{\aMs{4}{\ell_5}{3}} \,.
\end{align}
where the cut solutions read 
\begin{equation}
\ell_6^\ast    = -\frac{\lambda_3\ \ell_5\cdot\lam{4}}{\ab{34}}\,,\quad
\ell_{5,1}^{\ast}  =-\frac{\ab{34}}{\ab{31}}\lam{1}\lamt{4}-k_1-k_2\,, \quad
\ell_{5,2}^{\ast} =-\frac{\ab{43}}{\ab{41}}\lam{1}\lamt{3}-k_1-k_2 \,.
\end{equation}
Using these basis integrals, the full two-loop four-point
amplitude can be written as a linear combination dressed with the
appropriate color and Parke-Taylor factors,
\begin{align}
\label{eqn:TwoL}
{\cal M}_4^{2\hbox{-}\rm loop}  = \frac{1}{4} \sum_{S_4} 
\bigg[& c^{\rm (p)}_{1234} \,  a^{\rm (p)} \PT(1234) \int d{\cal I}^{\rm (p)}  \\
 + \null & c^{(\rm np)}_{1234} \, \left(a_1^{\rm (np)}\PT(1243) 
\int d{\cal I}^{\rm (np)}_{1} + a^{\rm (np)}_2 \PT(1234) 
\int d{\cal I}^{\rm (np)}_{2}\right)\bigg] \,,	\nonumber
\end{align}
where we sum over all 24 permutations of the external legs $S_4$.  The
overall $1/4$ divides out the symmetry factor for each diagram to
remove the overcount from the permutation sum. The planar and nonplanar double-box color factors are
\begin{align}
c^{\rm (p)}_{1234} & = \f^{a_1 a_7 a_9} \f^{a_2 a_5 a_7} \f^{a_5 a_6 a_8} 
\f^{a_9 a_8 a_{10}} \f^{a_3 a_{11} a_6} \f^{a_4 a_{10} a_{11}} \,,\nonumber \\
c^{\rm (np)}_{1234} & = \f^{a_1 a_7 a_8} \f^{a_2 a_5 a_7} \f^{a_5 a_{11} a_6} 
\f^{a_8 a_9 a_{10}} \f^{a_3 a_6 a_9} \f^{a_4 a_{10} a_{11}} \,,
\end{align}
where the $\f^{abc} = i \sqrt{2} f^{abc}$ are appropriately normalized color structure constants.

Matching the amplitude on unitarity cuts determines the coefficients to be
\begin{equation}
\label{eqn:TwoFourSolution}
a^{\rm (p)} = 1\,, \qquad a_1^{\rm (np)} = -1\,, \qquad a_2^{\rm (np)} =-1\,,
\end{equation}
so that the amplitude in \eqn{TwoL} is equivalent to the one presented in
Ref.~\cite{Log}. The trivial difference is that there the two pieces 
$d{\cal I}^{\rm (np)}_{1}$ and $d{\cal I}^{\rm (np)}_{2}$ are
combined into one numerator.

\subsection{Three-loop four-point amplitude}
\label{subsec:ThreeLoop4pt}

Now consider the three-loop four-point amplitude.  This amplitude has
been discussed already in various
papers~\cite{GravityThreeLoop,BCJLoop,ColorKinematics,ThreeLoopPaper}.
Here we will express the amplitude in a pure
integrand basis.  In order to find such a basis we follow the strategy of
Ref.~\cite{ThreeLoopPaper}, wherein integrands with only logarithmic
singularities were identified.  We proceed in the same way, but
at the end impose the additional requirement that the leading
singularities be $\pm 1$ or $0$.
The construction of diagram numerators which lead to pure integrands is very
similar to the previous representation of Ref.~\cite{ThreeLoopPaper},
so we will only summarize the construction here. 

The construction starts from a general $\NeqFour$ SYM power
counting of loop momenta. For a given loop variable we require the
overall scaling of a given integrand to behave like a box in that
variable.  For example, if there is a pentagon subdiagram for loop
variable $\ell$, we allow a nontrivial numerator in $\ell$, $N \sim
\rho_1\ell^2 + \rho_2(\ell\cdot Q) + \rho_3$, where $Q$ is some complex
momentum (not necessarily massless). Similarly, if there is a hexagon
subdiagram in loop variable $\ell$, we allow $N\sim \rho_1(\ell^2)^2 +
\rho_2 (\ell^2)(\ell\cdot Q)+ \rho_3 (\ell\cdot Q_1)(\ell\cdot Q_2) +
\dots$, and so on. Our conventions require that the overall mass
dimension of $d\mathcal{I}_j$ is zero%
\footnote{This mass dimension is different than in
  Ref.~\cite{ThreeLoopPaper}, where we factored out the totally crossing
  symmetric $\mathcal{K} = st \PT(1234) = su \PT(1243)=tu\PT(1324)$.}
  which fixes the mass dimension of the $\rho_j$.

In Ref.~\cite{ThreeLoopPaper}, we then directly constructed
the amplitude by constraining the 
ansatz numerators to obey the symmetry of the diagrams
and to vanish on poles at infinity and double (or multiple) poles.
We now take a slightly different approach and instead of constructing
the amplitude directly, focus on constructing the pure integrand basis.

\subsubsection{{\it Basis of unit leading singularity numerators}}

The next step in constructing the pure integrand basis is to require the elements
have unit leading singularities. We write each basis element
as an ansatz that has the same power counting as the diagram numerators.
We then constrain the elements
so that \textit{any} leading singularity --- codimension $4L$ residue --- is either $\pm1$ or 0.

The resulting basis elements differ slightly from those of Ref.~\cite{ThreeLoopPaper}. Terms that were
originally grouped so that the numerator obeyed diagram symmetry are
now split to make the unit leading singularity property manifest. 
This is exactly the same reason we rewrote \eqn{OlddLog} as \eqn{num1} in 
the two-loop four-point example. Additionally,
the basis elements are scaled by products $st$, $su$, or $tu$ to account
for differing normalizations.  The results of our construction 
of basis numerators yielding pure integrands are summarized in \tab{Basis3Loop4ptParents}.

\begin{table}[tb]
\centering
\vskip-.8cm
\begin{tabular}[tb]
{ >{\centering\arraybackslash} m{0.02\textwidth} 
 >{\centering\arraybackslash} m{0.16\textwidth}  
 >{$}l<{$} 
}
 & \hskip -1. cm {\small Diagram} & {\small \textrm{Numerators}} \\
\hline \hline\\[-.3cm]
(a) & \includegraphics{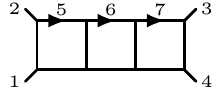} & 
\small
\begin{array}{r l}
N^{{\rm(a)}}_{1} & =  s^3 t \,, \twhite{\Bigg|}\\
\end{array}
\\
(b)& \includegraphics[scale=1.05]{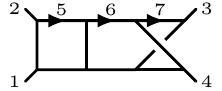} & 
\small
\begin{array}{r l}
 N^{{\rm(b)}}_{1} & = s^2u (\ell_6-k_3)^2
 \,, 
\hskip .6 cm 
N^{{\rm(b)}}_{2} =
 N^{{\rm(b)}}_{1} \big|_{3 \lra 4}
\,,
\\
\end{array}
\\
(c) & \includegraphics[scale=1.05]{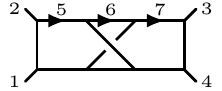} & 
\small
\begin{array}{r l}
N^{{\rm(c)}}_{1} & =
	s^2u (\ell_5-\ell_7)^2
 \,, 
\hskip .6 cm 
N^{{\rm(c)}}_{2} = N^{{\rm(c)}}_{1} \big|_{1 \lra 2}
\,,\\
\end{array}
\\
(d) & \includegraphics[scale=1.05]{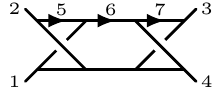} & 
\small
\begin{array}{r l}
N^{{\rm(d)}}_{1} & =
	su \Big[ (\ell_6-k_1)^2(\ell_6+k_3)^2 -\ell^2_6(\ell_6-k_1-k_2)^2	\Big]
\,,  \\[-.1cm]
\vphantom{\Biggl|}
	N^{{\rm(d)}}_{2} & = N^{{\rm(d)}}_{1} \big|_{3 \lra 4} \,, \hskip .6 cm 
	N^{{\rm(d)}}_{3} = N^{{\rm(d)}}_{1} \big|_{1 \lra 2} \,, \hskip .6 cm 
	N^{{\rm(d)}}_{4} = N^{{\rm(d)}}_{1} \big|_{\substack{ 1 \lra 2 \\ 3 \lra 4}} \,
\,,  \\
\end{array}
\\[-.3cm]
(e) & \includegraphics[scale=1.05]{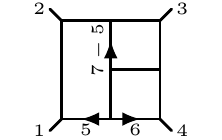} & 
\small
\begin{array}{r l}
N^{{\rm(e)}}_{1} & = 
	s^2t(\ell_{5}+k_{4})^{2}
 \,,  \\
\end{array}
\\
(f) & \includegraphics[scale=1.05]{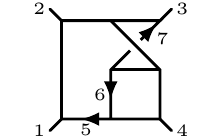} & 
\small
\begin{array}{r l}
N^{{\rm(f)}}_{1}  & =
	st (\ell_5+k_4)^2(\ell_5+k_3)^2	
 \,,
\hskip .6 cm 

N^{{\rm(f)}}_{2} = 	su (\ell_5+k_4)^2(\ell_5+k_4)^2
\,, \\
\end{array}
\\
(g) & \includegraphics[scale=1.05]{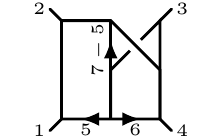} & 
\small
\begin{array}{r l}
N^{{\rm(g)}}_{1} & = 
	s^2t (\ell_5+\ell_6+k_3)^2 					
 \,, \vphantom{\Big|} \\
N^{{\rm(g)}}_{2}  & = 
	st (\ell_5+k_3)^2(\ell_6+k_1+k_2)^2		
 \,, \vphantom{\Big|} \hskip .3 cm
N^{{\rm(g)}}_{3}  =  
N^{{\rm(g)}}_{2}  \big|_{3 \lra 4}
 \,, \\               
\end{array}
\\[-.3cm]
& & \\
(h) & \includegraphics[scale=1.05]{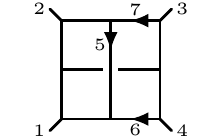} & 
\small
\begin{array}{r l}
 N^{{\rm(h)}}_{1} & =
   st \Big[(\ell_6+\ell_7)^2(\ell_5+k_2+k_3)^2 -\ell^2_5 (\ell_6+\ell_7-k_1-k_2)^2  \\
	  & \qquad -(\ell_5+\ell_6)^2(\ell_7+k_2+k_3)^2
	 -(\ell_5+\ell_6+k_2+k_3)^2 \ell^2_7											  \\
												& \qquad -(\ell_6+k_1+k_4)^2(\ell_5-\ell_7)^2
																 -(\ell_5-\ell_7+k_2+k_3)^2 \ell^2_6
									 \Big]
 \,,\\
N^{{\rm(h)}}_{2} & =
	tu\Big[
	[(\ell_5-k_1)^2+(\ell_5-k_4)^2][(\ell_6+\ell_7-k_1)^2+(\ell_6+\ell_7-k_2)^2] \\
	& \qquad -4\, \ell^2_5(\ell_6+\ell_7-k_1-k_2)^2		                     \\
	&\qquad  -(\ell_7+k_4)^2(\ell_5+\ell_6-k_1)^2	
						  -(\ell_7+k_3)^2(\ell_5+\ell_6-k_2)^2	\\
			&\qquad	-(\ell_6+k_4)^2(\ell_5-\ell_7+k_1)^2
								-(\ell_6+k_3)^2(\ell_5-\ell_7+k_2)^2
		\Big]
\,, \\
N^{{\rm(h)}}_{3} & = N^{{\rm(h)}}_{1} \big|_{2 \lra 4} \,, \hskip 0.4 cm
N^{{\rm(h)}}_{4} = N^{{\rm(h)}}_{2} \big|_{2 \lra 4}
\,, \\

\end{array}
\\[-.3cm]
& & \\
(i) & \includegraphics[scale=1.05]{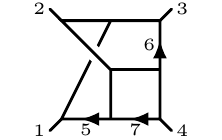} & 
\small
\begin{array}{r l}
N^{{\rm(i)}}_{1} & =
	tu (\ell_6+k_4)^2(\ell_5-k_1-k_2)^2
 \,,  \hskip 1.9cm 
N^{{\rm(i)}}_{2} =  N^{{\rm(i)}}_{1} \big|_{1 \lra 3}
\\
N^{{\rm(i)}}_{3} & =
	st \big[ (\ell_6+k_4)^2(\ell_5-k_1-k_3)^2 - \ell^2_5 (\ell_6-k_2)^2\big]
 \,,  \hskip 0.3cm N^{{\rm(i)}}_{4} =  N^{{\rm(i)}}_{3} \big|_{1 \lra 3}
 \\
\end{array}
\\
(j) & \includegraphics[scale=1.05]{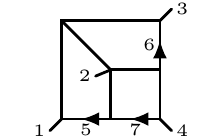} & 
\small
\begin{array}{r l}
N^{{\rm(j)}}_{\textcolor{white}{1}} & =
	stu
 \,.  \\
\end{array}
\\
(k) & \includegraphics[scale=1.05]{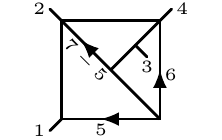} & 
\small
\begin{array}{r l}
 N^{{\rm(k)}}_{\textcolor{white}{1}} & =
  su 
 \,,
\\
\end{array}
\\ [-.5cm]
$\null$ & $\null$\\
\end{tabular}
\caption{The basis of numerators for pure integrands for the three-loop four-point amplitude.
The notation $N \big|_{i \lra j}$ is defined in the text.
\label{tab:Basis3Loop4ptParents}  }
\vskip -1.7 cm
\end{table}

In \tab{Basis3Loop4ptParents} we use the relabeling convention
$N \big|_{i \lra j}$: ``redraw the graph associated with numerator $N$ 
with the indicated exchanges of external momenta $i,j$ and also relabel 
loop momenta accordingly.''
As a simple example look at $N^{(i)}_1\Big|_{1\lra3}$,
\begin{align}
	N^{{\rm(i)}}_{1} & = tu (\ell_6+k_4)^2(\ell_5-k_1-k_2)^2
	\,,  \hskip 1.9cm 
	N^{{\rm(i)}}_{2} =  N^{{\rm(i)}}_{1} \big|_{1 \lra 3}\,.
\end{align}
Under this relabeling, the Mandelstam variables $s$ and $t$ transform 
into one another $s=(k_1+k_2)^2 \lra (k_3+k_2)^2 = t$ and $u$ stays invariant.
Besides changing the external labels, we are instructed to relabel the loop momenta as well.
In the chosen example, this corresponds to interchanging $\ell_5\lra \ell_6$, so that
\begin{align}
 N^{{\rm(i)}}_{2} =  N^{{\rm(i)}}_{1} \big|_{1 \lra 3}
 		=  su (\ell_5+k_4)^2(\ell_6-k_3-k_2)^2\,.
\end{align}

\subsubsection{{\it Matching the amplitude}}
\label{sect:Matching}

The three-loop four-point amplitude is assembled from
the basis numerators as
\begin{equation}
{\cal M}_4^{3\hbox{-}\rm loop}  = \sum_{S_4}\sum_{x}\frac{1}{{\cal S}_{x}} 
	\int d^4\ell_5 d^4 \ell_6 d^4 \ell_7
	\frac{\mathcal{N}^{(x)}}{\prod_{\alpha_x} p^2_{\alpha_x}}\,,
\label{eqn:ThreeFourBuildAmp}
\end{equation}
analogously to \eqn{TwoL}.
Now the sum over $x$ runs over all diagrams in the basis listed in
\tab{Basis3Loop4ptParents}, the sum over $S_4$ is a sum over all 24
permutations of the external legs, and $\mathcal{S}_{x}$ is the
symmetry factor of diagram $x$ determined by counting the
number of automorphisms of diagram $x$.  The
product over $\alpha_x$ indicates the product of Feynman propagators
$p^2_{\alpha_x}$ of diagram $x$, as read from the graphs in
\tab{Basis3Loop4ptParents}.  The Parke-Taylor factors, color factors,
and coefficients are absorbed in ${\cal N}^{(x)}$, which we list in
\tab{ThreeLoopFourPointSol}.

For four external particles, there are only two independent Parke-Taylor factors.
We abbreviate these as
\begin{equation}
\PT_1 = \PT(1234) \,, \qquad \PT_2 = \PT(1243) \,.
\label{eqn:ThreeLoopFourPointPTBasis}
\end{equation}
The third possible factor, $\PT(1423)$, is related to the
other two by a $U(1)$ decoupling
identity or dual Ward identity~\cite{Mangano:1990by}
\begin{equation}
\PT(1423) = - \PT(1234) - \PT(1243)\,,
\end{equation}
and is therefore linearly dependent on $\PT_1$ and $\PT_2$.

When checking cuts of the amplitude, certain cuts may combine
contributions from different terms in the permutation
sum of \eqn{ThreeFourBuildAmp}, resulting in a cut expression that involves
diagrams that are relabellings of those in \tab{Basis3Loop4ptParents}. In that case,
the procedure is to relabel the numerators, propagators, Parke-Taylor factors, and
color factors given in the tables into the cut labels. The resulting Parke-Taylor
factors may not be in the original basis of Parke-Taylor factors; however every Parke-Taylor
in the relabeled expression can be expanded in the original Parke-Taylor basis. 

The diagrams with 10 propagators contain only three-point vertices and
therefore have unique color factors included in ${\cal N}^{(x)}$.  For
the two diagrams with less than 10 propagators, we include in our
ansatz for ${\cal N}$ all independent color factors from all
10-propagator diagrams that are related to the lower-propagator
diagrams by collapsing internal legs.  For example, three
10-propagator diagrams are related to diagram (j) in this way, with
color factors $c^{\rm (i)}_{1234}$, $c^{\rm (i)}_{1243}$ and
$c^{\rm (i)}_{3241}$, where
\begin{equation}
 c^{\rm (i)}_{1234} =
\f^{a_1 a_8 a_5}  \f^{a_6 a_2 a_9} \f^{a_3 a_{11} a_{10}} \f^{a_{12} a_4 a_{13}}
\f^{a_9 a_{10} a_8} \f^{a_{11} a_{12} a_{14}} \f^{a_{13} a_5 a_7} \f^{a_{14} a_7 a_6}\,,
\end{equation}
is the standard color factor in terms of appropriately normalized
structure constants, and the others $c$'s are relabellings of 1234
of this color factor.  
The Jacobi relation between the three color
factors allows us to eliminate, say $c^{\rm (i)}_{1243}$. This is
exactly what we do for diagram (j).  In diagram (k), there are nine
contributing parent diagrams.  Typically there are four independent
color factors in the solution to the set of six Jacobi relations, but
in this case two of the color factors that contribute happen to be
identical up to a sign, and thus there are only three independent color
factors.

\begin{table}[tb]
\centering
\small
\begin{tabular}[tb]
{
>{$}r<{$} 
>{$}l<{$} 
>{$}l<{$} 
}
& \hskip -1. cm {\small \textrm{Color\ Dressed\ Numerators}} & {\small \hskip 1cm \textrm{PT\ Matrices}} \\
\hline \hline\\[-.3cm]
\mathcal{N}^{\rm (a)} =
& \hspace{-.25cm}
c^{\rm (a)}_{1234} {\displaystyle \sum_{1 \le \sigma \le 2 }}
N^{\rm (a)}_1 {\tt a}_{1\sigma}^{\rm (a)} \PT_\sigma \,,
& \hspace{1 cm} {\tt a}^{\rm (a)}_{1\sigma} = 
\bigl(
\begin{array}{cc}
1 & 0 \\
\end{array}
\bigr) \,,
\\
\mathcal{N}^{\rm (b)} =
& \hspace{-.25cm}
c^{\rm (b)}_{1234} {\displaystyle \sum_{\substack{ 1 \le \nu \le 2 \\ 1 \le \sigma \le 2 }}}
N^{\rm (b)}_\nu {\tt a}_{\nu\sigma}^{\rm (b)} \PT_\sigma \,,
& \hspace{1 cm}
{\tt a}^{\rm (b)}_{\nu\sigma} = 
(-1) \left(
\begin{array}{cc}
0 & 1 \\
1 & 0 \\
\end{array}
\right) ,
\\
\mathcal{N}^{\rm (c)} =
& \hspace{-.25cm}
c^{\rm (c)}_{1234} {\displaystyle \sum_{\substack{1 \le \nu \le 2 \\ 1 \le \sigma \le 2 }}}
N^{\rm (c)}_\nu {\tt a}_{\nu\sigma}^{\rm (c)} \PT_\sigma \,,
&\hspace{1 cm}
{\tt a}^{\rm (c)}_{\nu\sigma} = 
(-1)\left(
\begin{array}{cc}
0 & 1 \\
1 & 0 \\
\end{array}
\right) ,
\\
\mathcal{N}^{\rm (d)} =
& \hspace{-.25cm}
c^{\rm (d)}_{1234} {\displaystyle \sum_{\substack{1 \le \nu \le 4 \\ 1 \le \sigma \le 2 }}}
N^{\rm (d)}_\nu {\tt a}_{\nu\sigma}^{\rm (d)} \PT_\sigma \,,
&\hspace{1 cm}
{\tt a}^{\rm (d)}_{\nu\sigma} = 
\left(
\begin{array}{cccc}
0 & 1 & 1 & 0 \\
1 & 0 & 0 & 1 \\
\end{array}
\right)^T\! ,
\\
\mathcal{N}^{\rm (e)} =
& \hspace{-.25cm}
c^{\rm (e)}_{1234} {\displaystyle \sum_{1 \le \sigma \le 2 }}
N^{\rm (e)}_1 {\tt a}_{1\sigma}^{\rm (e)} \PT_\sigma \,,
& \hspace{1 cm}
{\tt a}^{\rm (e)}_{1\sigma} = 
\bigl(
\begin{array}{cc}
1 & 0 \\
\end{array}
\bigr) \,,
\\
\mathcal{N}^{\rm (f)} =
& \hspace{-.25cm}
c^{\rm (f)}_{1234} {\displaystyle \sum_{\substack{ 1 \le \nu \le 2 \\ 1 \le \sigma \le 2 }}}
N^{\rm (f)}_\nu {\tt a}_{\nu\sigma}^{\rm (f)} \PT_\sigma \,,
& \hspace{1 cm}
{\tt a}^{\rm (f)}_{\nu\sigma} = 
(-1)\left(
\begin{array}{cc}
1 & 0 \\
0 & 1 \\
\end{array}
\right) ,
\\
\mathcal{N}^{\rm (g)} =
& \hspace{-.25cm}
c^{\rm (g)}_{1234} {\displaystyle \sum_{\substack{1 \le \nu \le  3 \\ 1 \le \sigma \le 2 }}}
N^{\rm (g)}_\nu {\tt a}_{\nu\sigma}^{\rm (g)} \PT_\sigma \,,
&\hspace{1 cm}
{\tt a}^{\rm (g)}_{\nu\sigma} = 
\left(
\begin{array}{cccc}
-1 & 1 & 0 \\
0 & 0 & 1 \\
\end{array}
\right)^T \! ,
\\
\mathcal{N}^{\rm (h)} =
& \hspace{-.25cm}
c^{\rm (h)}_{1234} {\displaystyle \sum_{\substack{1 \le \nu \le 4 \\ 1 \le \sigma \le 2 }}}
N^{\rm (h)}_\nu {\tt a}_{\nu\sigma}^{\rm (h)} \PT_\sigma \,,
&\hspace{1 cm}
{\tt a}^{\rm (h)}_{\nu\sigma} = 
\dfrac{1}{2} \left(
\begin{array}{cccc}
1 & 1 & 1 & 0 \\
0 & 1 & 0 & -1 \\
\end{array}
\right)^T \! ,
\\
\mathcal{N}^{\rm (i)} =
& \hspace{-.25cm}
c^{\rm (i)}_{1234} {\displaystyle \sum_{\substack{1 \le \nu \le 4 \\ 1 \le \sigma \le 2 }}}
N^{\rm (i)}_\nu {\tt a}_{\nu\sigma}^{\rm (i)} \PT_\sigma \,,
&\hspace{1 cm}
{\tt a}^{\rm (i)}_{\nu\sigma} = 
(-1) \left(
\begin{array}{cccc}
1 & 0 & -1 & 1 \\
1 & 1 & 0 & 0 \\
\end{array}
\right)^T \! ,
\\
\mathcal{N}^{\rm (j)} =
& \hspace{-.25cm}
c^{\rm (i)}_{1234} {\displaystyle \sum_{ 1 \le \sigma \le 2 }} N^{\rm (j)}_1 {\tt a}_{1\sigma,(1234)}^{\rm (j)} \PT_\sigma
&\hspace{1 cm}
{\tt a}^{\rm (j)}_{1\sigma,(1234)} =
\bigl(
\begin{array}{cc}
1 & 1 \\
\end{array}
\bigr) \,,
\\
&\hspace{-.25cm} \null +
c^{\rm (i)}_{3241} {\displaystyle \sum_{ 1 \le \sigma \le 2 }} N^{\rm (j)}_1 {\tt a}_{1\sigma,(3241)}^{\rm (j)} \PT_\sigma \,,
&\hspace{1 cm}
{\tt a}^{\rm (j)}_{1\sigma,(3241)} =
\bigl(
\begin{array}{cc}
-1 & 0 \\
\end{array}
\bigr) \,,
\\
\mathcal{N}^{\rm (k)} =
& \hspace{-.25cm}
c^{\rm (g)}_{1234} {\displaystyle \sum_{ 1 \le \sigma \le 2 }} N^{\rm (k)}_1 {\tt a}_{1\sigma,(1234)}^{\rm (k)} \PT_\sigma
&\hspace{1 cm}
{\tt a}_{1\sigma,(1234)}^{\rm (k)} = 
\bigl(
\begin{array}{cc}
-2 & 0 \\
\end{array}
\bigr)\,,
\\
& \hspace{-.25cm} \null
+ c^{\rm (g)}_{4312} {\displaystyle \sum_{ 1 \le \sigma \le 2 }} N^{\rm (k)}_1 {\tt a}_{1\sigma,(4312)}^{\rm (k)} \PT_\sigma
&\hspace{1 cm}
{\tt a}_{1\sigma,(4312)}^{\rm (k)} = 0	\,,
\\
& \hspace{-.25cm} \null
+
c^{\rm (f)}_{2431} {\displaystyle \sum_{ 1 \le \sigma \le 2 }} N^{\rm (k)}_1 {\tt a}_{1\sigma,(2431)}^{\rm (k)} \PT_\sigma \,.
&\hspace{1 cm}
{\tt a}_{1\sigma,(2431)}^{\rm (k)} = 0	\,. 
\\
\end{tabular}
\vspace{.2cm}
\caption[]{ The three-loop four-point numerators that contribute to
  the amplitude.  The $N^{\rm (x)}_{\nu}$ are listed in
  \tab{Basis3Loop4ptParents}.  The four-point Parke-Taylor factors $\PT_\sigma$
  are listed in \eqn{ThreeLoopFourPointPTBasis}.  The numerators
  including color factors are denoted as $\mathcal{N}^{(x)}$.  The
  symbol `$T$' denotes a transpose.
}
\label{tab:ThreeLoopFourPointSol}
\vskip -.5 cm 
\end{table}

In Ref.~\cite{ThreeLoopPaper}, the final representation of the
amplitude contained arbitrary free parameters associated with the
color Jacobi identity that allowed contact terms to be moved between
parent diagrams without altering the amplitude.  Here we removed
this freedom by assigning the contact terms to their own diagrams
and keeping only a basis of independent color factors for each.

One advantage of the Parke-Taylor expansion of the amplitude is that
we can compactly express the solution to the cut equations in the set
of matrices listed on the right hand side of
\tab{ThreeLoopFourPointSol}. For example, ${\cal N}^{\rm (i)}$ can be
read off from the table as
\begin{align}
{\cal N}^{\rm (i)} = c^{\rm (i)}_{1234}  (-1)
\big(
N^{\rm (i)}_1 (\PT_1 + \PT_2) + N^{\rm (i)}_2 \PT_2 - N^{\rm (i)}_3 \PT_1 + N^{\rm (i)}_4 \PT_1
\big) \,.
\end{align}
This expression supplies the Parke-Taylor and color dependence required for \eqn{ThreeFourBuildAmp},
in agreement with the general form of \eqn{gen3}.
%
\subsection{Two-loop five-point amplitude}
\label{subsec:2loop5ptAmpl}
The integrand for the two-loop five-point amplitude was first obtained in
Ref.~\cite{HenrikJJ} in a format that makes the duality between color
and kinematics manifest.  Here we find a pure integrand representation.
An additional feature of our representation is that it is manifestly free
of spurious poles in external kinematics.

\subsubsection{{\it Basis of unit leading-singularity numerators}}
\label{subsubsec:2Loop5ptBasis}

\begin{table}[tb]
\begin{center}
\hspace{-.5cm}
\begin{tabular}[tb]
{ >{\centering\arraybackslash} m{0.02\textwidth} 
 >{\centering\arraybackslash} m{0.16\textwidth}  
 >{$}l<{$} 
}
  & \hskip -1. cm  {\small Diagram} & {\small \textrm{Numerators}} \\
\hline \hline\\[-.1cm]
(a) & \includegraphics[scale=0.08]{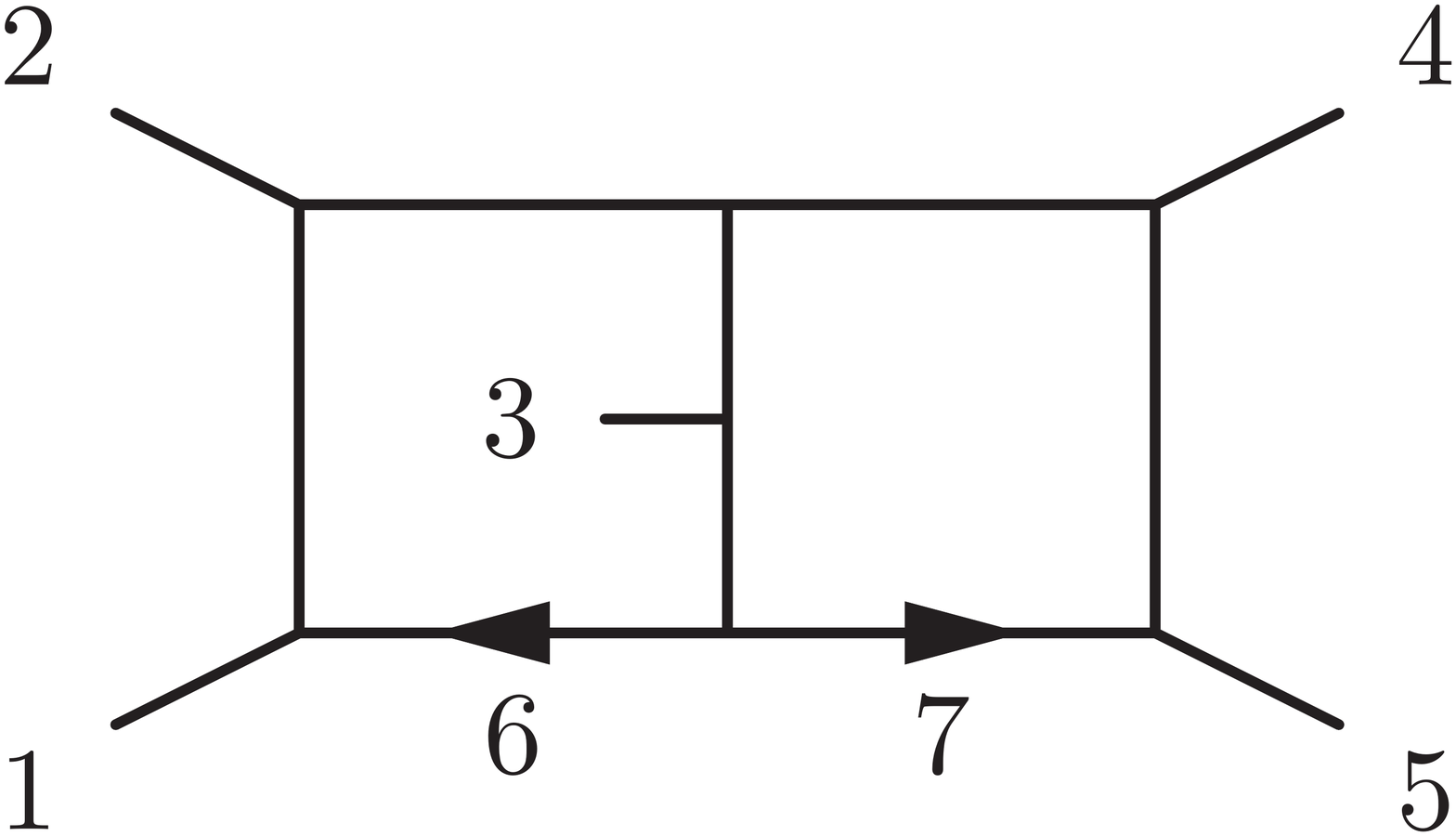} & 
\small
\begin{array}{r l}
 N^{{\rm(a)}}_1 & =
  \ab{13}\ab{24}
			\Bigl[\sqb{24}\sqb{13} \left(\ell_7+\frac{\sqb{45}}{\sqb{24}} \lam{5}\lamt{2}\right)^2
				\Bigl(\ell_6-\frac{Q_{12}\cdot\lamt{3} \ \lamt{1}}{\sqb{13}}\Bigr)^2	\\
				&\hspace{1.8cm}									
				 -\sqb{14}\sqb{23}  \Bigl(\ell_7+\frac{\sqb{45}}{\sqb{14}} \lam{5}\lamt{1}\Bigr)^2
				    \Bigl(\ell_6-\frac{Q_{12}\cdot\lamt{3} \ \lamt{2}}{\sqb{23}}\Bigr)^2 
					\Bigr] \,,\\
  \vphantom{\bigg|}
  N^{{\rm(a)}}_2 & = N^{{\rm(a)}}_1 \big|_{\substack{ 1 \lra 2 \\ 4 \lra 5}} \,, \hskip.65cm 
  N^{{\rm(a)}}_3   = N^{{\rm(a)}}_1 \big|_{\substack{ 2 \lra 4 \\ 1 \lra 5}} \,, \hskip.65cm 
  N^{{\rm(a)}}_4   = N^{{\rm(a)}}_1 \big|_{\substack{ 1 \lra 4 \\ 2 \lra 5}}
 \,,\\
 \vphantom{\bigg|}
  N^{{\rm(a)}}_5 & = \overline{N}^{{\rm(a)}}_1\,, \hskip.4cm 
  N^{{\rm(a)}}_6   = \overline{N}^{{\rm(a)}}_2\,, \hskip.4cm 
  N^{{\rm(a)}}_7   = \overline{N}^{{\rm(a)}}_3\,, \hskip.4cm 
  N^{{\rm(a)}}_8   = \overline{N}^{{\rm(a)}}_4\,, \hskip.4cm
\end{array}
\\
(b) & \includegraphics[scale=0.08]{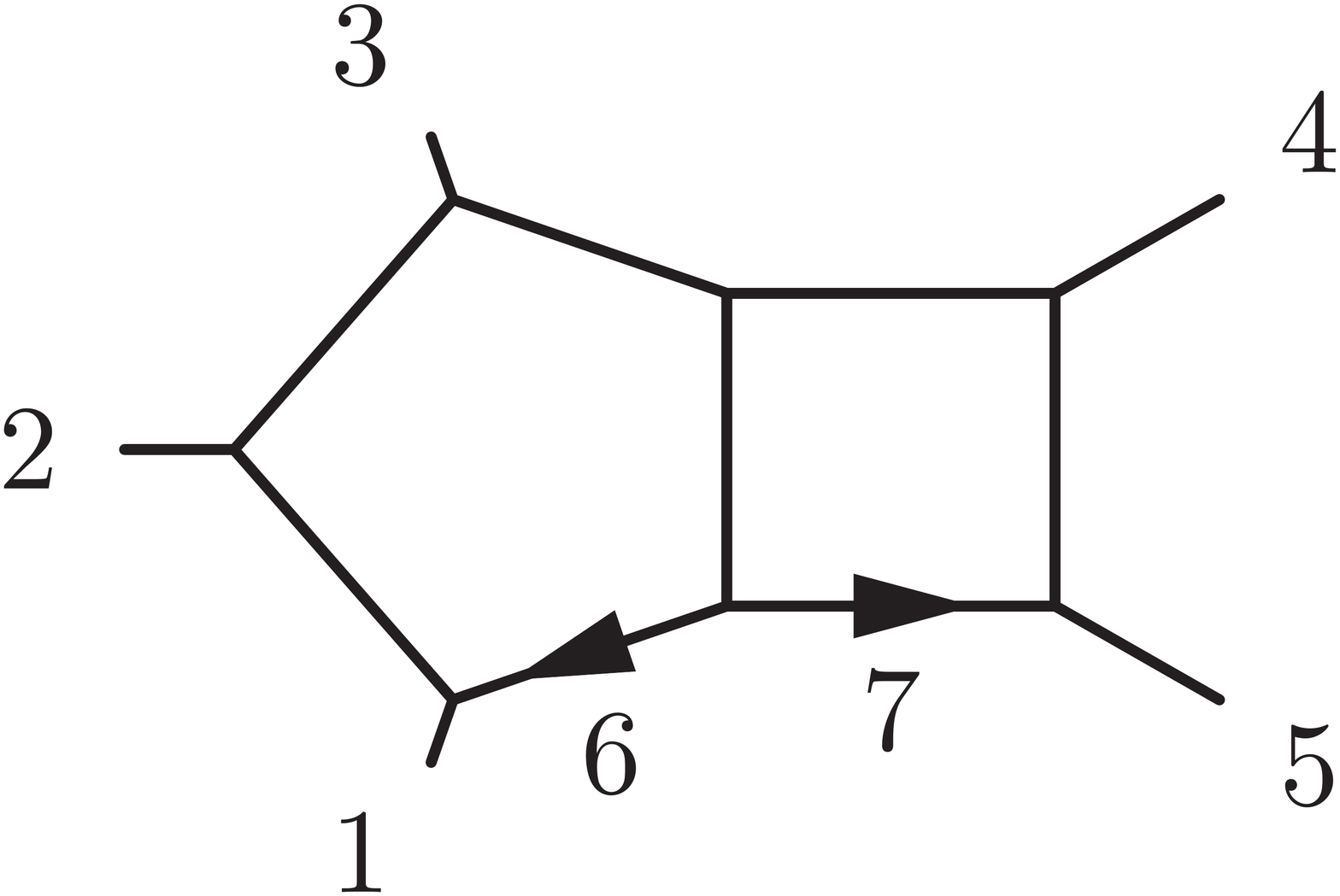} & 
\small
\begin{array}{r l}
N^{{\rm(b)}}_1 & =
	\ab{15}\sqb{45}\ab{43}\sab{45}\sqb{13} \left(\ell_6+\frac{Q_{45}\cdot \lamt{3} \ \lamt{1}}{\sqb{13}}\right)^2
 \,,  \\
N^{{\rm(b)}}_2 & =
	\overline{N}^{{\rm(b)}}_1
 \,,  \\
\end{array}
\\
(c) & \includegraphics[scale=0.08]{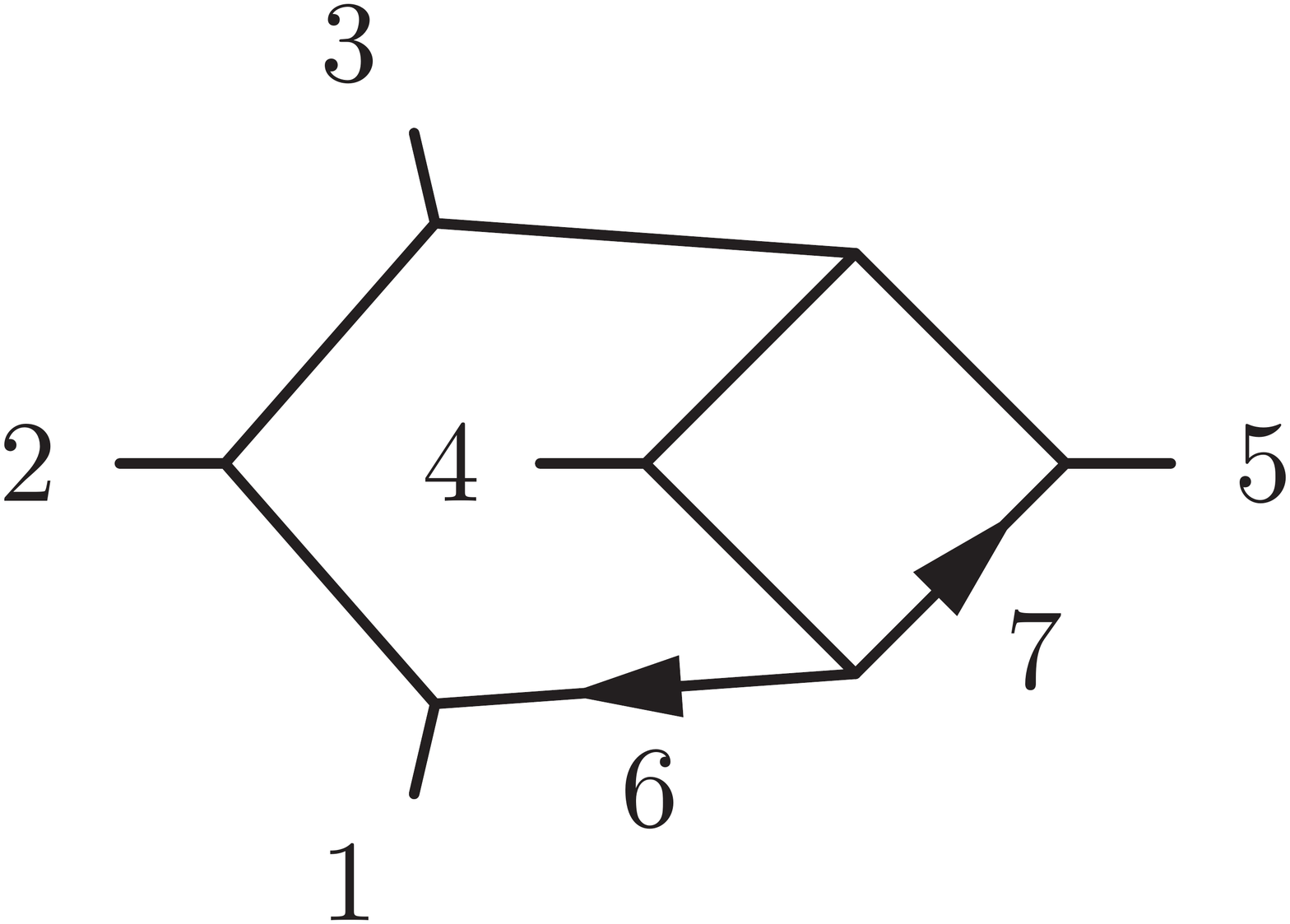} & 
\small
\begin{array}{r l}
N^{{\rm(c)}}_1 & =
	\sqb{13} \left(\ell_6 + \frac{Q_{45}\cdot\lamt{3}\ \lamt{1}}{\sqb{13}}\right)^2 \ab{15}\sqb{54}\ab{43}(\ell_6+k_4)^2 
 \,,  \\
N^{{\rm(c)}}_2 & =
 N^{{\rm(c)}}_1 \big|_{4 \lra 5}
 \,, \hskip.4cm
 N^{{\rm(c)}}_3 =
	\overline{N}^{{\rm(c)}}_1
 \,,  \hskip.4cm
N^{{\rm(c)}}_4 =
	\overline{N}^{{\rm(c)}}_2
 \,,  \\
\end{array}
\\
(d) & \includegraphics[scale=0.08]{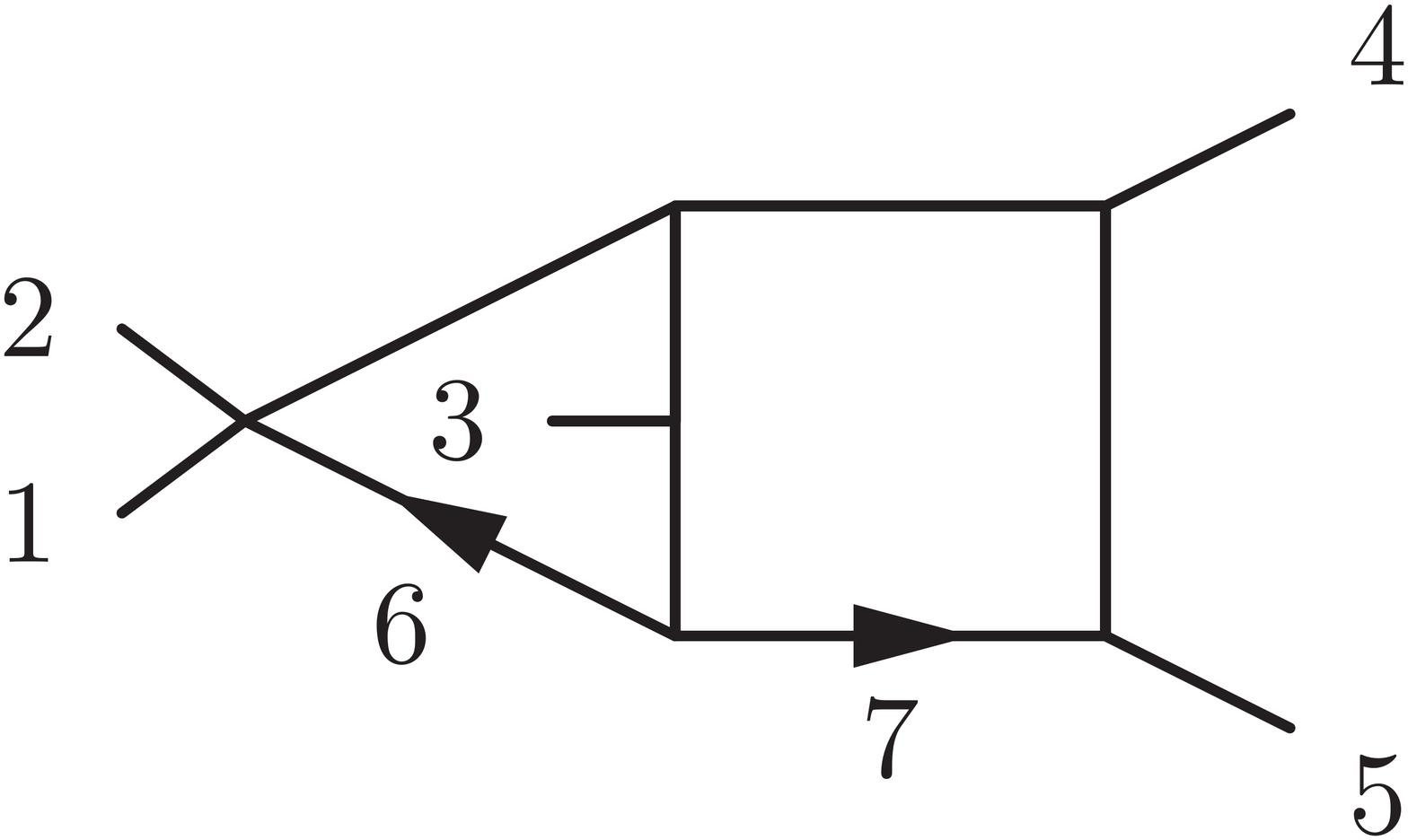} & 
\small
\begin{array}{r l}
 N^{{\rm(d)}}_1 & =
  \sab{34}(\sab{34}+\sab{35}) \left(\ell_7-k_5+\frac{\ab{35}}{\ab{34}}\lam{4}\lamt{5}\right)^2
 \,,\\
 N^{{\rm(d)}}_2 & = N^{{\rm(d)}}_1 \big|_{4 \lra 5}
 \,, \hskip0.6cm
 N^{{\rm(d)}}_3  =
  \overline{N}^{{\rm(d)}}_1
 \,, \hskip.6cm
 N^{{\rm(d)}}_4  =
  \overline{N}^{{\rm(d)}}_2
 \,,\\
\end{array}
\\[-.1cm]
(e) & \includegraphics[scale=0.08]{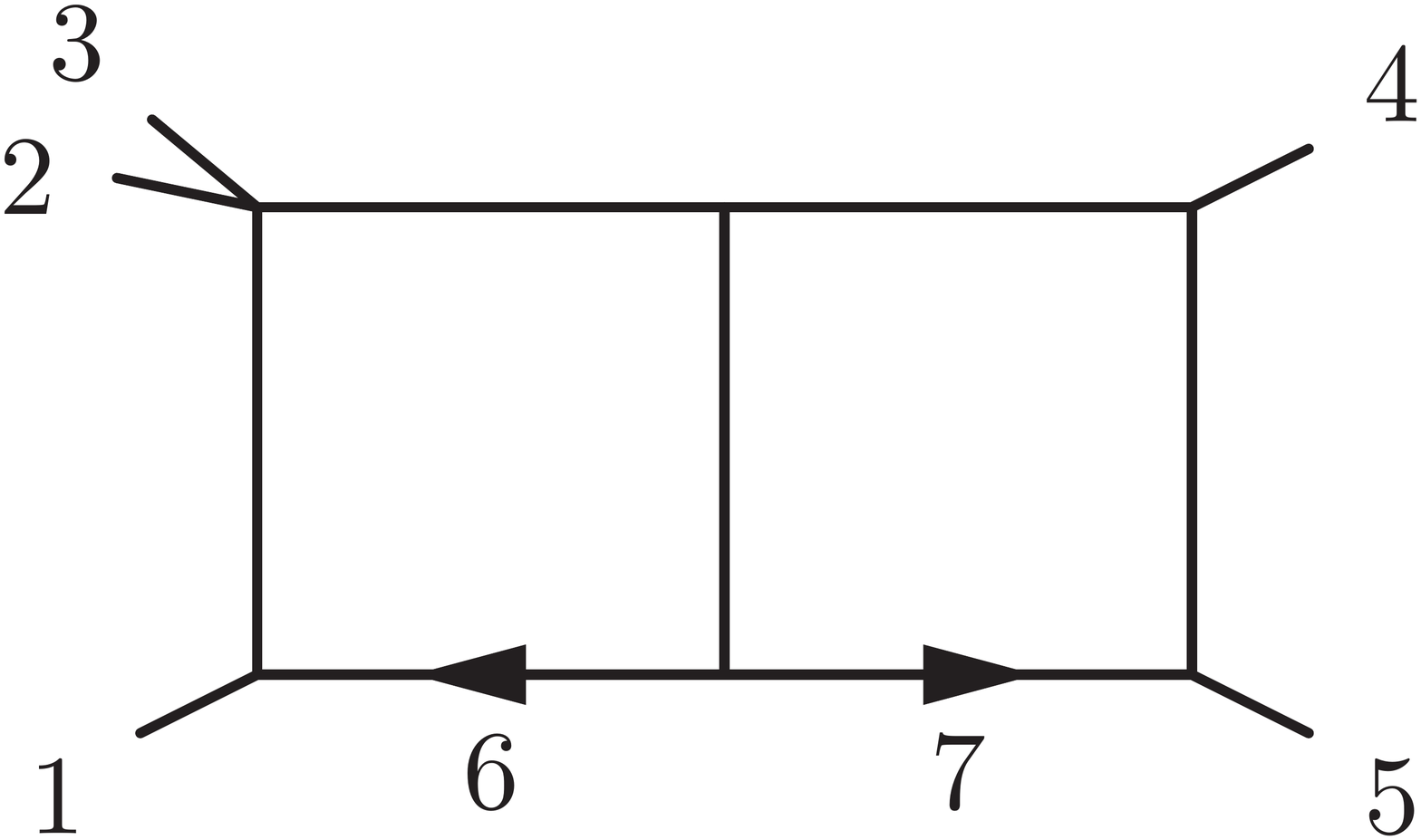} & 
\small
\begin{array}{r l}
N^{{\rm(e)}}_1 & =	\sab{15}^{\textcolor{white}{1}} \sab{45}^2
 \,,  \\
\end{array}
\\[-.1cm]
(f) & \includegraphics[scale=0.08]{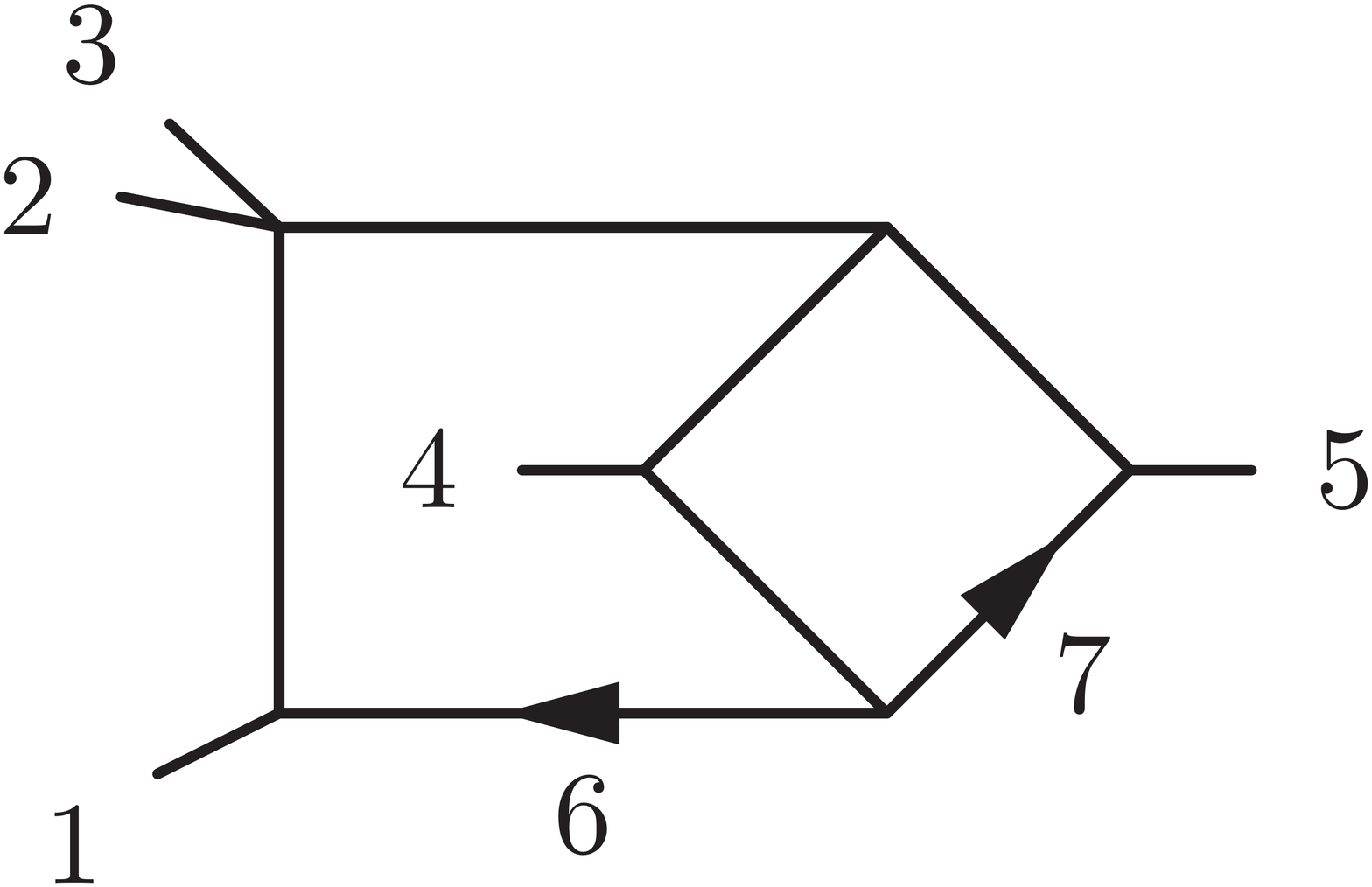} & 
\small
\begin{array}{r l}
N^{{\rm(f)}}_1 & =
	\sab{14}\sab{45}(\ell_6+k_5)^2
 \,,  \hskip0.6cm N^{{\rm(f)}}_2  = N^{{\rm(f)}}_1 \big|_{4 \lra 5}
 \,,  \\
\end{array}
\\[-.1cm]
(g) & \includegraphics[scale=0.08]{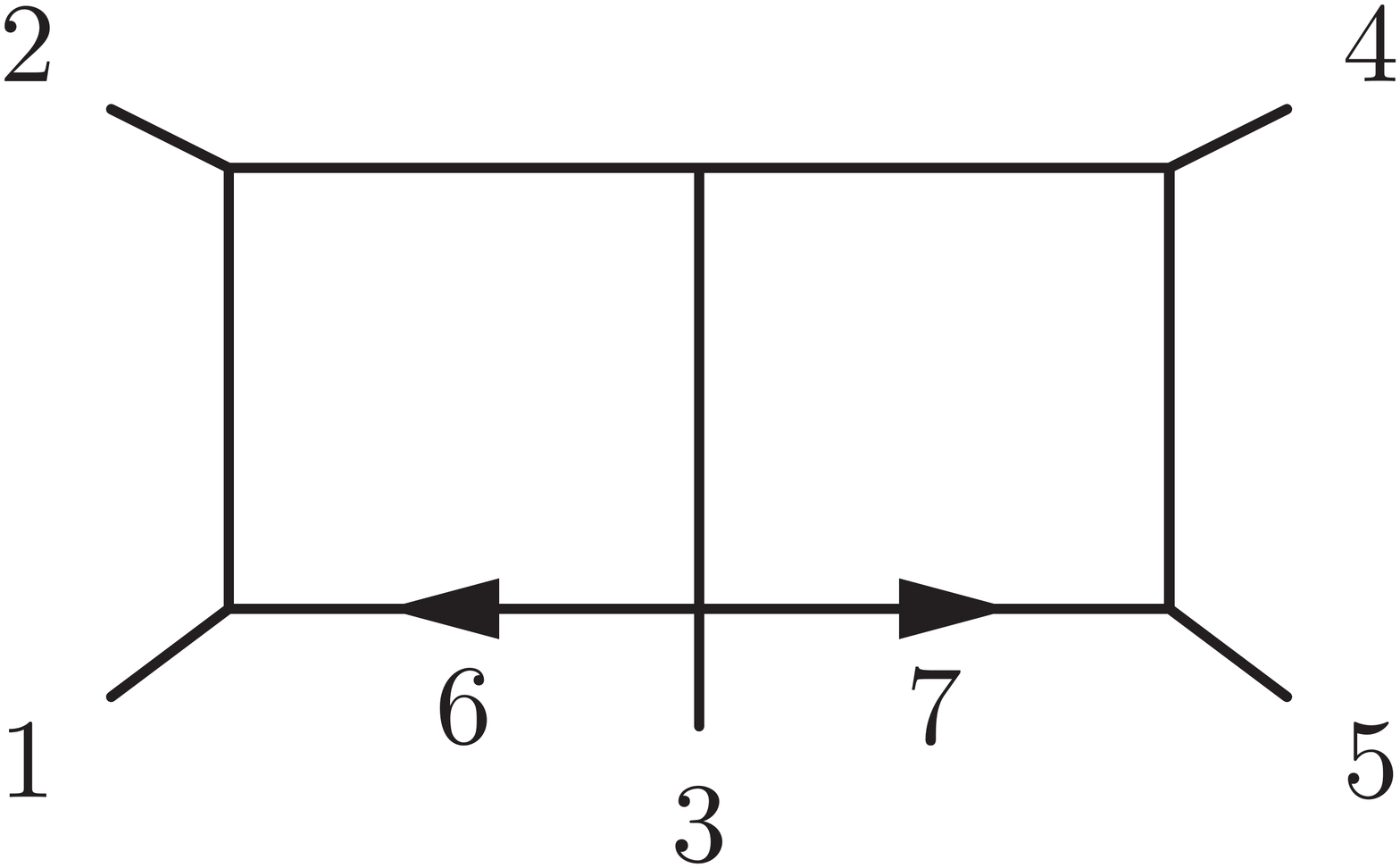} & 
\small
\begin{array}{r l}
N^{{\rm(g)}}_1 & =
	\sab{12}\sab{45}\sab{24}
 \,,  \\
\end{array}
\\[-.1cm]
(h) & \includegraphics[scale=0.08]{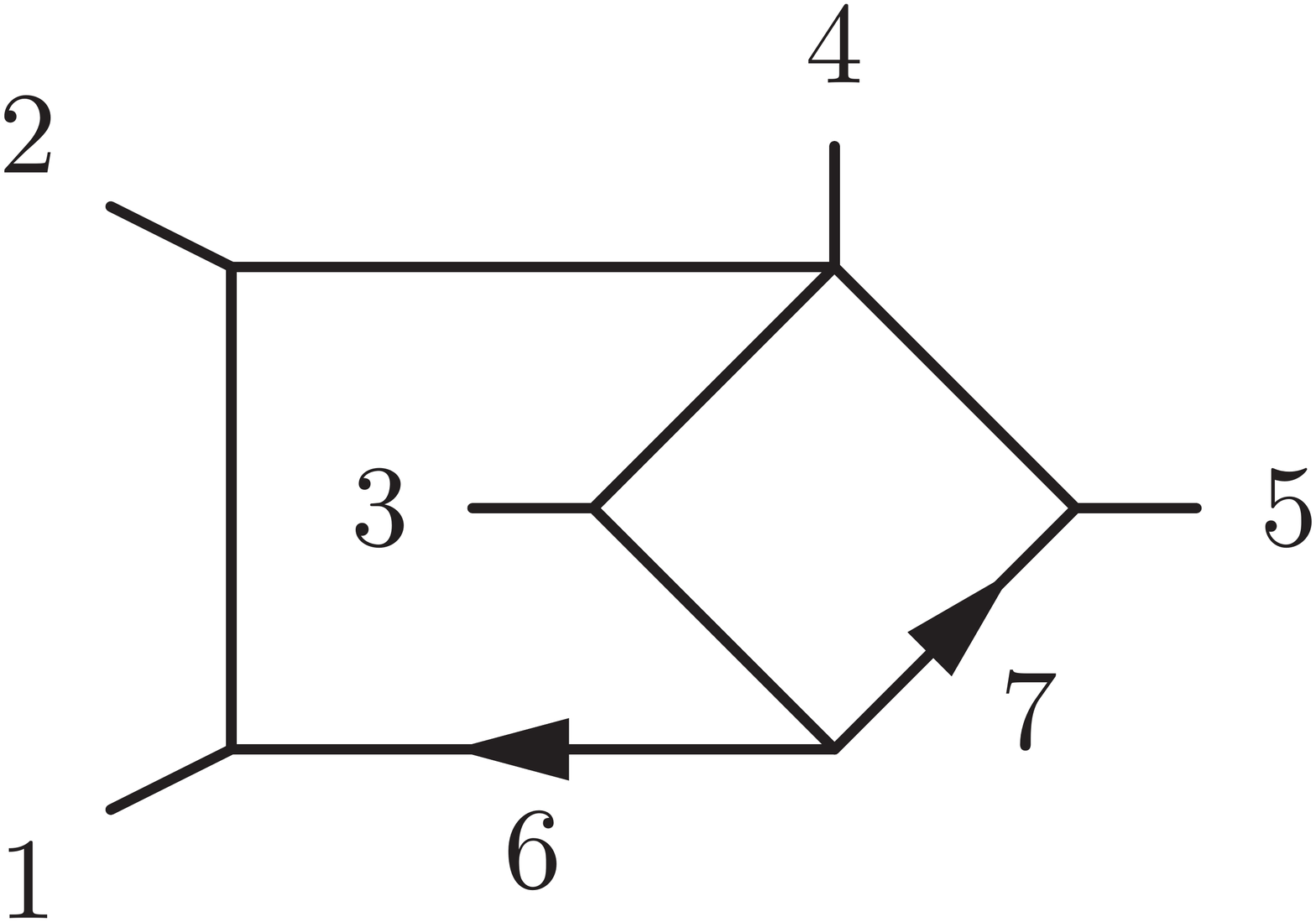} & 
\small
\begin{array}{r l}
N^{{\rm(h)}}_1 & =
	\ab{15}\sqb{35}\ab{23}\sqb{12} \left(\ell_6-\frac{\ab{12}}{\ab{32}}\lam{3}\lamt{1}\right)^2
 \,, \hskip 0.5cm
N^{{\rm(h)}}_2 = N^{{\rm(h)}}_1 \big|_{3 \lra 5}
 \,, \\
N^{{\rm(h)}}_3 & =
	\sab{12}\ab{13}\sqb{15}\aMs{5}{\ell_6}{3}
 \,, \quad 
N^{{\rm(h)}}_4   = \sab{12}\sqb{13}\ab{15}\aMs{3}{\ell_6}{5},  \\
 N^{{\rm(h)}}_5 & = \overline{N}^{{\rm(h)}}_1\,, \hskip.5cm
 N^{{\rm(h)}}_6   = \overline{N}^{{\rm(h)}}_2\,, \hskip.5cm
\\
\end{array}
\\[-.1cm]
(i) & \includegraphics[scale=0.08,angle=00]{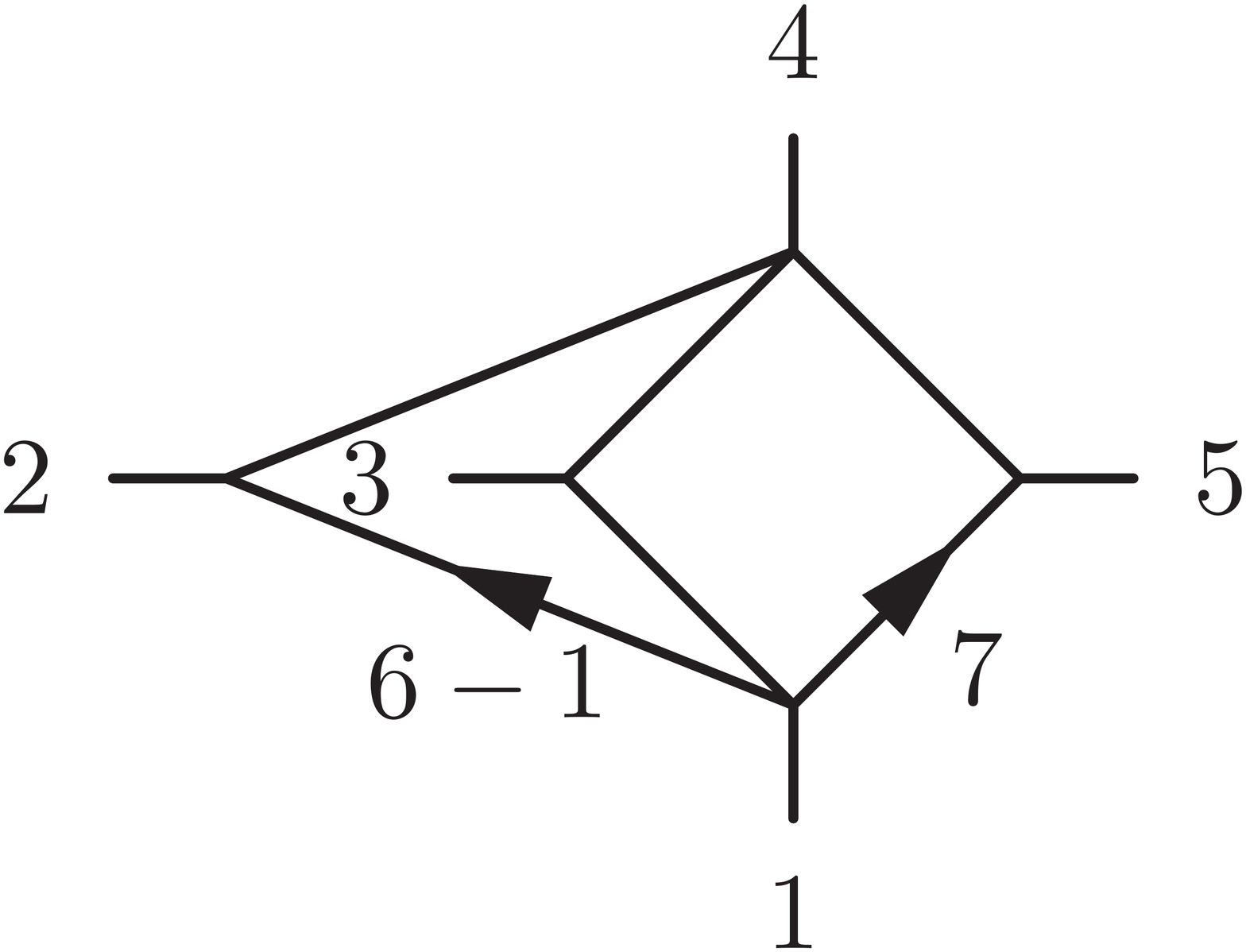} & 
\small
\begin{array}{r l}
N^{{\rm(i)}}_1 & =
	\aMs{2}{4}{3}\aMs{3}{5}{2}-\aMs{3}{4}{2}\aMs{2}{5}{3}
 \,.  \\
\end{array}
\end{tabular}
\vskip -.2cm
\caption{The parent diagram numerators that give pure integrands for the two-loop five-point amplitude.
  Each basis diagram is consistent with requiring logarithmic singularities and no poles at infinity. 
  	The overline notation means $\sqb{\cdot} \lra \ab{\cdot}$.
\label{tab:Basis2Loop5ptParents}
  }
\end{center}
\vskip -1. cm 
\end{table}

Following the three-loop four-point case,
our first step is to construct a pure integrand basis.
Constructing this basis is similar to constructing the
three-loop four-point amplitude in Ref.~\cite{ThreeLoopPaper} and
summarized in Sec.~\ref{subsec:ThreeLoop4pt}. Although deriving the
numerators for the two-loop five-point case is in principle
straightforward, it does require a nontrivial amount of algebra,
which we suppress. We again split the basis
elements according to diagram topologies and distinguish between
parent diagrams and contact diagrams. The numerators of
each pure integrand are given in \tab{Basis2Loop5ptParents}.

\Tab{Basis2Loop5ptParentsExtra} contains an additional pure
integrand. However we do not include it in our basis because it is
linearly dependent on two other basis elements: $ N^{(\rm h)}_1 -
N^{(\rm h)}_2 + N^{(\rm j)}_1(\ell_6-k_1)^2= 0 $.  In our result, we
choose $N^{(\rm h)}_1$ and $N^{(\rm h)}_2$ as our linearly independent
pure integrands, and only mention $N^{(\rm j)}_1$ because it might be
an interesting object in future studies.

\begin{table}[tb]
\begin{center}
\hspace{-.5cm}
\begin{tabular}[tb]
{ >{\centering\arraybackslash} m{0.02\textwidth}
 >{\centering\arraybackslash} m{0.16\textwidth}  
 >{$}l<{$} 
}
  & \hskip -1. cm  {\small Diagram} & {\small \textrm{Numerators}} \\
\hline \hline\\[-.2cm]
\vskip -.15cm
(j) &
\vskip -.15cm
\includegraphics[scale=0.08]{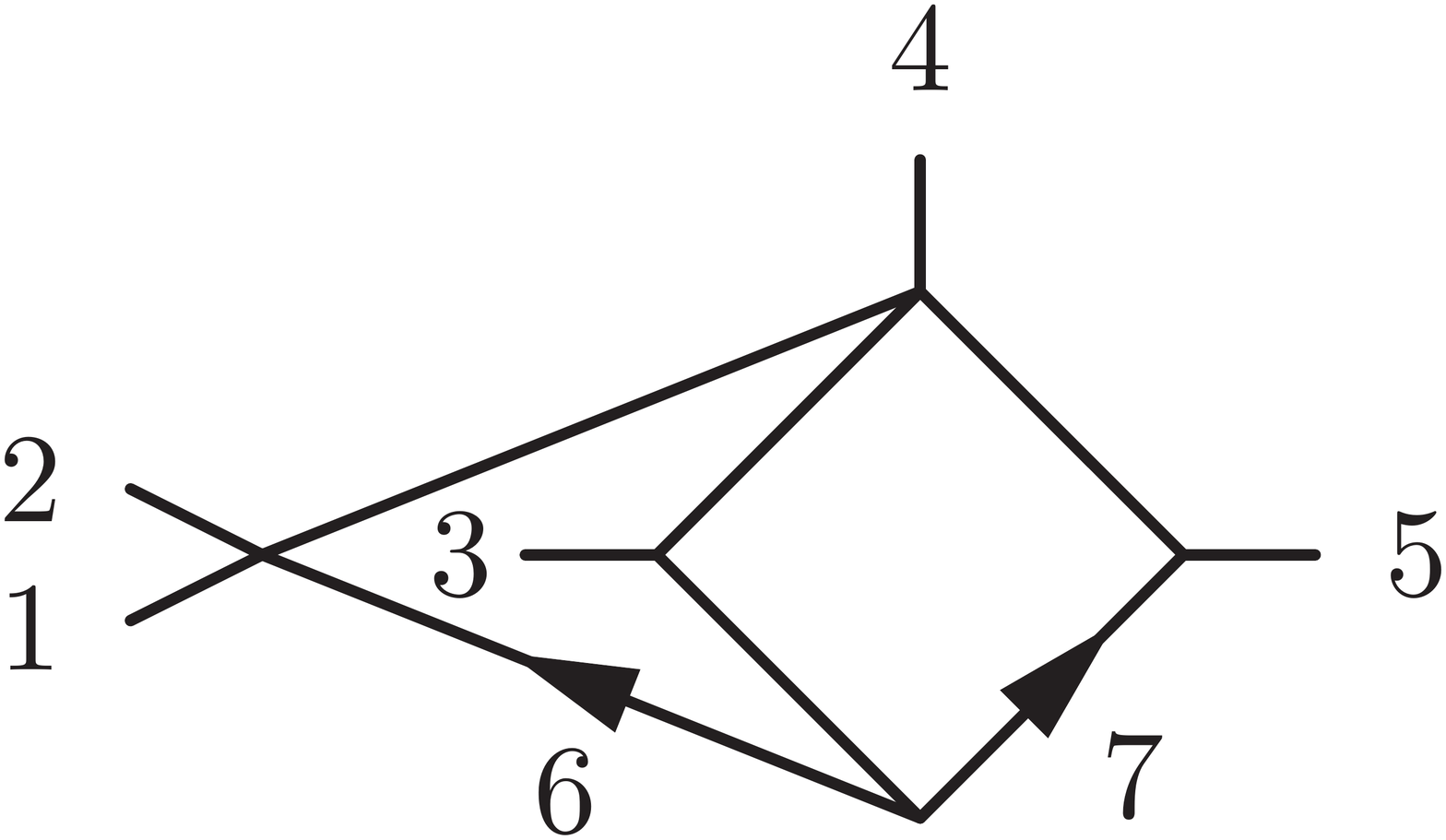} &
\small
\begin{array}{r l}
N^{{\rm(j)}}_1 & =
        \sab{12}\sab{35} = ( N^{{\rm(h)}}_2 - N^{{\rm(h)}}_1) / (\ell_6 - k_1)^2
 \,,  \\
\end{array}
\end{tabular}
\vskip -.2cm
\caption{A diagram and numerator that gives a pure integrand.  However, as indicated in the table and explained in the text,
it is not a an independent basis element. 
\label{tab:Basis2Loop5ptParentsExtra}
  }
\end{center}
\end{table}

In contrast to the three-loop four-point basis, 
in the two-loop five-point case it is useful to allow spinor helicity variables associated
with external momenta. Specifically, several of the expressions in \tab{Basis2Loop5ptParents} have
the structure $(\ell + \alpha \lam{i}\lamt{j})^2$, where $\alpha$ is such that both mass dimension and little group 
weights are consistent. For example, the penta-box numerator
\begin{equation}
N^{\rm (b)}_1 \sim
\left(\ell_6 + \frac{Q_{45}\cdot\lamt{3}\ \lamt{1}}{\sqb{13}} \right)^2 =
(\ell_6 - \ell_6^\ast)^2 \,,
\end{equation}
is a ``chiral'' numerator 
that manifestly vanishes on the $\overline{\text{MHV}}$ solution $\ell_6 = \ell^{\ast}_6$~\cite{ArkaniHamed:2010gh}.
As a shorthand notation, we use $Q_{ij} = k_i+k_j$.

\subsubsection{{\it Matching the amplitude}}
\label{subsubsec:2Loop5ptAmplitude}

Following the construction of the pure integrand basis in 
Sect.~\ref{subsubsec:2Loop5ptBasis} we are ready to build 
up the amplitude. In complete analogy to \eqn{ThreeFourBuildAmp}, 
the two-loop five-point amplitude is assembled from the basis numerators as,
\begin{equation}
{\cal M}_5^{2\hbox{-}\rm loop}  = \sum_{S_5}\sum_{x}\frac{1}{{\cal S}_{x}} 
		\int d^4 \ell_6 d^4 \ell_7
					\frac{\mathcal{N}^{(x)}}{\prod_{\alpha_x} p^2_{\alpha_x}}\,,
\label{eqn:TwoFiveBuildAmp}
\end{equation}
where the sum over $x$ runs over all diagrams in the basis listed in \tab{Basis2Loop5ptParents}, 
the sum over $S_5$ is a sum over all 120 permutations of the external legs,
and $\mathcal{S}_{x}$ is the symmetry factor of diagram $x$. 
The product over $\alpha_x$ indicates the product of Feynman propagators $p^2_{\alpha_x}$ of diagram $x$, 
as read from the graphs in \tab{Basis2Loop5ptParents}.

We refer the reader to the discussion in Sect.~\ref{sect:Matching}
for explicit examples on how to read \tab{TwoLoopFivePointSol}.
We choose the following set of independent five-point Parke-Taylor basis elements:
\begin{equation}
\begin{array}{c}
\PT_1 = \PT(12345) \,, \qquad
\PT_2 = \PT(12354) \,, \qquad
\PT_3 = \PT(12453) \,, \\
\PT_4 = \PT(12534)  \,, \qquad
\PT_5 = \PT(13425)  \,, \qquad
\PT_6 = \PT(15423) \,.
\end{array}
\label{eqn:TwoLoopFivePointPTBasis}
\end{equation}
The basis elements $\overline{N}^{(x)}$ in \tab{TwoLoopFivePointSol} do not contribute to the MHV amplitude so those data are omitted from the ${\tt a}_{\nu\sigma}^{(x)}$.

\begin{table}[tb]
\centering
\begin{tabular}
{
>{$}r<{$} 
>{$}l<{$} 
>{$}l<{$} 
}
& \hskip -1. cm {\small \textrm{Color Dressed\ Numerators}} & {\small \hskip 1cm \textrm{PT\ Matrices}} \\
\hline \hline\\[-.3cm]
\mathcal{N}^{\rm (a)} =
& \hspace{-.25cm}
c^{\rm (a)}_{12345} {\displaystyle \sum_{\substack{1 \le \nu \le 4 \\ 1 \le \sigma \le 6 }}}
N^{\rm (a)}_\nu {\tt a}_{\nu\sigma}^{\rm (a)} \PT_\sigma \,,
& \hspace{.5cm}
{\tt a}^{\rm (a)}_{\nu\sigma} = 
\dfrac{1}{4}
\left(
\begin{array}{cccccc}
 -1 & 0 & 1 & 0 & 0 & 2 \\
 1 & 0 & -1 & 0 & 0 & 2 \\
 -3 & 0 & -1 & 0 & 0 & 2 \\
 -1 & 0 & -3 & 0 & 0 & 2 \\
\end{array}
\right) ,
\\\hspace{.5cm}
\mathcal{N}^{\rm (b)} =
& \hspace{-.25cm}
c^{\rm (b)}_{12345} {\displaystyle \sum_{1 \le \sigma \le 6 }}
N^{\rm (b)}_1 {\tt a}_{1\sigma}^{\rm (b)} \PT_\sigma \,,
&\hspace{.5cm}
{\tt a}^{\rm (b)}_{1\nu} = 
\left(
\begin{array}{cccccc}
-1 & 0 & 0 & 0 & 0 & 0 \\
\end{array}
\right) ,
\\
\mathcal{N}^{\rm (c)} =
& \hspace{-.25cm}
c^{\rm (c)}_{12345} {\displaystyle \sum_{\substack{1 \le \nu \le 2 \\ 1 \le \sigma \le 6 }}}
N^{\rm (c)}_\nu {\tt a}_{\nu\sigma}^{\rm (c)} \PT_\sigma \,,
&\hspace{.5cm}
{\tt a}^{\rm (c)}_{\nu\sigma} = 
\left(
\begin{array}{cccccc}
-1 & 0 & 0 & 0 & 0 & 0 \\
0 & -1 & 0 & 0 & 0 & 0 \\
\end{array}
\right) ,
\\
\vphantom{ \Bigg| } 
\mathcal{N}^{\rm (d)} =  &\hspace{-.25cm}	\mathcal{N}^{\rm (e)} = \mathcal{N}^{\rm (f)} = 0 \,, &
\\
\mathcal{N}^{\rm (g)} =
& \hspace{-.25cm}
c^{\rm (a)}_{12345} {\displaystyle \sum_{ 1 \le \sigma \le 6 }} N^{\rm (g)}_1 {\tt a}_{1\sigma,(12345)}^{\rm (g)} \PT_\sigma
&\hspace{.5cm}
{\tt a}^{\rm (g)}_{1\sigma,(12345)} =
\dfrac{1}{4} \left(
\begin{array}{cccccc}
1 & 0 & 3 & 0 & 0 & -2 \\
\end{array}
\right) ,
\\
& \null +
c^{\rm (b)}_{31245} {\displaystyle \sum_{ 1 \le \sigma \le 6 }} N^{\rm (g)}_1 {\tt a}_{1\sigma,(31245)}^{\rm (g)} \PT_\sigma \,,
&\hspace{.5cm}
{\tt a}^{\rm (g)}_{1\sigma,(31245)} =
\left(
\begin{array}{cccccc}
 0 & 0 & -1 & 0 & 0 & 0 \\
\end{array}
\right) ,
\\
\mathcal{N}^{\rm (h)} =
& \hspace{-.25cm}
c^{\rm (a)}_{12345} {\displaystyle \sum_{ \substack{ 1 \le \nu \le 4 \\ 1 \le \sigma \le 6} }} N^{\rm (h)}_\nu {\tt a}_{\nu\sigma,(12345)}^{\rm (h)} \PT_\sigma
& \hspace{.5 cm}
	{\tt a}_{\nu\sigma,(12345)}^{\rm (h)}  = 
		   \dfrac{1}{4} \begin{pmatrix}
 		 4 & 0 & 4 & 0 & 0 & -4 \\
		 2 & 0 & 3 & 0 & 1 & -2 \\
		 -2 & 0 & -3 & 0 & -1 & 2 \\
		 4 & 0 & 4 & 0 & 0 & -4 \\
		\end{pmatrix} ,
\\
& \null
+ c^{\rm (a)}_{12543} {\displaystyle \sum_{  \substack{ 1 \le \nu \le 4 \\1 \le \sigma \le 6} }} N^{\rm (h)}_\nu {\tt a}_{\nu\sigma,(12543)}^{\rm (h)} \PT_\sigma \,,
& \hspace{.5cm}
{\tt a}_{\nu\sigma,(12543)}^{\rm (h)}  = {\tt a}_{\nu\sigma,(12345)}^{\rm (h)}\,,
\\
\mathcal{N}^{\rm (i)} = & 
\hspace{-.2cm}   
c^{\rm (a)}_{12345} {\displaystyle \sum_{ 1 \le \sigma \le 6 }} N^{\rm (i)}_1 {\tt a}_{1\sigma,(12345)}^{\rm (i)} \PT_\sigma
&\hspace{.5cm}
{\tt a}^{\rm (i)}_{1\sigma,(12345)} = 
2 \left(
\begin{array}{cccccc}
0 & 0 & -1 & 0 & 0 & 1 \\
\end{array}
\right) ,
\\
&  
\null + c^{\rm (a)}_{13245} {\displaystyle \sum_{ 1 \le \sigma \le 6 }} N^{\rm (i)}_1 {\tt a}_{1\sigma,(13245)}^{\rm (i)} \PT_\sigma
&\hspace{.5cm} 
{\tt a}^{\rm (i)}_{1\sigma,(13245)} = 
2 \left(
\begin{array}{cccccc}
 0 & 0 & 0 & 0 & 0 & 1 \\
\end{array}
\right) ,
\\
& \null
 + c^{\rm (a)}_{12543} {\displaystyle \sum_{ 1 \le \sigma \le 6 }} N^{\rm (i)}_1 {\tt a}_{1\sigma,(12543)}^{\rm (i)} \PT_\sigma
&\hspace{.5cm}
{\tt a}^{\rm (i)}_{1\sigma,(12543)} = 
2 \left(
\begin{array}{cccccc}
1 & 0 & 1 & 0 & 1 & -1 \\
\end{array}
\right) ,
\\
& \null
 + c^{\rm (a)}_{15243} {\displaystyle \sum_{ 1 \le \sigma \le 6 }} N^{\rm (i)}_1 {\tt a}_{1\sigma,(15243)}^{\rm (i)} \PT_\sigma \,,
& \hspace{.5cm}
{\tt a}^{\rm (i)}_{1\sigma,(15243)} = 
2 \left(
\begin{array}{cccccc}
0 & 0 & 0 & 0 & -1 & 0 \\
\end{array}
\right).
\\
\end{tabular}
\caption[]{
The two-loop five-point numerators that contribute to the amplitude.
The $N^{(x)}_{\nu}$ are listed in \tab{Basis2Loop5ptParents}.
The five-point $\PT_\sigma$ are listed in \eqn{TwoLoopFivePointPTBasis}. We denote the 
numerators including color information as $\mathcal{N}^{(x)}$. 
}
\label{tab:TwoLoopFivePointSol}
\end{table}

\section{Zeros of the Integrand}
\label{sec:Zeros}

In the previous section we gave explicit examples of the expansion of
the amplitude, \eqn{gen2}, in terms of a basis of pure integrands,
giving new nontrivial evidence that the analytic consequences of dual
conformal symmetry hold beyond the planar sector.  In this section we take a
further step and present evidence that the amplituhedron concept,
which is a complete and self-contained geometric definition of the
integrand, may exist beyond the planar sector as well.

As already mentioned in previous sections, beyond the planar limit we
currently have no alternative other than to use diagrams representing
local integrals, \eqn{gen2}, as a starting point for defining nonplanar
integrands.  The lack of global variables makes it unclear how to
directly test for a geometric construction analogous to the amplituhedron in
the nonplanar sector.  However, as discussed in
\sect{DualPicture}, in the planar sector the (dual) amplituhedron
construction implies that all coefficients in the expansion
in \eqn{exp1} are determined by zero conditions, up to an overall
normalization. We expect that if an analogous geometric construction
exists in the nonplanar sector, then zero conditions should also
determine the amplitude. This can be tested directly. Indeed, we
conjecture that for the representation in \eqn{gen2}:

\begin{center}
\framebox[1.05\width]{All coefficients $d_j$ 
 are fixed by {\it zero conditions}, up to overall normalization.}
\end{center}
This is the direct analog of the corresponding planar statement in \sect{DualPicture}.  
In the MHV case, which we consider here, the coefficients $d_j$ are linear combinations
of Parke-Taylor factors, so that only numerical coefficients
$a_{\sigma,k,j}$ in \eqn{gen3} need to be determined. 
The above conjecture is a statement that we can obtain these coefficients
using only zero conditions, up to an overall constant.
Here we confirm this proposal for all amplitudes constructed in
the previous section.

\begin{figure}[tb]
\begin{center}
\includegraphics[scale=1]{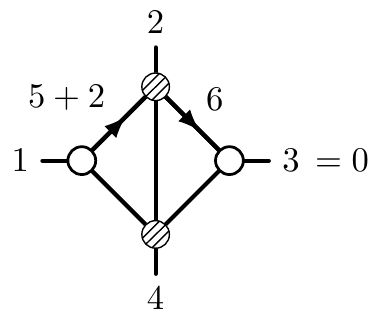}
\caption[]{The two-loop four-point MHV amplitude vanishes on this
  cut. The four-point trees in the diagram have $k=2$,
  so the overall helicity counting is $k=1$.}
\vskip -.2 cm 
\label{fig:IllegalCut}
\end{center}
\end{figure}

As a simple first example, consider the two-loop four-point
amplitude. The integrand is given as a linear combination of planar
and nonplanar double boxes, c.f. Sect.~\ref{subsec:TwoLoop4ptAmpl}.
The only required condition to determine the 
unknown conditions is the cut in \fig{IllegalCut}.

\begin{figure}[t]
\centering
\includegraphics[scale=1.2]{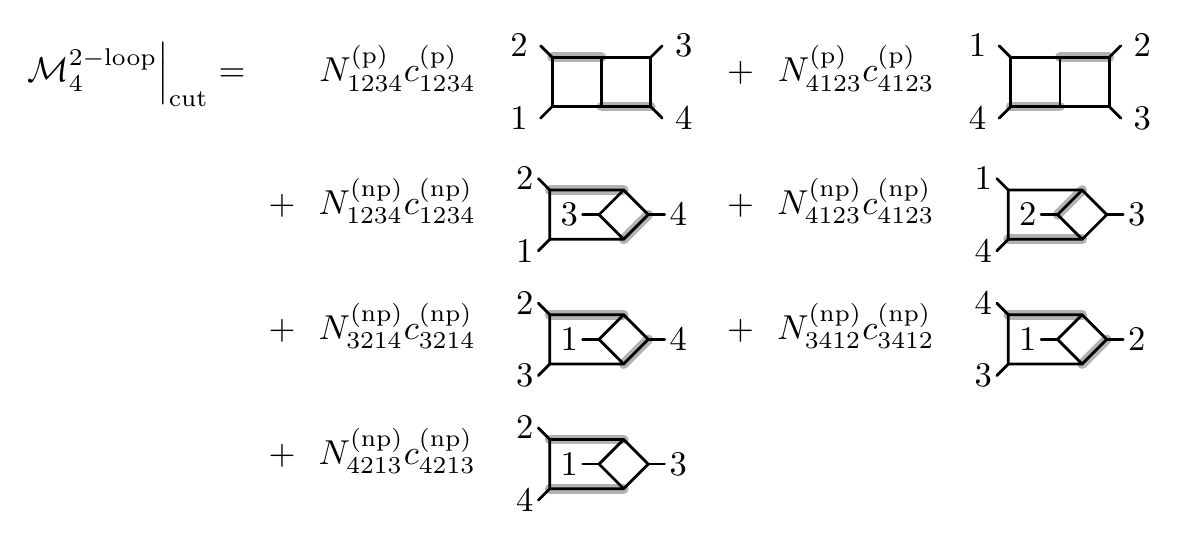}
\caption[]{The two-loop four-point amplitude evaluated on the cut
of \fig{IllegalCut}.  In each diagram the two shaded propagators are
uncut, and every other propagator is cut. \Eqn{TwoLoopVanishingCut} gives the value of the cut.}
\label{fig:2loopNPZeroCutDiags}
\end{figure}

In the full amplitude, we have contributions from the planar and
nonplanar double boxes in \fig{2loop4ptParents} and their permutations
of external legs.  All permutations of diagrams that contribute to the cut 
in \fig{IllegalCut} are shown in \fig{2loopNPZeroCutDiags}, along with their
numerators and color factors. For convenience, we indicate the permutation 
labels of external legs of the seven contributing diagrams. There are only seven diagrams
rather than nine because two of the nine diagrams have triangle
subdiagrams, and so have vanishing numerators in $\NeqFour$ SYM.

For the cut in \fig{IllegalCut}, five propagators are put on-shell so
that the cut solution depends on $\alpha$, $\gamma$, and $\delta$,
three unfixed parameters of the loop momenta.
Explicitly, the cut solution is
\begin{align}
\ell_{5}^*+k_{2} & = \lam{1} \left[ \alpha\lamt{1}+\frac{1}{\delta\ab{13}\sqb{23}}
							\big(\delta t
						 -\alpha(s+\delta u + \gamma\ab{13}\sqb{12})
										\big)\lamt{2}
									 \right] \,, 
\nonumber \\
\ell_{6}^* & = \lam{3} \left[ \delta\lamt{3}+\gamma\lamt{2} \right].
\end{align}
On this $k=1$ cut, the MHV amplitude vanishes for any values of $\alpha$,
$\gamma$, $\delta$. By cutting the Jacobian 
\begin{equation}
J = \gamma \big(\delta t - \alpha (s+\delta u + \gamma\ab{13}\sqb{12})\big)\,,
\end{equation}
the amplitude remains zero, and this condition simplifies.
Specifically, this allows us to localize $\ell_{5}+k_{2}$
to be collinear with $k_{1}$ and to localize $\ell_{6}$ to be collinear with
$k_{3}$. This is equivalent to taking further residues of the already-cut integrand at
$\gamma=0,\ \alpha =  {\delta t}/{(s+\delta u)}$. 
On this cut, the solution for the loop momenta simplifies,
\begin{align}
\ell_{5}^*+k_{2} & =  \frac{\delta t}{s+\delta u}\lam{1}\lamt{1}\,,
\qquad
\ell_{6}^* = \delta\lam{3}\lamt{3}\,,
\label{eqn:2loop4point1paramZeroCutSol}
\end{align}
with the overall Jacobian $J'=s+u\delta$. Even in this simplified
setting with one parameter $\delta$ left, the single zero cut
condition \fig{IllegalCut} is sufficient to fix the integrand up to an overall constant.

The numerators for the pure integrands, using the labels in \fig{2loop4ptParents}, are
given in \eqn{num1}.  Including labels for the external legs 
to help us track relabellings, these are
\begin{align}
N_{1234}^{(\rm p, 1)}  = s^{2}t,\quad
N_{1234}^{(\rm np, 1)} = su\left(\ell_{5}-k_{3}\right)^{2},\quad
N_{1234}^{(\rm np, 2)} = st\left(\ell_{5}-k_{4}\right)^{2} \,.
\label{eqn:TwoFourNumsWithExt}
\end{align}
As noted near \eqn{ThreeLoopFourPointPTBasis} there are only two 
Parke-Taylor factors independent under $U(1)$ relations for
four-particle scattering, namely $\PT_1 = \mathrm{PT}\left(1234\right)$ and 
$\PT_2 = \mathrm{PT}\left(1243\right)$. Therefore the numerator ansatz for the
planar diagram is
\begin{equation}
N_{1234}^{(\rm p)}=
	\left(a_{1,1}^{(\rm p)}\ \PT_1
       +a_{1,2}^{(\rm p)}\ \PT_2
	\right) N_{1234}^{(\rm p, 1)}.
\end{equation}
For the nonplanar diagram, there are two
pure integrands, each of which gets decorated with the two independent
Parke-Taylor factors, so that the ansatz takes the form
\begin{align}
N_{1234}^{(\rm np)} =  \Big[
													&\left(a_{1,1}^{(\rm np)}\ \PT_1
															 + a_{1,2}^{(\rm np)}\ \PT_2
													 \right)N_{1234}^{(\rm np, 1)}  \nonumber \\ 
											  + &\left(a_{2,1}^{(\rm np)}\ \PT_1
															  +a_{2,2}^{(\rm np)}\ \PT_2
													 \right)N_{1234}^{(\rm np, 2)} 
											\Big]\,,
\end{align}
and both numerators are then decorated with corresponding color
factors $c_{1234}^{(\rm p)}$, $c_{1234}^{(\rm np)}$ and 
propagators. The $a_{i,j}^{(x)}$ coefficients are determined by
demanding the integrand vanishes on the cut solution in
\eqn{2loop4point1paramZeroCutSol}.

Explicitly, the zero condition from the cut corresponding to \fig{2loopNPZeroCutDiags} is:
\begin{eqnarray}
0 & = & \left(\frac{c_{1234}^{(\rm p)} N_{1234}^{\left(\rm p\right)}}
                             {\ell_{5}^{2}\left(\ell_{6}-k_{3}-k_{4}\right)^{2}}
   + \frac{c_{4123}^{(\rm p)} N_{4123}^{\left(\rm p\right)}}
                                  {\left(\ell_{5}-k_{3}\right)^{2}\left(\ell_{6}+k_{2}\right)^{2}}			
   + \frac{c_{1234}^{(\rm np)} N_{1234}^{\left(\rm np\right)}}
                            {\ell_{5}^{2}\left(\ell_{5}-\ell_{6}-k_{4}\right)^{2}}
			\right.\nonumber \\
		&  & \left.\left.
  \null + \frac{c_{4123}^{(\rm np)} N_{4123}^{\left(\rm np\right)}}
           {\left(\ell_{5}-k_{3}\right)^{2}\left(\ell_{5}-\ell_{6}+k_{2}\right)^{2}}
   + \frac{c_{3214}^{(\rm np)} N_{3214}^{\left(\rm np\right)}}
     {\left(\ell_{6}+k_{2}\right)^{2}\left(\ell_{5}-\ell_{6}-k_{4}\right)^{2}}
		\right.\right. \nonumber\\
	&  & \left. \left.
  \null +\frac{c_{3412}^{(\rm np)}N_{3412}^{(\rm np)}}
     {\left(\ell_{6}-k_{3}-k_{4}\right)^{2}\left(\ell_{5}-\ell_{6}+k_{2}\right)^{2}}
	+\frac{c_{4213}^{(\rm np)} N_{4213}^{(\rm np)}}
      {\left(\ell_{5}-\ell_{6}+k_{2}\right)^{2}\left(\ell_{5}-\ell_{6}-k_{4}\right)^{2}}\right)\right|_{\ell_5^\ast\,, \ell_6^\ast} \hspace{-.2 cm} .
\label{eqn:TwoLoopVanishingCut}
\end{eqnarray}
The sum runs over the seven contributing diagrams, following the order
displayed in \fig{2loopNPZeroCutDiags}. The denominators are the two
propagators that are left uncut in each diagram when performing this
cut.  One of the terms in the cut equation, for example, is
\begin{eqnarray}
\label{eqn:ExampleOnCut}
\frac{N_{3214}^{(\rm np)}}{\left(\ell_{6}+k_{2}\right)^{2}\left(\ell_{5}-\ell_{6}-k_{4}\right)^{2}} 
& = & \frac{1} {\left(\ell_{6}+k_{2}\right)^{2}\left(\ell_{5}-\ell_{6}-k_{4}\right)^{2}}
			\\
& & \times \left[\left(a_{1,1}^{(\rm np)}\mathrm{PT}\left(3214\right)
					     +a_{1,2}^{(\rm np)}\mathrm{PT}\left(4213\right)
					\right)tu\left(\ell_{6}+k_{1}+k_{2}\right)^{2}\right. \nonumber \\
 &  & \left.
       \null +\left(a_{2,1}^{(\rm np)}\mathrm{PT}\left(3214\right)
				      +a_{2,2}^{(\rm np)}\mathrm{PT}\left(4213\right)
				 \right)st\left(\ell_{6}+k_{2}+k_{4}\right)^{2}
	\right]	\,.	\nonumber
\end{eqnarray}
This has been relabeled from the master labels of \eqn{TwoFourNumsWithExt} to the labels of the third
nonplanar diagram in \fig{2loopNPZeroCutDiags}, including the two uncut propagators. Specifically
$\ell_{5}\mapsto-\ell_{6}-k_{2}$ and $\ell_{6}\mapsto-\ell_{5}-k_{1}$
is the relabeling for this diagram. A key simplifying feature is that
the $a_{i,j}^{(x)}$ coefficients do not change under this relabeling
so as to maintain crossing symmetry; the same four coefficients
contribute to all five of the nonplanar double boxes that appear, for
example. 
As discussed in Sect.~\ref{sect:Matching}, the Parke-Taylor factors that appear in
\eqn{ExampleOnCut} do not necessarily need to be in the chosen basis,
although here $\PT(3214) = \PT_1$ and $\PT(4213) = \PT_2$.

The single zero condition \eqn{TwoLoopVanishingCut}
determines five of the six $a_{ij}^{(x)}$
parameters. This is, consistent with our conjecture above, 
the maximum amount of information that we can extract 
from all zero conditions. To do so in this example, we reduce to 
the two-member Parke-Taylor basis mentioned before, and also use 
Jacobi identities to reduce the seven contributing color factors 
to a basis of four. Since the remaining Parke-Taylor and color factors
are independent, setting the coefficients of $\PT \cdot c$ to zero
yields eight potentially independent 
equations for the six coefficients. It turns out only five 
are independent:
\begin{eqnarray}
&&a_{1,2}^{\left(\rm p\right)}=a_{1,1}^{\left(\rm p\right)}+3a_{1,1}^{\left(\rm np\right)}
+a_{2,1}^{\left(\rm np\right)}= a_{1,1}^{\left(\rm p\right)}+a_{1,1}^{\left(\rm np\right)}+a_{2,1}^{\left(\rm np\right)}=0, \\
&&2a_{1,2}^{\left(\rm p\right)}-a_{1,1}^{\left(\rm np\right)}+a_{1,2}^{\left(\rm np\right)}
	  -a_{2,1}^{\left(\rm np\right)}+a_{2,2}^{\left(\rm np\right)}=
   a_{1,2}^{\left(\rm p\right)}+a_{1,1}^{\left(\rm np\right)}+a_{1,2}^{\left(\rm np\right)}
       -a_{2,1}^{\left(\rm np\right)}+3a_{2,2}^{\left(\rm np\right)}=0.\nonumber
\end{eqnarray}
The solution for this system is 
\begin{equation}
a_{1,2}^{\left(\rm p\right)}  = a_{1,1}^{\left(\rm np\right)} = 
a_{2,2}^{\left(\rm np\right)}  =  0\,, \hskip 1 cm 
a_{1,2}^{\left(\rm np\right)} = a_{2,1}^{\left(\rm np\right)}   = -a_{1,1}^{\left(\rm p\right)} \,,
\end{equation}
and any of $a_{1,2}^{\left(\rm np\right)}$, $a_{2,1}^{\left(\rm np\right)}$, or $a_{1,1}^{\left(\rm p\right)}$
is the overall undetermined parameter. 
This matches the result in \eqn{TwoFourSolution}, if we take 
$a_{1,1}^{\rm (p)} = 1$. This last condition is exactly the overall 
scale that the zero conditions cannot determine.

\begin{figure}
\centering
\includegraphics[scale=1]{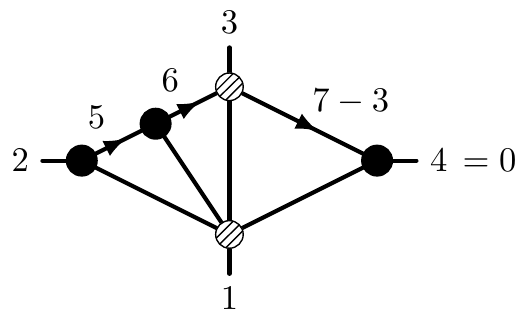}
\caption[]{The three-loop four-point MHV amplitude vanishes on this
  cut. The five-point tree at the bottom of the diagram has $k=2$ or
  $k=3$, so the overall helicity counting is $k=3$ or $k=4$.}
   \label{fig:3loop4ptNPzeroCut}
\end{figure}

\begin{figure}
\centering
\includegraphics[scale=1]{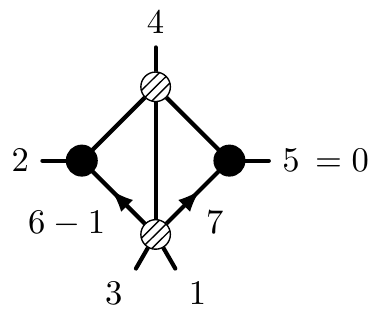}
\caption[]{
The two-loop five-point MHV amplitude vanishes on this
  cut. The five-point tree at the bottom of the diagram has $k=2$ or
  $k=3$, so the overall helicity counting is $k=3$ or $k=4$.}
\label{fig:2loop5ptNPzeroCut}
\end{figure}

Finally, we confirmed that the three-loop four-point and two-loop
five-point amplitudes can also be uniquely determined via a zero cut
condition up to a single overall constant.  We used the cut in
\fig{3loop4ptNPzeroCut} to determine the arbitrary parameters in 
the three-loop four-point
amplitude, and we used the cut in \fig{2loop5ptNPzeroCut} to determine the
parameters of the two-loop five-point amplitude.  We also confirmed in both cases that
using one cut where the amplitude does not vanish is sufficient to determine
the overall unfixed parameter to the correct value.  To confirm that
the so-constructed amplitudes are correct, we verified a 
complete set of unitarity cuts needed to fully determine the amplitudes,
matching to the corresponding cuts of previously known results in
Refs.~\cite{GravityThreeLoop,BCJLoop,HenrikJJ}.

We thus conclude that in all three examples that we analyzed, the
coefficients in the expansion \eqn{gen3} are determined up to one
constant by zero conditions. The set of relations is more complicated
in the three-loop four-point and two-loop five-point examples than in
the two-loop four-point example, but in all cases all coefficients are
determined as simple rational numbers without any kinematic dependence,
leaving one overall coefficient free.

\medskip

While far from a proof, these results point to the existence of an
amplituhedron-like construction in the nonplanar sector of the
theory. As discussed in \sect{DualPicture}, in the planar sector 
the existence of such a
construction implies that homogeneous conditions determine the
amplitudes up to an overall normalization.  This is indeed what we
have found in the various nonplanar examples studied above: the
homogeneous requirements of only logarithmic singularities, no poles
at infinity and vanishing of unphysical cuts do determine the
amplitudes.  In any case, the notion that homogeneous conditions 
fully determine amplitudes opens a door to applying these ideas
to other theories where no geometric properties are expected.  Of
course, we ultimately would like a direct amplituhedron-like geometric
formulation of $\NeqFour$ SYM amplitudes, including the nonplanar
contributions.  As a next step we would need sensible global
variables that allow us to define a unique integrand.

\section{Conclusion}
\label{sec:Conclusion}

In this paper we found evidence that an
amplituhedron-like construction of nonplanar $\NeqFour$ SYM theory scattering
amplitudes may exist. We did so by
checking the expected consequences of such a construction: that the integrand should be
determined by homogeneous conditions, such as vanishing on certain
cut solutions.  We also gave additional nontrivial
evidence for the conjecture that only logarithmic singularities appear
in nonplanar amplitudes~\cite{Log,ThreeLoopPaper}, which is another 
characteristic feature of planar amplitudes resulting from the 
amplituhedron construction.

An important complication is that unlike in the planar sector, there is
no unique integrand of scattering amplitudes which can be directly
interpreted as a volume of some space.  This forced us to chop up the
amplitude into local diagrams containing only Feynman propagators.
As pointed out in Ref.~\cite{Log} and further developed in
Ref.~\cite{ThreeLoopPaper}, analytic properties that follow from dual
conformal invariance can be imposed on such local diagrams.  We developed the
notion of a pure integrand basis: a basis of integrands with
only logarithmic singularities, no poles at infinity and 
only unit leading singularities. The first property is motivated by the 
analogous statement for on-shell diagrams in ${\cal N}=4$ SYM. 
If, like in the planar case, we understood how to formulate nonplanar
recursion relations, we expect that it would then be possible to
express nonplanar amplitudes directly as sums of on-shell
diagrams~\cite{OnshellDiagrams,NonPlanarOnshellToAppear,
  Franco:2015rma} and manifestly expose their $\dlog$ structure.
The latter two properties lift the exact content of dual conformal 
symmetry in the planar sector to the nonplanar one.

We constructed a pure integrand basis
for each of the two-loop four-point, three-loop four-point and
two-loop five-point amplitudes, and showed
that the amplitudes could be expanded in their respective bases.
This confirmed that the three example
amplitudes share the three properties of the pure integrands. 
Our pure integrand representations here are closely related to
Refs.~\cite{Log,ThreeLoopPaper} for four-point amplitudes at two- and
three-loops, while our representation of the two-loop five-point amplitude 
has completely novel properties compared to the result in Ref.~\cite{HenrikJJ}.
The fact that we exposed analytic properties in the nonplanar sector 
similar to those connected to dual conformal symmetry in the
planar sector suggests that an analog of dual conformal symmetry
may exist in the nonplanar sector. (For Yangian symmetry a similar
statement is less clear.)  

One particularly bold future goal is to lift 
the amplituhedron~\cite{Arkani-Hamed:2013jha} paradigm from the planar 
sector to the nonplanar sector of $\NeqFour$ SYM.  
The amplituhedron provides a geometric picture of the planar 
integrand where all standard physical principles like locality
and unitarity are derived. In such a picture, traditional ways of organizing
amplitudes, be it via Feynman diagrams, unitarity cuts, or even
on-shell diagrams, are consequences of amplituhedron
geometry, rather than a priori organizational principles.
The amplituhedron reverses traditional logic, as logarithmic singularities 
and dual conformal symmetry, rather than locality and unitarity,
are fundamental inputs into the definition of the amplituhedron. 
The definition then invokes intuitive 
geometric ideas about the inside of a projective triangle,
generalized to the more complicated setting of Grassmannian geometry.

We would like to carry this geometric picture over to the nonplanar
sector.  However, a lack of global variables limits us to demanding
that the amplitude be a sum of local integrals.  This already imposes
locality and some unitarity constraints.  Nevertheless, after imposing
special analytic structures on the basis integrals --- unit logarithmic
singularities with no poles at infinity --- one can extract the
``remaining'' geometric information. Motivated by the discussion in
Ref.~\cite{Arkani-Hamed:2014dca}, in this paper we conjectured that
this remaining information is a set of {\it zero conditions},
i.e.~cuts on which the amplitude vanishes. This is exactly the
statement which we successfully carried over to the nonplanar sector
and tested in examples in \sect{Zeros}.  Here we propose that
after constructing a pure integrand basis, zero conditions are
sufficient for finding the complete amplitude.

This provides nontrivial evidence that an amplituhedron-like
construction might very well exist beyond the planar limit for
amplitudes in ${\cal N}=4$ SYM theory. However there are still 
many obstacles including, among other things, a choice of good variables and
a geometric space in which nonplanar scattering amplitudes are defined as volumes.
If such a nonplanar amplituhedron really exists, it would be phrased in terms of
very interesting mathematical structures going beyond those of the planar amplituhedron.

If our zero condition conjecture indeed holds, how might it extend to
other theories? The most naive possibility is that $\NeqFour$ SYM
amplitudes are the most constrained amplitudes and so need no
inhomogeneous conditions except for overall normalization, while
amplitudes in other theories, with less supersymmetry for example, do
need additional inhomogeneous information. Even in such theories the
zero conditions would still constrain the amplitudes, and it would be
interesting to see which and how many additional inhomogeneous conditions are
required to completely determine the amplitudes.

It may be possible to link the $\NeqFour$ SYM results we presented
here directly to identical helicity amplitudes in quantum
chromodynamics (QCD) via dimension shifting
relations~\cite{DimShift,DimShift2}. These relations were recently
employed to aid in the construction of a representation of the
two-loop five-point identical helicity QCD amplitude where
the duality between color and kinematics holds~\cite{OConnellAllPlus}.
It should also be possible to find a new representation of the identical
helicity QCD amplitude in terms of the $\NeqFour$ SYM representation
we gave here.

Another line of research is to concentrate on individual integrals
rather than on the full amplitude.  After integration, integrands with
only logarithmic singularities are expected to have uniform maximum
transcendental weight at the loop order of the
integrand~\cite{KLOV}. This provides a nice connection between
properties of the integrand and conjectured properties of final
integrated amplitudes.  On the practical level, having a good basis of
master integrals under integral reduction is important for many
problems, including applications to phenomenology.  As explained in
Refs.~\cite{Henn,HennSmirnov}, uniformly transcendental integrals obey
relatively simple differential equations, making them easy to work
with \cite{Caron-Huot:2014lda, Henn:2014qga}. This also makes our
basis of pure integrands useful for five-point scattering in NNLO
QCD. For a recent discussion of the planar case see
Ref.~\cite{Gehrmann:2015bfy}.

As already noted in Ref.~\cite{ThreeLoopPaper}, the types of
gauge-theory results described here can have a direct bearing on
issues in quantum gravity, through the double-copy relation of
Yang-Mills theories to gravity~\cite{BCJLoop}.  We expect that
developing a better understanding of the nonplanar sector of
$\NeqFour$ SYM will aid our ability to construct corresponding gravity
amplitudes, where no natural separation of planar and nonplanar
contributions exists.

In summary, we have presented evidence that nonplanar integrands of
$\NeqFour$ SYM share important analytic structure with planar ones.  We have also
presented evidence for a geometric structure similar to the
amplituhedron~\cite{Arkani-Hamed:2013jha} based on the idea that such
a structure implies that zero conditions are sufficient to fix the
amplitude, up to an overall normalization.  While there is much more
to do, these results suggest that the full theory
has structure at least as rich as the planar theory.

\subsection*{Acknowledgments}
This work was supported in part by the US Department of Energy under
Award Number DE-{S}C0009937 and DE-SC0011632.  J.~T. is supported in
part by the David and Ellen Lee Postdoctoral Scholarship.  E.~H. is
supported in part by a Dominic Orr Graduate Fellowship.  Z.~B. is
grateful to the Simons Foundation for support.


\bibliographystyle{JHEP}
\newpage
\phantomsection         
            \bibliography{References}

\providecommand{\href}[2]{#2}\begingroup\raggedright\begin{thebibliography}{10}

\bibitem{DualConformalMagic}
J.~Drummond, J.~Henn, V.~Smirnov, and E.~Sokatchev, {\it {Magic Identities for
  Conformal Four-Point Integrals}},  {\em JHEP} {\bf 0701} (2007) 064,
  [\href{http://arxiv.org/abs/hep-th/0607160}{{\tt hep-th/0607160}}].

\bibitem{Alday:2007hr}
L.~F. Alday and J.~M. Maldacena, {\it {Gluon Scattering Amplitudes at Strong
  Coupling}},  {\em JHEP} {\bf 0706} (2007) 064,
  [\href{http://arxiv.org/abs/0705.0303}{{\tt arXiv:0705.0303}}].

\bibitem{Drummond:2008vq}
J.~Drummond, J.~Henn, G.~Korchemsky, and E.~Sokatchev, {\it {Dual
  Superconformal Symmetry of Scattering Amplitudes in $N=4$ Super-Yang-Mills
  Theory}},  {\em Nucl. Phys.} {\bf B828} (2010) 317--374,
  [\href{http://arxiv.org/abs/0807.1095}{{\tt arXiv:0807.1095}}].

\bibitem{Drummond:2009fd}
J.~M. Drummond, J.~M. Henn, and J.~Plefka, {\it {Yangian Symmetry of Scattering
  Amplitudes in $N=4$ Super Yang-Mills Theory}},  {\em JHEP} {\bf 0905} (2009)
  046, [\href{http://arxiv.org/abs/0902.2987}{{\tt arXiv:0902.2987}}].

\bibitem{Beisert:2003yb}
N.~Beisert and M.~Staudacher, {\it {The $N=4$ SYM Integrable Super Spin
  Chain}},  {\em Nucl. Phys.} {\bf B670} (2003) 439--463,
  [\href{http://arxiv.org/abs/hep-th/0307042}{{\tt hep-th/0307042}}].

\bibitem{Beisert:2006ez}
N.~Beisert, B.~Eden, and M.~Staudacher, {\it {Transcendentality and Crossing}},
   {\em J. Stat. Mech.} {\bf 0701} (2007) P01021,
  [\href{http://arxiv.org/abs/hep-th/0610251}{{\tt hep-th/0610251}}].

\bibitem{Drummond:2007aua}
J.~Drummond, G.~Korchemsky, and E.~Sokatchev, {\it {Conformal Properties of
  Four-Gluon Planar Amplitudes and Wilson Loops}},  {\em Nucl. Phys.} {\bf
  B795} (2008) 385--408, [\href{http://arxiv.org/abs/0707.0243}{{\tt
  arXiv:0707.0243}}].

\bibitem{Brandhuber:2007yx}
A.~Brandhuber, P.~Heslop, and G.~Travaglini, {\it {MHV Amplitudes in $N=4$
  Super Yang-Mills and Wilson Loops}},  {\em Nucl. Phys.} {\bf B794} (2008)
  231--243, [\href{http://arxiv.org/abs/0707.1153}{{\tt arXiv:0707.1153}}].

\bibitem{DualConfWI}
J.~Drummond, J.~Henn, G.~Korchemsky, and E.~Sokatchev, {\it {Conformal Ward
  Identities for Wilson loops and a Test of the Duality with Gluon
  Amplitudes}},  {\em Nucl. Phys.} {\bf B826} (2010) 337--364,
  [\href{http://arxiv.org/abs/0712.1223}{{\tt arXiv:0712.1223}}].

\bibitem{Mason:2010yk}
L.~Mason and D.~Skinner, {\it {The Complete Planar S-matrix of $N=4$ SYM as a
  Wilson Loop in Twistor Space}},  {\em JHEP} {\bf 1012} (2010) 018,
  [\href{http://arxiv.org/abs/1009.2225}{{\tt arXiv:1009.2225}}].

\bibitem{CaronHuot:2010ek}
S.~Caron-Huot, {\it {Notes on the Scattering Amplitude / Wilson Loop Duality}},
   {\em JHEP} {\bf 1107} (2011) 058,
  [\href{http://arxiv.org/abs/1010.1167}{{\tt arXiv:1010.1167}}].

\bibitem{Alday:2010zy}
L.~F. Alday, B.~Eden, G.~P. Korchemsky, J.~Maldacena, and E.~Sokatchev, {\it
  {From Correlation Functions to Wilson Loops}},  {\em JHEP} {\bf 1109} (2011)
  123, [\href{http://arxiv.org/abs/1007.3243}{{\tt arXiv:1007.3243}}].

\bibitem{Basso:2013vsa}
B.~Basso, A.~Sever, and P.~Vieira, {\it {Spacetime and Flux Tube S-Matrices at
  Finite Coupling for $N=4$ Supersymmetric Yang-Mills Theory}},  {\em Phys.
  Rev. Lett.} {\bf 111} (2013), no.~9 091602,
  [\href{http://arxiv.org/abs/1303.1396}{{\tt arXiv:1303.1396}}].

\bibitem{Basso:2014hfa}
B.~Basso, J.~Caetano, L.~Cordova, A.~Sever, and P.~Vieira, {\it {OPE for all
  Helicity Amplitudes}},  {\em JHEP} {\bf 08} (2015) 018,
  [\href{http://arxiv.org/abs/1412.1132}{{\tt arXiv:1412.1132}}].

\bibitem{Basso:2015uxa}
B.~Basso, A.~Sever, and P.~Vieira, {\it {Hexagonal Wilson Loops in Planar
  $\mathcal{N}=4$ SYM Theory at Finite Coupling}},
  \href{http://arxiv.org/abs/1508.0304}{{\tt arXiv:1508.0304}}.

\bibitem{Dixon:2013eka}
L.~J. Dixon, J.~M. Drummond, M.~von Hippel, and J.~Pennington, {\it {Hexagon
  Functions and the Three-Loop Remainder Function}},  {\em JHEP} {\bf 12}
  (2013) 049, [\href{http://arxiv.org/abs/1308.2276}{{\tt arXiv:1308.2276}}].

\bibitem{Dixon:2014iba}
L.~J. Dixon and M.~von Hippel, {\it {Bootstrapping an NMHV Amplitude through
  Three Loops}},  {\em JHEP} {\bf 10} (2014) 065,
  [\href{http://arxiv.org/abs/1408.1505}{{\tt arXiv:1408.1505}}].

\bibitem{Dixon:2015iva}
L.~J. Dixon, M.~von Hippel, and A.~J. McLeod, {\it {The Four-Loop Six-Gluon
  NMHV Ratio Function}},  \href{http://arxiv.org/abs/1509.0812}{{\tt
  arXiv:1509.0812}}.

\bibitem{Goncharov:2010jf}
A.~B. Goncharov, M.~Spradlin, C.~Vergu, and A.~Volovich, {\it {Classical
  Polylogarithms for Amplitudes and Wilson Loops}},  {\em Phys. Rev. Lett.}
  {\bf 105} (2010) 151605, [\href{http://arxiv.org/abs/1006.5703}{{\tt
  arXiv:1006.5703}}].

\bibitem{Golden:2013xva}
J.~Golden, A.~B. Goncharov, M.~Spradlin, C.~Vergu, and A.~Volovich, {\it
  {Motivic Amplitudes and Cluster Coordinates}},  {\em JHEP} {\bf 1401} (2014)
  091, [\href{http://arxiv.org/abs/1305.1617}{{\tt arXiv:1305.1617}}].

\bibitem{Drummond:2014ffa}
J.~M. Drummond, G.~Papathanasiou, and M.~Spradlin, {\it {A Symbol of
  Uniqueness: The Cluster Bootstrap for the 3-Loop MHV Heptagon}},  {\em JHEP}
  {\bf 03} (2015) 072, [\href{http://arxiv.org/abs/1412.3763}{{\tt
  arXiv:1412.3763}}].

\bibitem{Parker:2015cia}
D.~Parker, A.~Scherlis, M.~Spradlin, and A.~Volovich, {\it {Hedgehog Bases for
  A$_{n}$ Cluster Polylogarithms and an Application to Six-point Amplitudes}},
  {\em JHEP} {\bf 11} (2015) 136, [\href{http://arxiv.org/abs/1507.0195}{{\tt
  arXiv:1507.0195}}].

\bibitem{OnshellDiagrams}
N.~Arkani-Hamed, J.~L. Bourjaily, F.~Cachazo, A.~B. Goncharov, A.~Postnikov,
  and J.~Trnka, {\it {Scattering Amplitudes and the Positive Grassmannian}},
  \href{http://arxiv.org/abs/1212.5605}{{\tt arXiv:1212.5605}}.

\bibitem{ArkaniHamed:2009dn}
N.~Arkani-Hamed, F.~Cachazo, C.~Cheung, and J.~Kaplan, {\it {A Duality For The
  S Matrix}},  {\em JHEP} {\bf 1003} (2010) 020,
  [\href{http://arxiv.org/abs/0907.5418}{{\tt arXiv:0907.5418}}].

\bibitem{ArkaniHamed:2009vw}
N.~Arkani-Hamed, F.~Cachazo, and C.~Cheung, {\it {The Grassmannian Origin Of
  Dual Superconformal Invariance}},  {\em JHEP} {\bf 1003} (2010) 036,
  [\href{http://arxiv.org/abs/0909.0483}{{\tt arXiv:0909.0483}}].

\bibitem{Mason:2009qx}
L.~Mason and D.~Skinner, {\it {Dual Superconformal Invariance, Momentum
  Twistors and Grassmannians}},  {\em JHEP} {\bf 0911} (2009) 045,
  [\href{http://arxiv.org/abs/0909.0250}{{\tt arXiv:0909.0250}}].

\bibitem{ArkaniHamed:2009dg}
N.~Arkani-Hamed, J.~Bourjaily, F.~Cachazo, and J.~Trnka, {\it {Unification of
  Residues and Grassmannian Dualities}},  {\em JHEP} {\bf 1101} (2011) 049,
  [\href{http://arxiv.org/abs/0912.4912}{{\tt arXiv:0912.4912}}].

\bibitem{ArkaniHamed:2009sx}
N.~Arkani-Hamed, J.~Bourjaily, F.~Cachazo, and J.~Trnka, {\it {Local Spacetime
  Physics from the Grassmannian}},  {\em JHEP} {\bf 1101} (2011) 108,
  [\href{http://arxiv.org/abs/0912.3249}{{\tt arXiv:0912.3249}}].

\bibitem{ArkaniHamed:2010kv}
N.~Arkani-Hamed, J.~L. Bourjaily, F.~Cachazo, S.~Caron-Huot, and J.~Trnka, {\it
  {The All-Loop Integrand For Scattering Amplitudes in Planar $N=4$ SYM}},
  {\em JHEP} {\bf 1101} (2011) 041, [\href{http://arxiv.org/abs/1008.2958}{{\tt
  arXiv:1008.2958}}].

\bibitem{Huang:2013owa}
Y.-T. Huang and C.~Wen, {\it {ABJM Amplitudes and the Positive Orthogonal
  Grassmannian}},  {\em JHEP} {\bf 1402} (2014) 104,
  [\href{http://arxiv.org/abs/1309.3252}{{\tt arXiv:1309.3252}}].

\bibitem{Huang:2014xza}
Y.-t. Huang, C.~Wen, and D.~Xie, {\it {The Positive Orthogonal Grassmannian and
  Loop Amplitudes of ABJM}},  \href{http://arxiv.org/abs/1402.1479}{{\tt
  arXiv:1402.1479}}.

\bibitem{Kim:2014hva}
J.~Kim and S.~Lee, {\it {Positroid Stratification of Orthogonal Grassmannian
  and ABJM Amplitudes}},  {\em JHEP} {\bf 1409} (2014) 085,
  [\href{http://arxiv.org/abs/1402.1119}{{\tt arXiv:1402.1119}}].

\bibitem{ElvangGrassmannian}
H.~Elvang, Y.-t. Huang, C.~Keeler, T.~Lam, T.~M. Olson, et~al., {\it
  {Grassmannians for Scattering Amplitudes in 4d $\mathcal{N}=4$ SYM and 3d
  ABJM}},  \href{http://arxiv.org/abs/1410.0621}{{\tt arXiv:1410.0621}}.

\bibitem{Arkani-Hamed:2013jha}
N.~Arkani-Hamed and J.~Trnka, {\it {The Amplituhedron}},  {\em JHEP} {\bf 10}
  (2014) 030, [\href{http://arxiv.org/abs/1312.2007}{{\tt arXiv:1312.2007}}].

\bibitem{Arkani-Hamed:2013kca}
N.~Arkani-Hamed and J.~Trnka, {\it {Into the Amplituhedron}},  {\em JHEP} {\bf
  12} (2014) 182, [\href{http://arxiv.org/abs/1312.7878}{{\tt
  arXiv:1312.7878}}].

\bibitem{Franco:2014csa}
S.~Franco, D.~Galloni, A.~Mariotti, and J.~Trnka, {\it {Anatomy of the
  Amplituhedron}},  \href{http://arxiv.org/abs/1408.3410}{{\tt
  arXiv:1408.3410}}.

\bibitem{Bai:2014cna}
Y.~Bai and S.~He, {\it {The Amplituhedron from Momentum Twistor Diagrams}},
  {\em JHEP} {\bf 02} (2015) 065, [\href{http://arxiv.org/abs/1408.2459}{{\tt
  arXiv:1408.2459}}].

\bibitem{Arkani-Hamed:2014dca}
N.~Arkani-Hamed, A.~Hodges, and J.~Trnka, {\it {Positive Amplitudes In The
  Amplituhedron}},  {\em JHEP} {\bf 08} (2015) 030,
  [\href{http://arxiv.org/abs/1412.8478}{{\tt arXiv:1412.8478}}].

\bibitem{Lam:2014jda}
T.~Lam, {\it {Amplituhedron Cells and Stanley Symmetric Functions}},
  \href{http://arxiv.org/abs/1408.5531}{{\tt arXiv:1408.5531}}.

\bibitem{Bai:2015qoa}
Y.~Bai, S.~He, and T.~Lam, {\it {The Amplituhedron and the One-loop
  Grassmannian Measure}},  \href{http://arxiv.org/abs/1510.0355}{{\tt
  arXiv:1510.0355}}.

\bibitem{Ferro:2015grk}
L.~Ferro, T.~Lukowski, A.~Orta, and M.~Parisi, {\it {Towards the Amplituhedron
  Volume}},  \href{http://arxiv.org/abs/1512.0495}{{\tt arXiv:1512.0495}}.

\bibitem{Lusztig}
G.~Lusztig, {\it Total positivity in partial flag manifolds},  {\em Represent.
  Theory} {\bf 2} (1998) 70--78.

\bibitem{postnikov}
A.~{Postnikov}, {\it {Total Positivity, Grassmannians, and Networks}},  {\em
  ArXiv Mathematics e-prints} (Sept., 2006)
  [\href{http://arxiv.org/abs/math/0609764}{{\tt math/0609764}}].

\bibitem{postnikov2}
A.~{Postnikov}, D.~{Speyer}, and L.~{Williams}, {\it {Matching Polytopes, Toric
  Geometry, and the Non-negative Part of the Grassmannian}},  {\em ArXiv
  e-prints} (June, 2007) [\href{http://arxiv.org/abs/0706.2501}{{\tt
  arXiv:0706.2501}}].

\bibitem{lauren}
L.~K. {Williams}, {\it {Enumeration of Totally Positive Grassmann Cells}},
  {\em ArXiv Mathematics e-prints} (July, 2003)
  [\href{http://arxiv.org/abs/math/0307271}{{\tt math/0307271}}].

\bibitem{goncharov}
A.~B. {Goncharov} and R.~{Kenyon}, {\it {Dimers and Cluster Integrable
  Systems}},  {\em ArXiv e-prints} (July, 2011)
  [\href{http://arxiv.org/abs/1107.5588}{{\tt arXiv:1107.5588}}].

\bibitem{knutson}
A.~{Knutson}, T.~{Lam}, and D.~{Speyer}, {\it {Positroid Varieties: Juggling
  and Geometry}},  {\em ArXiv e-prints} (Nov., 2011)
  [\href{http://arxiv.org/abs/1111.3660}{{\tt arXiv:1111.3660}}].

\bibitem{BCJ}
Z.~Bern, J.~Carrasco, and H.~Johansson, {\it {New Relations for Gauge-Theory
  Amplitudes}},  {\em Phys. Rev.} {\bf D78} (2008) 085011,
  [\href{http://arxiv.org/abs/0805.3993}{{\tt arXiv:0805.3993}}].

\bibitem{BCJLoop}
Z.~Bern, J.~J.~M. Carrasco, and H.~Johansson, {\it {Perturbative Quantum
  Gravity as a Double Copy of Gauge Theory}},  {\em Phys. Rev. Lett.} {\bf 105}
  (2010) 061602, [\href{http://arxiv.org/abs/1004.0476}{{\tt
  arXiv:1004.0476}}].

\bibitem{Log}
N.~Arkani-Hamed, J.~L. Bourjaily, F.~Cachazo, and J.~Trnka, {\it {Singularity
  Structure of Maximally Supersymmetric Scattering Amplitudes}},  {\em Phys.
  Rev. Lett.} {\bf 113} (2014), no.~26 261603,
  [\href{http://arxiv.org/abs/1410.0354}{{\tt arXiv:1410.0354}}].

\bibitem{ThreeLoopPaper}
Z.~Bern, E.~Herrmann, S.~Litsey, J.~Stankowicz, and J.~Trnka, {\it {Logarithmic
  Singularities and Maximally Supersymmetric Amplitudes}},  {\em JHEP} {\bf 06}
  (2015) 202, [\href{http://arxiv.org/abs/1412.8584}{{\tt arXiv:1412.8584}}].

\bibitem{Parke:1986gb}
S.~J. Parke and T.~R. Taylor, {\it {An Amplitude for $n$ Gluon Scattering}},
  {\em Phys. Rev. Lett.} {\bf 56} (1986) 2459.

\bibitem{Mangano:1987xk}
M.~L. Mangano, S.~J. Parke, and Z.~Xu, {\it {Duality and Multi-Gluon
  Scattering}},  {\em Nucl. Phys.} {\bf B298} (1988) 653.

\bibitem{NonPlanarOnshellToAppear}
N.~Arkani-Hamed, J.~L. Bourjaily, F.~Cachazo, A.~Postnikov, and J.~Trnka, {\it
  {On-Shell Structures of MHV Amplitudes Beyond the Planar Limit}},  {\em JHEP}
  {\bf 06} (2015) 179, [\href{http://arxiv.org/abs/1412.8475}{{\tt
  arXiv:1412.8475}}].

\bibitem{WithLance}
L.~J. Dixon, A.~J. McLeod, J.~Trnka, and M.~von Hippel
  \href{http://arxiv.org/abs/\hskip -.17 cm , to appear}{{\tt \hskip -.17 cm ,
  to appear}}.

\bibitem{GravityThreeLoop}
Z.~Bern, J.~Carrasco, L.~J. Dixon, H.~Johansson, D.~Kosower, and R.~Roiban,
  {\it {Three-Loop Superfiniteness of $N=8$ Supergravity}},  {\em Phys. Rev.
  Lett.} {\bf 98} (2007) 161303,
  [\href{http://arxiv.org/abs/hep-th/0702112}{{\tt hep-th/0702112}}].

\bibitem{HenrikJJ}
J.~J.~M. Carrasco and H.~Johansson, {\it {Five-Point Amplitudes in N=4
  Super-Yang-Mills Theory and $N=8$ Supergravity}},  {\em Phys. Rev.} {\bf D85}
  (2012) 025006, [\href{http://arxiv.org/abs/1106.4711}{{\tt
  arXiv:1106.4711}}].

\bibitem{DDimCheck}
Z.~Bern, J.~J. Carrasco, T.~Dennen, Y.-t. Huang, and H.~Ita, {\it {Generalized
  Unitarity and Six-Dimensional Helicity}},  {\em Phys. Rev.} {\bf D83} (2011)
  085022, [\href{http://arxiv.org/abs/1010.0494}{{\tt arXiv:1010.0494}}].

\bibitem{Bern:2008ap}
Z.~Bern, L.~Dixon, D.~Kosower, R.~Roiban, M.~Spradlin, et~al., {\it {The
  Two-Loop Six-Gluon MHV Amplitude in Maximally Supersymmetric Yang-Mills
  Theory}},  {\em Phys. Rev.} {\bf D78} (2008) 045007,
  [\href{http://arxiv.org/abs/0803.1465}{{\tt arXiv:0803.1465}}].

\bibitem{Bern:1994zx}
Z.~Bern, L.~J. Dixon, D.~C. Dunbar, and D.~A. Kosower, {\it {One-Loop $n$-Point
  Gauge-Theory Amplitudes, Unitarity and Collinear Limits}},  {\em Nucl. Phys.}
  {\bf B425} (1994) 217--260, [\href{http://arxiv.org/abs/hep-ph/9403226}{{\tt
  hep-ph/9403226}}].

\bibitem{Bern:1994cg}
Z.~Bern, L.~J. Dixon, D.~C. Dunbar, and D.~A. Kosower, {\it {Fusing Gauge
  Theory Tree amplitudes into Loop Amplitudes}},  {\em Nucl. Phys.} {\bf B435}
  (1995) 59--101, [\href{http://arxiv.org/abs/hep-ph/9409265}{{\tt
  hep-ph/9409265}}].

\bibitem{MaximalCuts}
Z.~Bern, J.~Carrasco, H.~Johansson, and D.~Kosower, {\it {Maximally
  Supersymmetric Planar Yang-Mills Amplitudes at Five Loops}},  {\em Phys.
  Rev.} {\bf D76} (2007) 125020, [\href{http://arxiv.org/abs/0705.1864}{{\tt
  arXiv:0705.1864}}].

\bibitem{BCFW}
R.~Britto, F.~Cachazo, B.~Feng, and E.~Witten, {\it {Direct Proof of Tree-level
  Recursion Relation in Yang-Mills Theory}},  {\em Phys. Rev. Lett.} {\bf 94}
  (2005) 181602, [\href{http://arxiv.org/abs/hep-th/0501052}{{\tt
  hep-th/0501052}}].

\bibitem{Hodges}
A.~Hodges, {\it {Eliminating Spurious Poles from Gauge-Theoretic Amplitudes}},
  {\em JHEP} {\bf 1305} (2013) 135, [\href{http://arxiv.org/abs/0905.1473}{{\tt
  arXiv:0905.1473}}].

\bibitem{Cachazo:2008vp}
F.~Cachazo, {\it {Sharpening the Leading Singularity}},
  \href{http://arxiv.org/abs/0803.1988}{{\tt arXiv:0803.1988}}.

\bibitem{Drummond:2007bm}
J.~M. Drummond, J.~Henn, G.~P. Korchemsky, and E.~Sokatchev, {\it {The Hexagon
  Wilson Loop and the BDS Ansatz for the Six-Gluon Amplitude}},  {\em Phys.
  Lett.} {\bf B662} (2008) 456--460,
  [\href{http://arxiv.org/abs/0712.4138}{{\tt arXiv:0712.4138}}].

\bibitem{ArkaniHamed:2010gh}
N.~Arkani-Hamed, J.~L. Bourjaily, F.~Cachazo, and J.~Trnka, {\it {Local
  Integrals for Planar Scattering Amplitudes}},  {\em JHEP} {\bf 06} (2012)
  125, [\href{http://arxiv.org/abs/1012.6032}{{\tt arXiv:1012.6032}}].

\bibitem{Witten:2003nn}
E.~Witten, {\it {Perturbative Gauge Theory as a String Theory in Twistor
  Space}},  {\em Commun.Math.Phys.} {\bf 252} (2004) 189--258,
  [\href{http://arxiv.org/abs/hep-th/0312171}{{\tt hep-th/0312171}}].

\bibitem{Bourjaily:2013mma}
J.~L. Bourjaily, S.~Caron-Huot, and J.~Trnka, {\it {Dual-Conformal
  Regularization of Infrared Loop Divergences and the Chiral Box Expansion}},
  {\em JHEP} {\bf 01} (2015) 001, [\href{http://arxiv.org/abs/1303.4734}{{\tt
  arXiv:1303.4734}}].

\bibitem{Bourjaily:2015jna}
J.~L. Bourjaily and J.~Trnka, {\it {Local Integrand Representations of All
  Two-Loop Amplitudes in Planar SYM}},  {\em JHEP} {\bf 08} (2015) 119,
  [\href{http://arxiv.org/abs/1505.0588}{{\tt arXiv:1505.0588}}].

\bibitem{BRY}
Z.~Bern, J.~Rozowsky, and B.~Yan, {\it {Two-Loop Four-Gluon Amplitudes in $N=4$
  SuperYang-Mills}},  {\em Phys.Lett.} {\bf B401} (1997) 273--282,
  [\href{http://arxiv.org/abs/hep-ph/9702424}{{\tt hep-ph/9702424}}].

\bibitem{KleissKuijf}
R.~Kleiss and H.~Kuijf, {\it {Multi-Gluon Cross-sections and Five-Jet
  Production at Hadron Colliders}},  {\em Nucl. Phys.} {\bf B312} (1989) 616.

\bibitem{BDDPR}
Z.~Bern, L.~J. Dixon, D.~Dunbar, M.~Perelstein, and J.~Rozowsky, {\it {On the
  Relationship Between Yang-Mills Theory and Gravity and its Implication for
  Ultraviolet Divergences}},  {\em Nucl. Phys.} {\bf B530} (1998) 401--456,
  [\href{http://arxiv.org/abs/hep-th/9802162}{{\tt hep-th/9802162}}].

\bibitem{ColorKinematics}
Z.~Bern, J.~Carrasco, L.~Dixon, H.~Johansson, and R.~Roiban, {\it {Simplifying
  Multiloop Integrands and Ultraviolet Divergences of Gauge Theory and Gravity
  Amplitudes}},  {\em Phys. Rev.} {\bf D85} (2012) 105014,
  [\href{http://arxiv.org/abs/1201.5366}{{\tt arXiv:1201.5366}}].

\bibitem{Mangano:1990by}
M.~L. Mangano and S.~J. Parke, {\it {Multiparton Amplitudes in Gauge
  Theories}},  {\em Phys. Rept.} {\bf 200} (1991) 301--367,
  [\href{http://arxiv.org/abs/hep-th/0509223}{{\tt hep-th/0509223}}].

\bibitem{Franco:2015rma}
S.~Franco, D.~Galloni, B.~Penante, and C.~Wen, {\it {Non-Planar On-Shell
  Diagrams}},  {\em JHEP} {\bf 06} (2015) 199,
  [\href{http://arxiv.org/abs/1502.0203}{{\tt arXiv:1502.0203}}].

\bibitem{DimShift}
Z.~Bern, L.~J. Dixon, D.~C. Dunbar, and D.~A. Kosower, {\it {One-Loop Selfdual
  and $N=4$ Super-Yang-Mills}},  {\em Phys. Lett.} {\bf B394} (1997) 105--115,
  [\href{http://arxiv.org/abs/hep-th/9611127}{{\tt hep-th/9611127}}].

\bibitem{DimShift2}
Z.~Bern, L.~J. Dixon, M.~Perelstein, and J.~S. Rozowsky, {\it {Multileg One
  Loop Gravity Amplitudes from Gauge Theory}},  {\em Nucl. Phys.} {\bf B546}
  (1999) 423--479, [\href{http://arxiv.org/abs/hep-th/9811140}{{\tt
  hep-th/9811140}}].

\bibitem{OConnellAllPlus}
G.~Mogull and D.~O'Connell, {\it {Overcoming Obstacles to Colour-Kinematics
  Duality at Two Loops}},  \href{http://arxiv.org/abs/1511.0665}{{\tt
  arXiv:1511.0665}}.

\bibitem{KLOV}
A.~Kotikov, L.~Lipatov, A.~Onishchenko, and V.~Velizhanin, {\it {Three-Loop
  Universal Anomalous Dimension of the Wilson Operators in $N=4$ SUSY
  Yang-Mills Model}},  {\em Phys.Lett.} {\bf B595} (2004) 521--529,
  [\href{http://arxiv.org/abs/hep-th/0404092}{{\tt hep-th/0404092}}].

\bibitem{Henn}
J.~M. Henn, {\it {Multiloop Integrals in Dimensional Regularization Made
  Simple}},  {\em Phys. Rev. Lett.} {\bf 110} (2013), no.~25 251601,
  [\href{http://arxiv.org/abs/1304.1806}{{\tt arXiv:1304.1806}}].

\bibitem{HennSmirnov}
J.~M. Henn, A.~V. Smirnov, and V.~A. Smirnov, {\it {Analytic Results for Planar
  Three-Loop Four-Point Integrals from a Knizhnik-Zamolodchikov Equation}},
  {\em JHEP} {\bf 1307} (2013) 128, [\href{http://arxiv.org/abs/1306.2799}{{\tt
  arXiv:1306.2799}}].

\bibitem{Caron-Huot:2014lda}
S.~Caron-Huot and J.~M. Henn, {\it {Iterative Structure of Finite Loop
  Integrals}},  {\em JHEP} {\bf 06} (2014) 114,
  [\href{http://arxiv.org/abs/1404.2922}{{\tt arXiv:1404.2922}}].

\bibitem{Henn:2014qga}
J.~M. Henn, {\it {Lectures on Differential Equations for Feynman Integrals}},
  \href{http://arxiv.org/abs/1412.2296}{{\tt arXiv:1412.2296}}.

\bibitem{Gehrmann:2015bfy}
T.~Gehrmann, J.~M. Henn, and N.~A. Lo~Presti, {\it {Analytic Form of the
  Two-Loop Planar Five-Gluon All-Plus-Helicity Amplitude in QCD}},
  \href{http://arxiv.org/abs/1511.0540}{{\tt arXiv:1511.0540}}.

\end{thebibliography}\endgroup
            \clearpage

\end{document}